\documentclass{article}
\usepackage{graphicx} 
\usepackage{authblk} 

\usepackage[page,toc,titletoc,title]{appendix}
\usepackage{tocloft}

\title{Coalitions, Stability, and Majority Jackpots: \\ Frameworks for Unifying Majoritarian and Proportional Allocation in Multiwinner Elections}

\author[1]{Kit Scriven\thanks{This work was primarily completed when the author was at UC Berkeley.}}
\author[2]{Nitin Kohli}

\affil[1]{London School of Economics}
\affil[2]{UC Berkeley}

\date{}

\usepackage{graphicx, 
mathtools,
amsmath,
comment,
algorithm,
algpseudocode,
amsthm,
amssymb,
dsfont,
tikz, pgfplots,
nicefrac}
\usepackage{stackengine}
\usepackage[perpage, symbol*]{footmisc}
\usepackage{newpxtext,newpxmath}
\usetikzlibrary{shapes.geometric}
\usetikzlibrary{shapes.misc}

\usepackage{enumitem}

\usepackage[margin=1.25in]{geometry}

\newcommand{\floor}[1]{\left\lfloor #1 \right\rfloor}
\newcommand{\ceil}[1]{\left\lceil #1 \right\rceil}
\newcommand{\set}[1]{\left\{ #1\right\}}
\newcommand{\paren}[1]{\left( #1\right)}
\newcommand{\abs}[1]{\left| #1\right|}
\newcommand{\norm}[1]{\left|\left| #1\right|\right|}
\newcommand{\bb}[1]{\mathds{#1}}
\newcommand{\bm}[1]{\mathbf{#1}}

\newcommand{\R}{\bb{R}}

\newcommand{\Z}{\bb{Z}}
\newcommand{\bigO}{\mathcal{O}}
\newcommand{\Rule}{\mathcal{R}}

\newcommand{\pty}[1]{\textit{#1}}

\newcommand{\lel}{\lessdot}
\newcommand{\gel}{\gtrdot}
\newcommand{\geqdot}{\stackunder[1pt]{$\gel$}{$-$}}
\newcommand{\leqdot}{\stackunder[1pt]{$\lel$}{$-$}}
\DeclareMathOperator{\topp}{top}

\DeclareMathOperator{\argmax}{argmax}

\usepackage{xcolor}
\usepackage[normalem]{ulem}

\newcommand\kitxout{\bgroup\markoverwith{\textcolor{teal}{\rule[0.5ex]{2pt}{0.4pt}}}\ULon}
\newcommand\nitinxout{\bgroup\markoverwith{\textcolor{blue}{\rule[0.5ex]{2pt}{0.4pt}}}\ULon}

\RequirePackage[colorlinks,citecolor=blue,urlcolor=blue,linkcolor=.]{hyperref}
\usepackage[numbers]{natbib}

\newtheorem{lem}{Lemma}

\newtheorem{thm}{Theorem}
\newtheorem{cor}{Corollary}
\newtheorem{prop}{Proposition}

\newtheorem{defn}{Definition}

\newtheorem{examplex}{Example}
\newtheorem*{examplenonum}{Example}

\newenvironment{ex}
  {\pushQED{\qed}\examplex}
  {\popQED\endexamplex}
\newenvironment{ex*}
  {\pushQED{\qed}\examplenonum}
  {\popQED\endexamplenonum}

\pgfplotsset{
    myaxis/.style={
        xlabel={$x$}, ylabel={$y$},
        xmin=0.01, xmax=1,
        ymin=0, ymax=1.1,
        axis lines=left,
        axis on top,
        grid=major,
        width=8cm, height=6cm,
        title style={at={(0.5,1.05)}, anchor=south, font=\bfseries},
        tick style={draw=none}
    }
}
\usetikzlibrary{patterns}
\pgfplotsset{compat=1.18}
\pgfplotsset{
    myaxis/.style={
        xlabel={$p$}, ylabel={$r$},
        xmin=0.01, xmax=1,
        ymin=0, ymax=1.1,
        xticklabels = { $\frac{1}{4}$, $\frac{1}{3}$, $\frac{1}{2}$,1},
        xtick = {1/4,1/3,1/2,1},
        yticklabels = { $\nicefrac{1}{4}$, $\nicefrac{1}{3}$, $\nicefrac{1}{2}$,1},
        ytick = {1/4,1/3,1/2,1},
        ylabel style={rotate=-90},
        axis lines=left,
        axis on top,
        grid=major,
        width=8cm, height=6cm,
        title style={at={(0.5,1.05)}, anchor=south, font=\bfseries},
        tick style={draw=none}
    }
}

\newcommand{\plotRegion}[2]{%
    \foreach \n in {1,...,100} {%
        \pgfmathsetmacro\xL{1/(\n+1)}
        \pgfmathsetmacro\xR{1/\n}
        \pgfmathtruncatemacro\yInt{floor(#1/\n)}
        \ifnum\yInt>0
            \pgfmathsetmacro\yMax{\yInt/#1}
            \addplot[fill=#2, opacity=0.35, draw=none, mark=none, forget plot]
                coordinates {(\xL,0) (\xR,0) (\xR,\yMax) (\xL,\yMax) (\xL,0)};
        \fi
    }%
}

\newcommand{\plotRegionHatched}[2]{%
    \foreach \n in {1,...,100} {%
        \pgfmathsetmacro\xL{1/(\n+1)}
        \pgfmathsetmacro\xR{1/\n}
        \pgfmathtruncatemacro\yInt{floor(#1/\n)}
        \ifnum\yInt>0
            \pgfmathsetmacro\yMax{\yInt/#1}
            \addplot[fill=#2, opacity=0.35, draw=none, mark=none, forget plot]
                coordinates {(\xL,0) (\xR,0) (\xR,\yMax) (\xL,\yMax) (\xL,0)};
            \addplot[pattern=north east lines, pattern color=black, draw=none, mark=none, forget plot]
                coordinates {(\xL,0) (\xR,0) (\xR,\yMax) (\xL,\yMax) (\xL,0)};
        \fi
    }%
}

\allowdisplaybreaks 

\begin{document}

\maketitle

\begin{abstract}
    Social choice literature on multiwinner elections tends to focus on characterizing two distinct tasks: majoritarian representation and proportional representation. However, many real world systems can be viewed as operating between these two extremes, such as majority jackpot systems used in several countries throughout history. Towards the goal of better understanding these systems from a social choice perspective, we introduce two frameworks that encompass existing desiderata for majoritarian and proportional settings: \emph{coalitions}, which encompasses properties such as fixed majority and solid coalitions; and \emph{stability}, which encompasses various generalizations of Condorcet's criterion for single-winner elections, such as Gehrlein stability and local stability, among others. We develop theory and tools that characterize the mathematical and computational aspects of both frameworks across a wide array of preference structures (such as the unrestricted domain over candidates and the restricted domain of party lists). Building on this foundation, we shed new light on majority jackpot electoral systems, which to date have received limited mathematical analysis in the social choice literature.
\end{abstract}


\newpage

{
  \hypersetup{linkcolor=black}
  \tableofcontents
}

\newpage


\section{Introduction}\label{sec:introduction}

Social choice theory studies how to aggregate individual preferences into a collective decision. Multiwinner elections -- where a committee, rather than a single winner, must be chosen -- are among its most widely applicable settings, underlying everything from parliamentary elections to shortlisting job candidates \cite{faliszewski2017}. The specific mechanics of an election (its \emph{mechanism}) can shape the resulting committee in ways that go beyond simply tallying votes, giving the mechanism designer an opportunity to build desirable properties into the committee by construction \cite{ kohli2021leveraging, kohli2018epsilon, mulligan2019thing}. In the multiwinner setting, some desiderata are adapted from criteria for single-winner elections, such as that of \citet{condorcet}, who argued that an election system should elect a candidate who beats every other candidate in a head-to-head match, provided one exists. Others are native to the multiwinner setting itself, reflecting the additional structure a committee provides \cite{faliszewski2017}.

Multiwinner election desiderata often fall into one of two categories: some desiderata favor the preferences of a majority of voters (\emph{majoritarianism}), while others aim to represent voters in proportion to their numbers (\emph{proportionality}). 

Majoritarian desiderata seek to ensure that whenever a sufficiently large majority of voters agree, their shared preferences are reflected in the committee. Desiderata such as the fixed majority criterion \cite{elkind2017properties} (which grants a majority the power to dictate the committee outright) and Gehrlein stability \cite{aziz2017condorcet, gehrlein1985, ratliff2003} (which requires every committee member to beat every non-member in a head-to-head vote) can both be viewed as forms of majoritarianism. 

Proportional representation, by contrast, seeks to divide seats among factions of voters according to their relative size, so that no single group -- majority or otherwise -- claims a disproportionate share of the committee. This goal garnered particular interest in the late nineteenth and early twentieth centuries, with Phragm\'en, Thiele, d'Hondt, Hare, and others designing systems to produce proportional outcomes \cite{hare1865,janson2016phragmen}, which reformers such as \citet{mill1903} and \citet{sterne1871} advocated for. Desiderata such as solid coalitions \cite{elkind2017properties} (which guarantees seats based on any sufficiently large group of like-minded voters) and local stability \cite{aziz2017condorcet} (which offers a similar guarantee based on pairwise comparisons) can likewise be viewed as forms of proportionality.

While much existing work in social choice falls into these two extremes, real-world election systems typically lie somewhere between them \cite{massicotte1999}. One such class -- the \emph{majority jackpot systems} used in Italy \cite{dalimonte2015,nohlen2010,britannica-italy}, San Marino \cite{sanmarino}, and Armenia \cite{armenia} -- is motivated by the argument that proportional representation can lead to instability and frequent elections that return similar results \cite{lardeyret1991,quade1991}, and that small deviations from proportionality can instead produce more stable governments with minimal misrepresentation \cite{bedock2014}.

Motivated by the desire to better understand multiwinner election systems that balance majoritarian and proportional goals,  we develop two new frameworks of desiderata that allow for more granular analysis of representation in committees, which we call \emph{coalitions} (Section \ref{sec:coalitions}) and \emph{stability} (Section \ref{sec:stability}). To showcase the expressiveness of our mathematical frameworks, we describe two applications. The first are complete party list elections, which are a subtype of party lists where voters rank political parties rather than individual candidates (Section \ref{sec:partylists}). The second are the majority jackpot systems mentioned above, which we formalize mathematically and relate to our two frameworks through associated desiderata (Section \ref{sec:majorityjackpot}). We find that our frameworks are flexible enough to capture, in just two parameters, multiple objectives that such systems aim to achieve.

The remainder of this manuscript is organized as follows. Section \ref{sec:related_works} describes related work directly relevant to our study. Section \ref{sec:prelims} provides the mathematical notation and preliminaries used throughout the paper. Sections \ref{sec:coalitions} and \ref{sec:stability} describe our coalitions and stability frameworks respectively. Sections \ref{sec:partylists} and \ref{sec:majorityjackpot} apply our  frameworks to provide new insights into complete party list elections and majority jackpot systems. Section \ref{sec:discussion} concludes with discussion. For clarity of exposition, we defer the presentation of our proofs to the Appendices.

\subsection{Related Works}
\label{sec:related_works}

A number of authors have mathematically described different criteria for multiwinner elections. One line of work adapts the criterion of \citet{condorcet} to the multiwinner setting. Some approaches compare entire committees to one another \citep{fishburn1981}, while others compare individual committee members to non-members in pairwise contests \citep{gehrlein1985, ratliff2003}. A second line of work formalizes proportional representation for ranked preferences \citep{dummett1984, ChamberlinCourant1983, monroe1995} and approval ballots \citep{aziz2017justified, sanchez2016pjr}. \citet{elkind2017properties} and \citet{faliszewski2016committee} provide useful surveys of many of these criteria, along with new results.

Of particular relevance to this study are the works of \citet{elkind2015condorcet} and \citet{aziz2017condorcet}, who synthesize these fields and introduce new Condorcet-like notions of proportional representation called local stability and full local stability. \citet{aziz2017condorcet} further discuss Gehrlein stability as majoritarian in spirit and local stability as suited to proportional representation. Our work in Sections \ref{sec:coalitions} and \ref{sec:stability} produces two mathematical frameworks that describe the broader landscape of criteria \emph{between} majoritarianism and proportionality in terms of two parameters $p$ and $r$. This not only encompasses many existing criteria, but also enables the description of new ones, allowing criteria to be compared directly against each other in this common language.

A common theme in the literature is to examine the ease of computation of different criteria. While some criteria can be computed in polynomial time -- such as unanimity \cite{elkind2017properties}, strong Gehrlein stability over unrestricted preferences and weak Gehrlein stability over single-peaked preferences \cite{aziz2017condorcet} -- many election rules and desirable criteria are computationally difficult to realize. For example, an election method proposed by Charles Dodgson was proven to be NP-hard to compute by \citet{bartholdi1989}, as were the methods of Chamberlin and Courant and Monroe by \citet{procaccia2008} and \citet{betzler2013}. We produce results on the computational tractability of our coalitions framework in Section \ref{sec:coalitions} and a subset of our stability framework in Section \ref{sec:stability}.

We also examine classes of restricted preferences called party list preferences and complete party list preferences. Party lists have mostly been discussed in empirical studies of real-world election systems, such as that of \citet{massicotte1999}. \citet{aziz2017condorcet} use these preferences as a case study in their exploration of Gehrlein stability and local stability. Ordinal voter preferences over party lists (which we call complete party list preferences) have been studied by \citet{vandeemen1993theo}, who identified paradoxes in how seats are allocated to lists, and by \citet{vandeemen1998emp}, who examined their prevalence in Dutch elections. In Section \ref{sec:partylists}, we discuss how our coalitions and stability frameworks interact with these restricted preference domains.

Our last line of inquiry considers majority jackpot systems, whose goal is to ensure some political party always gets a majority of seats in a committee. Such systems have been in use in San Marino \cite{sanmarino}, Armenia \cite{armenia}, and certain elections in Italy \cite{britannica-italy}, among others \cite{massicotte1999}. Despite their use in practice, there has been -- to the best of our knowledge -- limited formal mathematical analysis of majority jackpot systems. Most existing literature has focused on relating majority jackpot systems to other electoral systems \cite{bedock2014}, or on analyzing the details of a particular proposed majority jackpot system \cite{dalimonte2015}. \citet{speroni2019} are notable for proposing a new system that blends elements of both majority jackpot systems and proportional representation by using approval ballots over parties. In Section \ref{sec:majorityjackpot}, we contribute more detailed mathematical descriptions of majority jackpot elections, along with desiderata for them, including formalizations of the paradoxes identified by \citet{vandeemen1993theo}. This illustrates the applicability of our coalitions and stability frameworks to underexplored topics in social choice.

\section{Preliminaries}\label{sec:prelims}
In this section, we describe the mathematical primitives and notation used throughout this study. Following the social choice literature, we write $[z]$ in place of $\set{1,\dots,z}$ for any positive integer $z$. 

To study preference aggregation, we consider the setting of an election, with a set of $n > 1$ voters $V$ expressing strict ordinal preferences over a set of $m > 1$ candidates $C$. Formally, we say an election $E=(C,V, \succ)$ is a triple consisting of an indexed set of candidates $C = \{c_1,...,c_m\}$, $V$ an indexed set of voters $\set{v_1,\dots, v_n}$, and $\succ = (\succ_1,...,\succ_n)$ where $\succ_v$ denotes $v$'s strict preference ordering over the alternatives. 

We will write $a\succ_v b$ to denote that $v$ prefers candidate $a$ to candidate $b$, and $A\succ_v B$ to denote that for every $a\in A$ and $b\in B$, $a \succ_v b$. We let $\topp_\ell(v)$ be the set of the first $\ell$ elements of $\succ_v$. For shorthand, we will write $\topp(v)$ to be the sole element of $\topp_1(v)$. We will also write $\topp_\ell(V^*)$ to denote the collection of candidates that appear in the top $\ell$ positions of a subset of voters $V^* \subseteq V$; that is, $\topp_\ell(V^*) = \cup_{v \in V^*} \topp_\ell(v)$. In the case when $\ell=1$, we let $\topp(V^*) = \cup_{v \in V^*} \set{\topp(v)}$.


For a preference profile for a particular election, we will often display these as a table with the left side denoting the number of voters having this preference profile, and the right side denoting the profile itself. For example, 
\begin{center}
    \begin{tabular}{c|c}
        \# Voters & Preferences\\\hline
        30 & $a\succ b\succ c$ \\
        20 & $b\succ c\succ a$
    \end{tabular}
    \end{center}
displays an election where 30 voters submit preferences $(a,b,c)$ and 20 voters submit preferences $(b,c,a)$. 

We let $\mathcal{E}$ be the set of all finite elections, over any number $m \ge 2$ of candidates and any number $n \ge 2$ of voters. That is, $\mathcal{E}$ captures all possible preference profiles $\succ$ of any finite set $V$ of $n$ voters over any finite set $C$ of $m$ candidates.

A multiwinner election aims to find, for an election $E=(C,V,\succ)$, a \emph{committee} $S\subseteq C$ of a certain size $k \in \set{2,\dots, m}$ satisfying a desirable property $\chi$. For example, $\chi$ can be ensuring a majority party wins all $k$ seats, or ensuring that parties with $p\%$ of the vote make up $p\%$ of the committee. 

However, in general it is not guaranteed that an election can produce a desirable committee by some $\chi$. A simple example is if $\chi$ requires that every voter has their top preference on the committee; if $m>k$ and for every $c\in C$ some voter $v$ has $\topp(v)=c$, then $S$ must be a size $k$ set containing all $m$ candidates, which is clearly impossible. It is thus relevant to discuss which preference profiles allow us to find a committee which satisfies $\chi$. We introduce the concept of \emph{realizability} to convey this, formalized below.
\begin{defn}
    We say a condition $\chi$ is $k$-\emph{realizable} for an election $E=(C,V,\succ)\in \mathcal{E}$ if, given $E$, there exists some set $S\subseteq C$ with $\abs{S}=k$ satisfying $\chi$. Additionally, we say a condition $\chi$ is $k$-\emph{universally realizable} if for every $E\in \mathcal{E}$, $\chi$ is $k$-realizable for $E$.\footnote{While our focus is the multiwinner setting with $k \ge 2$, many of our results will still hold when $k = 1$.} 

    When $k$ is clear from context, we will simply refer to the notion of $k$-realizable as realizable (and similarly for universally realizable).
\end{defn}
In addition to the existence of committees with desirable properties, we are also interested in rules that find such committees or decide they do not exist. Formally, we say a \emph{multiwinner election rule} $\Rule(E,k)$ is a function that takes an election $E$ and an integer $k \in \set{2,\dots,m}$ and returns a size $k$ committee $S\subseteq C$.\footnote{Such rules are referred to as \emph{resolute} \cite{faliszewski2017} in the literature. For our analysis, we will not consider rules that return tying sets of committees.} We introduce the following useful definition.
\begin{defn}\label{def:rule_property}
    We say a rule $\Rule$ satisfies $\chi$ if $\Rule(E,k)$ satisfies $\chi$ whenever $\chi$ is $k$-\emph{realizable} for $E$.
\end{defn}

\section{Coalitions}\label{sec:coalitions}
Our first framework generalizes multiple desiderata that only account for the top $k$ preferences of voters. Two such properties, courtesy of \citet{elkind2017properties}, are the solid coalitions property and the fixed majority property. The solid coalitions property states that whenever $q$ or more voters agree on their top preference, the candidate in this top position is included in the final committee.

\begin{defn}[Solid Coalitions Property]\label{def:SC}
    A committee \(S\) satisfies the solid \(q\)-coalitions property if, for all groups of voters \(V^* \subseteq V\) with \(\abs{V^*}\geq q\), if $\topp(V^*)=\set{c}$, then \(c\in S\). When \(q=\nicefrac{n}{k}\), we have the Hare quota.
\end{defn}

On the other hand, the fixed majority property states that whenever a strict majority of voters agree on their top $k$ preferences (in any order), then these $k$ candidates will comprise the committee.

\begin{defn}[Fixed Majority]\label{def:fix_maj}
    A committee \(S\) satisfies the fixed majority condition if, when there is a group of voters \(V^* \subseteq V\) with \(\abs{V^*}>n/2\) where \(\abs{\topp_{k} (V^*)} = k\), then \(\topp_{k} V^* = S\).
\end{defn}

These properties are satisfied by different commonly used voting rules: the Single Non-Transferable Vote satisfies the solid coalitions property, and the Multiple Non-Transferable Vote (also called the Bloc Vote) satisfies the fixed majority property. 

Note that the solid coalitions property captures a notion of proportional representation, whereas the fixed majority property captures a notion of (winner-takes-all) majoritarianism.

To unify majoritarianism and proportionality in a single framework, we introduce a family of properties we call $(p,r)$-coalitions (Definition \ref{def:pr-coal}).\footnote{We choose the name \emph{coalitions} for our framework due to our definition's relation to proportionality for solid coalitions \cite{dummett1984} and the solid coalitions property \cite{elkind2017properties}} Conceptually, this definition specifies that whenever at least a $p$-fraction of voters agree on their top $\ceil{rk}$ candidates, those candidates are guaranteed a place in the committee. This allows us to encode an electoral system as an ordered pair $(p,r)\in (0,1]^2$ based on the proportion $r$ of the committee that represents a proportion $p$ of the voters.\footnote{Our two-parameter formulation is related to that of \citet{janson2018}, who reformulate classical proportionality criteria -- such as the Droop Proportionality Criterion and proportionality for solid coalitions -- as a single continuously-valued threshold (namely, the minimal fraction of voters agreeing on a common top-$\ell$ block of candidates needed to guarantee those $\ell$ candidates a place on the committee). Rather than fixing this fraction to one of the handful of values proposed in the classical literature, \citet{janson2018} treats it as a free parameter and computes its value for a range of specific voting rules (such as Block Vote and single non-transferable vote, amongst others). We differ from this idea in two ways. First, rather than computing the tightest threshold for one specific rule, we take the threshold (which we call $p$) as a free input that specifies a property directly: any committee, regardless of how it was produced, can be checked against a given $(p,r)$ on a given election. Second, we normalize the guarantee size by the committee size, setting $\ell = \ceil{rk}$, so that criteria for committees of different sizes can be compared directly on the same $(p,r)$ scale.}

\begin{defn}\label{def:pr-coal}
    A committee \(S\) satisfies the \((p,r)\)-coalitions property if, for all groups of voters \(V^* \subseteq V\) with \(\abs{V^*}\ge pn\) and \(\abs{\topp_{\ceil{rk}} V^*} = \ceil{rk}\), then \(\topp_{\ceil{rk}} V^*\subseteq S\).
\end{defn}

\subsection{Unpacking Our Coalitions Framework}

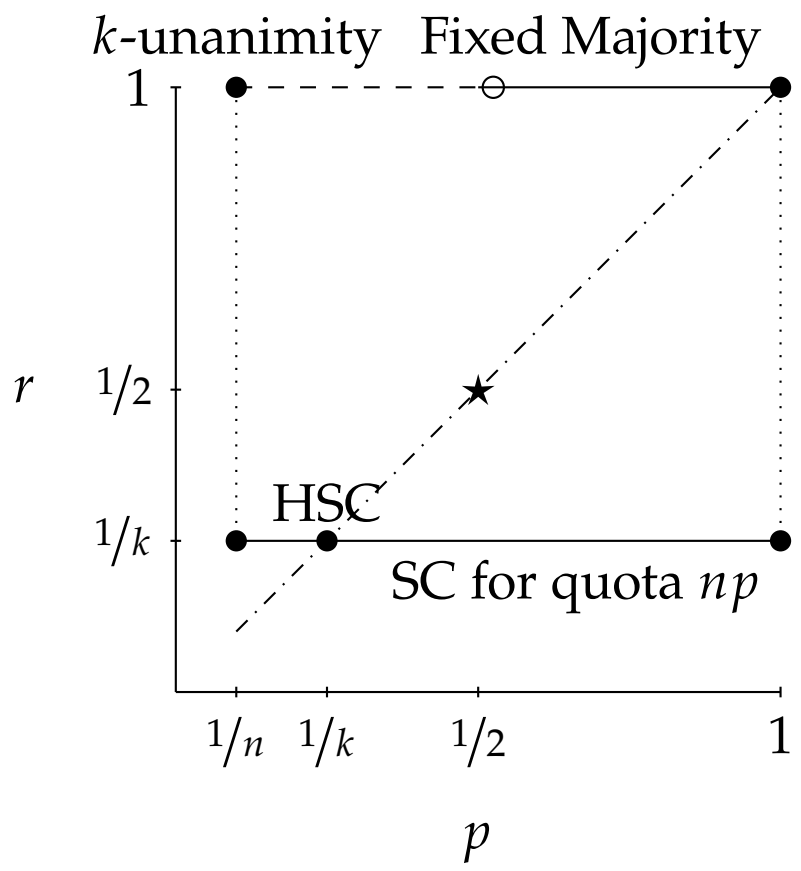
\begin{figure}
    \centering
    \begin{tikzpicture}
    \draw (0,0) -- (4,0) node[anchor=north] {};
    \draw (0,0) -- (0,4) node[anchor=east] {};
    \node at (-1,2) {$r$};
    \node at (2,-1) {$p$};
    \draw (4 cm,1pt) -- (4 cm,-1pt) node[anchor=north] {\(1\)};
    \draw (1pt, 4 cm) -- (-1pt,4 cm) node[anchor=east] {\(1\)};
    \draw (2 cm,1pt) -- (2 cm,-1pt) node[anchor=north] {\(\nicefrac{1}{2}\)};
    \draw (1pt, 2 cm) -- (-1pt,2 cm) node[anchor=east] {\(\nicefrac{1}{2}\)};
    \draw (1 cm,1pt) -- (1 cm,-1pt) node[anchor=north] {\(\nicefrac{1}{k}\)};
    \draw (0.4 cm, 1pt)--(0.4 cm, -1pt) node[anchor = north] {\(\nicefrac{1}{n}\)};
    \draw (1pt, 1 cm) -- (-1pt,1 cm) node[anchor=east] {\(\nicefrac{1}{k}\)};

    \fill (1 cm, 1 cm) circle[radius=2pt] node[anchor = south] {HSC};
    \fill (0.4 cm, 1 cm) circle[radius=2pt] node[anchor = north east] {};
    \fill (4 cm, 1 cm) circle[radius=2pt] node[anchor = north east] {};
    \draw (2.1 cm, 4 cm) circle[radius=2pt] node[anchor = north] {};
    \node at (2 cm, 2 cm) {$\star$};
    \fill (0.4 cm, 4 cm) circle[radius=2pt] node[align=center, anchor = south] {$k$-unanimity};
    \fill (4 cm, 4 cm) circle[radius=2pt] node[align=center, anchor = south] {};
    
    \draw (0.4 cm, 1cm) -- (4 cm, 1 cm) node[anchor=north east] {SC for quota $np$};
    \draw[dashed] (0.4 cm, 4 cm) -- (2 cm, 4cm) node[anchor=west] {};
    \draw (2 cm, 4 cm) -- (4 cm, 4cm) node[anchor=south east] {Fixed Majority};
    \draw[dash dot] (0.4 cm, 0.4 cm) -- (4 cm, 4cm);
    \draw[dotted] (4 cm, 4 cm) -- (4 cm, 1cm);
    \draw[dotted] (0.4 cm, 4 cm) -- (0.4 cm, 1cm);
\end{tikzpicture}
    \caption{A diagram of \((p,r)\)-coalitions notions (for the case when $n \ge k$). HSC is solid coalitions with Hare quota at $(\nicefrac{1}{k}, \nicefrac{1}{k})$, and SC is solid coalitions spanning $(p, \nicefrac{1}{k})$. The diagonal line represents Dummett's proportionality criterion. $k$-unanimity corresponds to $(\nicefrac{1}{n},1)$ (Corollary \ref{cor:coal_top_left_unanimity}). A point of interest we analyze later in Sections \ref{sec:partylists} and \ref{sec:majorityjackpot} is $(\nicefrac{1}{2},\nicefrac{1}{2})$-coalitions, labeled with $\star$.}
    \label{fig:prcoalitions}
\end{figure}

We present a visual representation of the $(p,r)$-coalitions framework in Figure \ref{fig:prcoalitions}. The $p$-axis lives on $(0,1]$, and captures the minimum size of a ``represented group.'' The $r$-axis lives on $(0,1]$, and interpolates between winner-takes-all majoritarianism ($r = 1$) and single member representation ($r = \nicefrac{1}{k}$).

\medskip

\noindent Propositions \ref{prop:sc-prcoal}, \ref{prop:fixmaj}, and \ref{prop:prdummett} express existing desiderata as points on our graph.\footnote{The point assigned to a criterion is not necessarily unique. Since $(p,r)$-coalitions depends on $r$ only through $\ceil{rk}$, every $r\in\left(\nicefrac{(\ell-1)}{k},\nicefrac{\ell}{k}\right]$ gives the same property as $r=\nicefrac{\ell}{k}$. For example, solid coalitions corresponds to every $r\in(0,\nicefrac{1}{k}]$. We present one representative point for each criterion.}

\begin{prop}\label{prop:sc-prcoal}
    A committee satisfies the \((\nicefrac{q}{n},\nicefrac{1}{k})\)-coalitions property if and only if it satisfies the solid \(q\)-coalitions property. 
\end{prop}

\begin{prop}\label{prop:fixmaj}
    A committee satisfies \((\nicefrac{1}{2}+\nicefrac{1}{2n},1)\)-coalitions if and only if it satisfies the fixed majority criterion.
\end{prop}

One of the earliest notions of proportional representation is from \citet{dummett1984}, called proportionality for solid coalitions, or sometimes \emph{Dummett's Proportionality for quota $q$}, which states that if $\ell q$ voters agree on their top $\ell$ preferences, then all $\ell$ of them will be elected. We note that our framework also encapsulates this notion as follows.
\begin{prop}\label{prop:prdummett}
    A committee $S$ satisfies Dummett's Proportionality criterion for quota $q$ if and only if it satisfies $(\nicefrac{\ell q}{n},\nicefrac{\ell}{k})$-coalitions for all $\ell\le k$.
\end{prop}

The definition of $(p,r)$-coalitions exhibits helpful structure that aids in analysis and computation. One such property is \emph{lifting}, which guarantees that if a set satisfies $(p,r)$-coalitions, it will also satisfy $(p,r)$-coalitions for all greater values of $p$.

\begin{prop}[\(p\)-lifting for \((p,r)\)-coalitions]\label{prop:p-coal-lifting}
    If \(S\) satisfies \((p,r)\)-coalitions, and if \(p'\ge p\), then \(S\) also satisfies \((p',r)\)-coalitions.
\end{prop}

\subsection{Realizability for Coalitions}

\begin{figure}
    \centering
    \begin{tikzpicture}
    \begin{axis}[
        myaxis,
    ]
        \plotRegionHatched{3}{gray}
        \plotRegion{30}{gray}

    \end{axis}
\end{tikzpicture}
    
    \caption{Universally realizable regions (in gray) for $(p,r)$-coalitions via Theorem \ref{thm:coalitions_ur}: $\ceil{rk}\floor{\nicefrac{1}{p}} \le k$. The hatched gray region corresponds to $k=3$, and the combination of the hatched and un-hatched gray regions corresponds to $k = 30$.}
    \label{fig:prcoal-ur}
\end{figure}
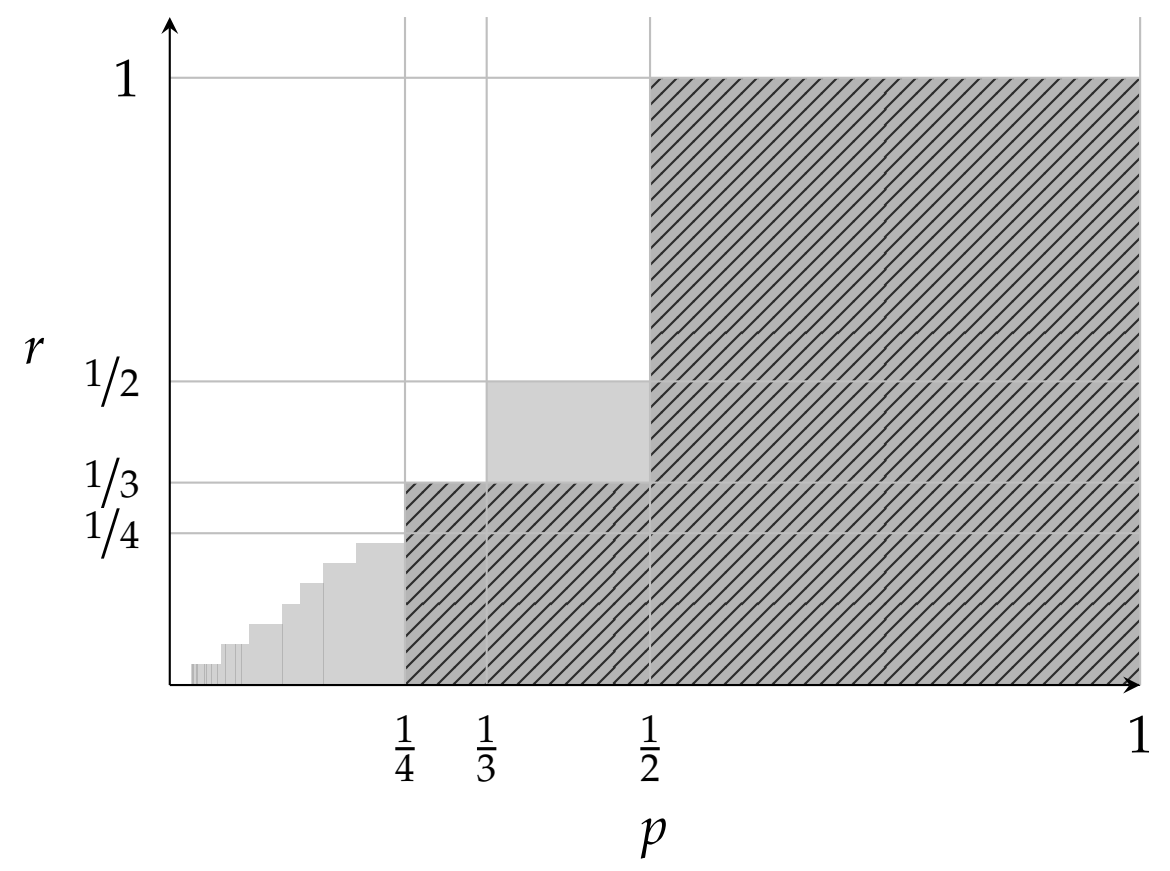

$(p,r)$-coalitions exhibits combinatorial structure which enables clean analysis of realizability and computational complexity. We begin by introducing helpful notation.

Let $\kappa$ denote the number of distinct top-$\ceil{rk}$ preference blocks that appear across all voters in $\succ$ (in any order); that is, 
$$\kappa = \abs{\set{\topp_{\ceil{rk}}(v) : v \in V }}$$ 
We can then partition $V$ into $\kappa$ subsets of voters: $V_1,\dots, V_\kappa$, where every voter in the same $V_i$ agrees on their top-$\ceil{rk}$ choices (in any order). Next, we define $\mathcal{V} $ as the subset of $\set{V_1,\dots,V_\kappa}$ that contain at least $pn$ voters.
$$
\mathcal{V} = \set{V_j : j \in [\kappa] \text{ and } |V_j| \ge pn }
$$

\subsubsection{Universal Realizability}

The following lemma and corollary are immediate consequences of the definition of $(p,r)$-coalitions (and hence their proofs are omitted). 
\begin{lem}
\label{lem:coals_iff} 
    $S$ satisfies $(p,r)$-coalitions if and only if $\cup_{V_j \in \mathcal{V}} \topp_{\ceil{rk}}(V_j) \subseteq S$.
\end{lem}

\begin{cor}
\label{cor:coalitions_checker}
$(p,r)$-coalitions is realizable for $E$ if and only if $|\cup_{V_j \in \mathcal{V}} \topp_{\ceil{rk}}(V_j)| \le k$.  
\end{cor}

Since $(p,r)$-coalitions requires us to take a union over $\mathcal{V}$ (Corollary \ref{cor:coalitions_checker}), it will be helpful to establish bounds on the maximum allowable size of $\mathcal{V}$. Let $\zeta = \abs{\mathcal{V}}$.

\begin{lem}\label{lem:coalitions_numunique}
     $\zeta \le \floor{p^{-1}}$ for every election.
\end{lem}

\noindent The above results facilitate the characterization of universal realizability of $(p,r)$-coalitions.

\begin{thm}\label{thm:coalitions_ur}
    For any fixed $k, p$, and $r$, $(p,r)$-coalitions is universally realizable if and only if $ \ceil{rk}\floor{\nicefrac{1}{p}} \le k$. In particular, when $r$ is an integer multiple of $\nicefrac{1}{k}$, this condition reduces to $r \le \floor{p^{-1}}^{-1}$.
\end{thm}

Figure \ref{fig:prcoal-ur} depicts the points $(p,r)$ of universal realizability with $k=3$ and $k = 30$ based on the inequality from Theorem \ref{thm:coalitions_ur}. Because the condition depends on $p$ only through $\floor{\nicefrac1p}$, each region's boundary is a step function of $p$ rather than a smooth curve: it remains flat on each interval $p \in (\nicefrac{1}{(z+1)}, \nicefrac1z]$, where $\lfloor 1/p\rfloor = z$ is constant, and jumps whenever $p$ crosses a reciprocal $\nicefrac1z$. 

As a consequence of Theorem \ref{thm:coalitions_ur}, since $\floor{p^{-1}}=1$ whenever $p>\nicefrac{1}{2}$, and $r\le 1$ by definition, we immediately deduce the following.
\begin{cor}
    When $p>\nicefrac{1}{2}$, $(p,r)$-coalitions is universally realizable for any $r$ and $k$. So in particular, the fixed majority criterion is universally realizable.
\end{cor}

\subsubsection{Regions of realizability only for homogeneous preferences}

We will now consider the case where $p=\nicefrac{1}{n}$.
We note the limitations of the universal realizability result of Theorem \ref{thm:coalitions_ur} in this situation. Plugging in $p = \nicefrac{1}{n}$ and $r= \nicefrac{w}{k}$ (with $w$ a positive integer) into Theorem \ref{thm:coalitions_ur}, it follows that $(\nicefrac{1}{n}, \nicefrac{w}{k})$-coalitions is universally realizable if and only if 
$$
r \le \floor{p^{-1}}^{-1} \iff \frac{w}{k} \le \frac{1}{n} \iff k\ge nw \ge n
$$
The usual assumption in elections is that $n> k$, so the preceding result from Theorem \ref{thm:coalitions_ur} does not provide much insight, except in a situation where a small group is making selections for a larger committee (e.g. a board of 3 directors shortlisting 5 CEO candidates).

Since universal realizability is essentially uninformative at $p=\nicefrac{1}{n}$ under standard assumptions, we consider a more targeted question: for a fixed set of candidates, voters, and committee size, \emph{which preference profiles $\succ$ make $(\nicefrac{1}{n}, r)$-coalitions realizable}?

Intuitively, $(p,r)$-coalitions indicates that if any group of at least $np$ voters agree on their top $\ceil{rk}$ candidates, then those candidates are all in the committee. So, when $p=\nicefrac{1}{n}$, every individual voter has their top $\ceil{rk}$ candidates in the committee. Since the committee size is fixed at $k$, achieving $(\nicefrac{1}{n}, r)$-coalitions demands a certain level of \emph{homogeneity} among preferences, formalized below.

\begin{defn}
    We say a collection of preference orderings $\succ = (\succ_1,...,\succ_n)$ are $(a,b)$-\emph{homogeneous} if 
    $$
    \left|\bigcup_{v \in V} \ \topp_{a}(\succ_v) \right| \le b
    $$
    That is, at most $b$ candidates show up in the top $a$ positions of all voters' profiles.
\end{defn}

\begin{thm}
\label{thm:coal_hom}
    $(\nicefrac{1}{n},r)$-coalitions is realizable on $E$ if and only if the preferences of $E$ are $(\ceil{rk},k)$-homogeneous.
\end{thm}

When $a = b$, the notion of homogeneity relates to a notion of $a$-\emph{unanimity}: that is, all $n$ voters agree on the collection of $a$ candidates in the first $a$ positions of their preference profiles (in any order). Hence, homogeneity can be  viewed as a generalization of unanimity. \citet{elkind2017properties} introduce a desideratum of \emph{unanimity} for multiwinner elections in terms of rules $\Rule$, which states that if all voters have such $k$-unanimous preferences, then $\Rule$ should elect these $k$ candidates.\footnote{For resolute rules, strong and weak unanimity are equivalent, so we do not differentiate them here.}

\begin{cor}
\label{cor:coal_top_left_unanimity}
    $(\nicefrac{1}{n}, 1)$-coalitions is realizable for an election $E=(C,V,\succ)$ if and only if $\succ$ is $k$-unanimous. Therefore, a rule $\Rule$ satisfies  $(\nicefrac{1}{n}, 1)$-coalitions if and only if it is unanimous.
\end{cor}

\subsection{Computational Considerations for Coalitions}

A desirable property of $(p,r)$-coalitions is that finding a set satisfying $(p,r)$-coalitions or determining that one does not exist can be done in polynomial time for any values of $k$, $p$, and $r$. We formalize the decision problem as follows.
\begin{defn}
Given an election $E=(C,V,\succ)$ with $\abs{C}=m$, $\abs{V}=n$ with a desired committee size $k$, and a desired $p$ and $r$, we define the problem $\textsc{Coalitions}(E,k,p,r)$ to output a subset of $C$  of size $k$ satisfying $(p,r)$-coalitions if and only if one exists, and $\bot$ otherwise.    
\end{defn}

We show in Theorem \ref{thm:coal_pnp} that $\textsc{Coalitions}$ can be solved in polynomial time using Algorithm \ref{alg:cocomhash}. We represent an election $E = (C, V, \succ)$ as an \(n\times m\) \emph{preference matrix} $M$, where \(M_{v,i}\) is the candidate in the \(i^{th}\) position of voter \(v\)'s preference profile $\succ_v$. 

At a conceptual level, Algorithm \ref{alg:cocomhash} constructs $\mathcal{V}$ using a hashmap to efficiently store the information required by the conditions of Lemma \ref{lem:coals_iff} and Corollary \ref{cor:coalitions_checker}. Notationally, we denote our hashmap as $H$ and initialize it with empty keys and values (denoted as $H=\mathrm{Hashmap}(\emptyset\to\emptyset)$). For any key $\eta$, we let $H(\eta)$ access the associated value of $\eta$. Upon completion, $H$ will have as keys all $\ceil{rk}$-sets in the top $\ceil{rk}$ positions of any voter, with the value being the number of voters with those top $\ceil{rk}$ preferences. $H^*$ is then the subset of $H$ with associated values no less than $np$.
 
Since two voters agree on their top-$\ceil{rk}$ choices whenever they share the same \emph{set} of candidates, regardless of order, our proof requires that $H$'s keys be compared as sets. This can be done in $\bigO(\ceil{rk})$ time via a single pass combining the $\ceil{rk}$ elements' individual hashes in an associative and commutative manner,\footnote{Such operations are available in common programming languages, such as \texttt{java.util.Set}'s \texttt{hashCode()} or Python's \texttt{frozenset}.} rather than by sorting the elements to obtain a canonical ordering. Correctness additionally requires that equality checks on these keys are likewise order-independent (i.e.\ true set equality, not sequence equality) so that two colliding hashes are confirmed as the same key only when the underlying sets genuinely match.\footnote{If only an order-\emph{dependent} (e.g.\ comparison-based) hash and equality check are available, sorting each extracted set before hashing and comparing achieves the same correctness at a cost of $\bigO(\ceil{rk}\max(1,\log\ceil{rk}))$ per voter instead of $\bigO(\ceil{rk})$. The rationale is that when $r$ and $k$ are such that $\ceil{rk} = 1$, sorting the top-1 element is constant time; on the other hand, when $r$ and $k$ are such that $\ceil{rk} > 1$, sorting the set in a canonical order can be done in $\bigO(\ceil{rk}\log\ceil{rk})$ time using a standard comparison sort. Since this only affects the cost of constructing $H$ and $U$, and not the $\bigO(k)$ cost of resolving the committee within the proof of Theorem \ref{thm:coal_pnp}, this gives an overall runtime of $\bigO(\max(n\ceil{rk}\max(1,\log(\ceil{rk})), k))$ for Algorithm \ref{alg:cocomhash} rather than $\bigO(\max(n\ceil{rk},k))$.}

\begin{algorithm}
\caption{Hashmap-based algorithm for electing a committee satisfying the \((p,r)\)-coalitions property, when one exists.}\label{alg:cocomhash}
    \begin{algorithmic}[1]
        \Procedure{CoalitionsCommitteeHashmap}{$M,k,p,r$}
        \State{\(H=\text{Hashmap}(\emptyset\to \emptyset)\)}
        \For{\(v\in [n]\)}
            \State{\(T_v = M[v,1:\ceil{rk}]\) as a set}\label{alg:cocomhash:extract}
            \If{\(T_v\in\text{Keys}(H)\)}\label{alg:cocomhash:inkeys}
                \State{\(H(T_v)\gets H(T_v)+1\)}
            \Else
                \State{\(H(T_v)\gets1\)}
            \EndIf
        \EndFor
        \State{\(H^* = \set{K\in \text{Keys}(H): H(K)\ge pn}\)}\label{alg:cocomhash:keycheck}
        \State{\(U=\bigcup_{K\in H^*} K\)}\label{alg:cocomhash:unionconstruct}
        \If{\(\abs{U}< k\)}\label{alg:cocomhash:union}
            \State{Let $W$ be any subset of $C\setminus U$ with $\abs{W} = k - \abs{U}$}\label{alg:cocomhash:select}
            \State{\Return{\(U\cup W\)}}\label{alg:cocomhash:tie}
        \ElsIf{\(\abs{U}=k\)}
            \State{\Return{\(U\)}}\label{alg:cocomhash:exact}
        \Else
            \State{\Return{\(\bot\)}}
            \Comment{This case is a failure to elect a committee satisfying \((p,r)\) coalitions.}\label{alg:cocomhash:fail}
        \EndIf
        \EndProcedure
    \end{algorithmic}
\end{algorithm}

\begin{thm}\label{thm:coal_pnp}
    For any $p,r \in (0,1]$, given the preference matrix $M$ of an election $E$, Algorithm \ref{alg:cocomhash} solves  $\textsc{Coalitions}(E,k,p,r)$ in $\bigO(\max(n\ceil{rk},k))$ time. 
\end{thm}

\section{Stability}\label{sec:stability}
The coalitions framework encompasses existing criteria with useful computational and structural properties; however, by definition it only incorporates properties which consider up to the top $k$ preferences of all voters. 

With our stability framework, we seek to capture properties that depend on the complete rankings of voters. Many such properties arise from extensions of the criterion of \citet{condorcet}, which considers whether a system will always elect a candidate who wins a pairwise election with all other candidates. There are multiple such variants, including those of \citet{fishburn1981, gehrlein1985, ratliff2003, elkind2015condorcet}, and \citet{aziz2017condorcet}.

These notions are called some variation of ``stability'': loosely speaking, all of these notions call a committee stable if ``enough'' voters prefer a subset of the committee of ``sufficient'' size to every non-committee member. The differences between the existing notions arise from what they consider ``enough'' and ``sufficient.'' The framework we develop in this section considers these to be a pair of variable parameters $(p,r)\in (0,1]^2$.


As with coalitions, we scaffold our discussion of our stability framework by examining a proportionality-based stability definition and a majoritarian-based stability definition. First, we consider local stability, which is a notion of representation of smaller groups, related to Condorcet winning sets \cite{elkind2015condorcet}.
\begin{defn}[Local Stability \cite{aziz2017condorcet}]
    A committee is  $q$-locally stable ($q$-LS) if 
    \[(\forall c\in C\setminus S) \ (\forall V^* \subseteq V : \abs{V^*}\ge q) \ (\exists s\in S) \quad  (\exists v\in V^*\ \text{ such that } s\succ_v c)  \]
    Equivalently, a committee is \emph{not} $q$-locally stable (or violates $q$-\pty{LS}) if
    \[(\exists c\in C\setminus S) \ (\exists V^*\subseteq V :\abs{V^*}\ge {q}) \ ({\forall} s\in S) \quad ({\forall}v\in V^* \text{ we have} \ c\succ_v s)\]
\end{defn}
That is, a committee $S$ violates $q$-local stability whenever there is a non-committee member $c\in C\setminus S$ where some group of $q$ voters all prefer $c$ to \emph{every} candidate on the committee. On the other hand, strong Gehrlein stability is a majoritarian notion \citep{aziz2017condorcet} . 

\begin{defn}[Strong Gehrlein Stability  \cite{gehrlein1985,ratliff2003}]
\label{defn:strong_g_stability}
    A committee \(S\) is strongly Gehrlein stable for an election $E=(C,V,\succ)$ if for any $s\in S$ and $c\in C\setminus S$, $s$ beats $c$ in the pairwise election by a strict majority; that is, \[\abs{\set{v\in V: s\succ_v c}}>\nicefrac{n}{2}\]
\end{defn}

We now show that strong Gehrlein stability can be rewritten with quantifiers in a form very similar to that of local stability.
\begin{prop}\label{prop:SGS-quantifiers}
    A set $S \subseteq C$ is strongly Gehrlein stable if and only if \[(\forall c\in C\setminus S) \ \left(\forall V^* \subseteq V : \abs{V^*}\ge \frac{n}{2}\right) \ (\forall s\in S) \quad  (\exists v\in V^* \text{ such that } s\succ_v c)  \] Equivalently, $S$ is not strongly Gehrlein stable if and only if \[(\exists c\in C\setminus S) \ \left(\exists V^*\subseteq V :\abs{V^*}\ge {\frac{n}{2}}\right) \;({\exists} s\in S) \quad ({\forall} v\in V^* \text{ we have }c\succ_v s)\]
\end{prop}

This reformulation of strong Gehrlein stability trades a direct statement about a strict majority ($>\nicefrac{n}{2}$) of voters who all prefer $s$ to $c$ (Definition \ref{defn:strong_g_stability}) for an equivalent statement about \emph{every}  weak majority of voters ($\ge\nicefrac{n}{2}$) containing a voter who prefers $s$ to $c$ (Proposition \ref{prop:SGS-quantifiers}).

There are two notable \emph{symbolic differences} between local stability and strong Gehrlein stability (as rewritten in Proposition \ref{prop:SGS-quantifiers}): the first is on the size of $V^*$, and the second is on the quantifier of $s \in S$. 

The innovation of our approach is to consider the difference in quantifier of $s \in S$ to be variations of the size of a subset $S^* \subseteq S$. In particular, our stability framework in Definition \ref{def:prstabind} below replaces ``$\forall s\in S$'' and ``$\exists s\in S$'' with the single template ``$\forall S^*\subseteq S$ with $|S^*|> t$, $\exists s\in S^*$'' for a flexible parameter $t \in [0,k-1]$. 
\begin{itemize}
    \item When $t=0$, this recovers ``$\forall s\in S$'' because checking the predicate ``$\exists v\in V^* \text{ such that } s\succ_v c$'' on every non-empty subset $S^*$ is equivalent to checking the predicate on every singleton $S^*=\{s\}$, which in turn is equivalent to checking it on every $s\in S$ directly.\footnote{We provide a formal proof of this line of reasoning in Lemma \ref{lem:r_one_line_equiv}.}
    \item  Alternatively, when $t=k-1$, this recovers ``$\exists s\in S$'' because $S^*=S$ is then the only valid subset $S^*$. 
\end{itemize}

This perspective hints at considering subsets $S^*$ of variable sizes to quantify the spectrum between majoritarian and proportional representation. With this insight, we create a new encompassing notion called $(p,r)$-stability, with $(p,r) \in (0,1]^2$.

\begin{defn}[\((p,r)\)-stability]\label{def:prstabind}
    A committee $S$ is $(p,r)$-stable if
\[(\forall c\in C\setminus S) \  (\forall V^*\subseteq V :\abs{V^*}\ge np)  \ (\forall S^*\subseteq S  : \abs{S^*}> k(1-r)) \quad (\exists v\in V^* \ \exists s\in S^* \text{ such that } s\succ_v c)\]
Equivalently, a committee $S$ violates \((p,r)\)-stability if 
    \[(\exists c\in C\setminus S) \  (\exists V^*\subseteq V :\abs{V^*}\ge np)  \ (\exists S^*\subseteq S  : \abs{S^*}> k(1-r)) \quad ({\forall} v\in V^* \ {\forall} s\in S^* \text{ we have } c\succ_v s)\]
\end{defn}

In the stability framework, $p$ sets the minimum fraction of voters in a group $V^*$ (that is, $\abs{V^*}\ge np$), and $r$ limits how much of the committee such a group can unanimously rank below a single outside candidate; $(p,r)$-stability requires that no group of at least $np$ voters unanimously ranks more than $k(1-r)$ committee members below any outside candidate $c\in C\setminus S$.

\subsection{Unpacking Our Stability Framework}

\begin{figure}
    \centering
    \begin{tikzpicture}
    \draw (0,0) -- (4,0) node[anchor=west] {};
    \draw (0,0) -- (0,4) node[anchor=south] {};
    \node at (2,-1) {$p$};
    \node at (-1,2) {$r$};
    \node at (2,2) {$\star$};
    \draw (4 cm,1pt) -- (4 cm,-1pt) node[anchor=north] {\(1\)};
    \draw (1pt, 4 cm) -- (-1pt,4 cm) node[anchor=east] {\(1\)};
    \draw (2 cm,1pt) -- (2 cm,-1pt) node[anchor=north] {\(\nicefrac{1}{2}\)};
    \draw (1pt, 2 cm) -- (-1pt,2 cm) node[anchor=east] {\(\nicefrac{1}{2}\)};
    \draw (1 cm,1pt) -- (1 cm,-1pt) node[anchor=north] {\(\nicefrac{1}{k}\)};
    \draw (0.4 cm, 1pt)--(0.4 cm, -1pt) node[anchor = north] {\(\nicefrac{1}{n}\)};
    \draw (1pt, 1 cm) -- (-1pt,1 cm) node[anchor=east] {\(\nicefrac{1}{k}\)};

    \fill (1 cm, 1 cm) circle[radius=2pt] node[align=center, anchor = south] {Hare-quota \\ LS};
    \fill (0.4 cm, 4 cm) circle[radius=2pt] node[align=center, anchor = south east] {$k$-unanimity};
    \fill (4 cm, 4 cm) circle[radius=2pt] node[align=center, anchor = west] {Candidate-wise\\undominated};
    \fill (2 cm, 4 cm) circle[radius=2pt] node[align=center, anchor = south] {Strong Gehrlein\\ Stability};
    \fill (2 cm, 1 cm) circle[radius=2pt] node[align=center, anchor = south] {CW};
    
    \draw (0.4 cm, 1cm) -- (4 cm, 1 cm) node[anchor=north east] {Weak $(1-\theta)$-WS / $q$-LS};
    \draw[dashed] (0.4 cm, 4 cm) -- (4 cm, 4cm) ;
    \draw[dotted] (4 cm, 4 cm) -- (4 cm, 1cm);
    \draw[dotted] (0.4 cm, 4 cm) -- (0.4 cm, 1cm);
\end{tikzpicture}
    \caption{A diagram of \((p,r)\)-stability notions. Strong Gehrlein stability and Hare-quota LS (local stability at quota $q=n/k$) sit at $(\nicefrac12,1)$ and $(\nicefrac1k,\nicefrac1k)$ respectively (Proposition \ref{prop:WGS-LS}); CW (weak Condorcet) sits at $(\nicefrac12,\nicefrac1k)$ (Proposition \ref{prop:theta1k}); and Candidate-wise undominated sits at $(1,1)$ (Proposition \ref{prop:candidate_undom_11_stable}). $k$-unanimity, at $(\nicefrac1n,1)$, is established later in Corollary \ref{cor:top_left_unanimity}. The horizontal line at $r=\nicefrac1k$ traces out both weak $(1-\theta)$-winning sets and $q$-local stability as $\theta$ and $q$ vary. A point on interest we analyze later in Sections \ref{sec:partylists} and \ref{sec:majorityjackpot} is $(\nicefrac{1}{2},\nicefrac{1}{2})$-stability, labeled with $\star$.}
    \label{fig:prstability}
\end{figure}
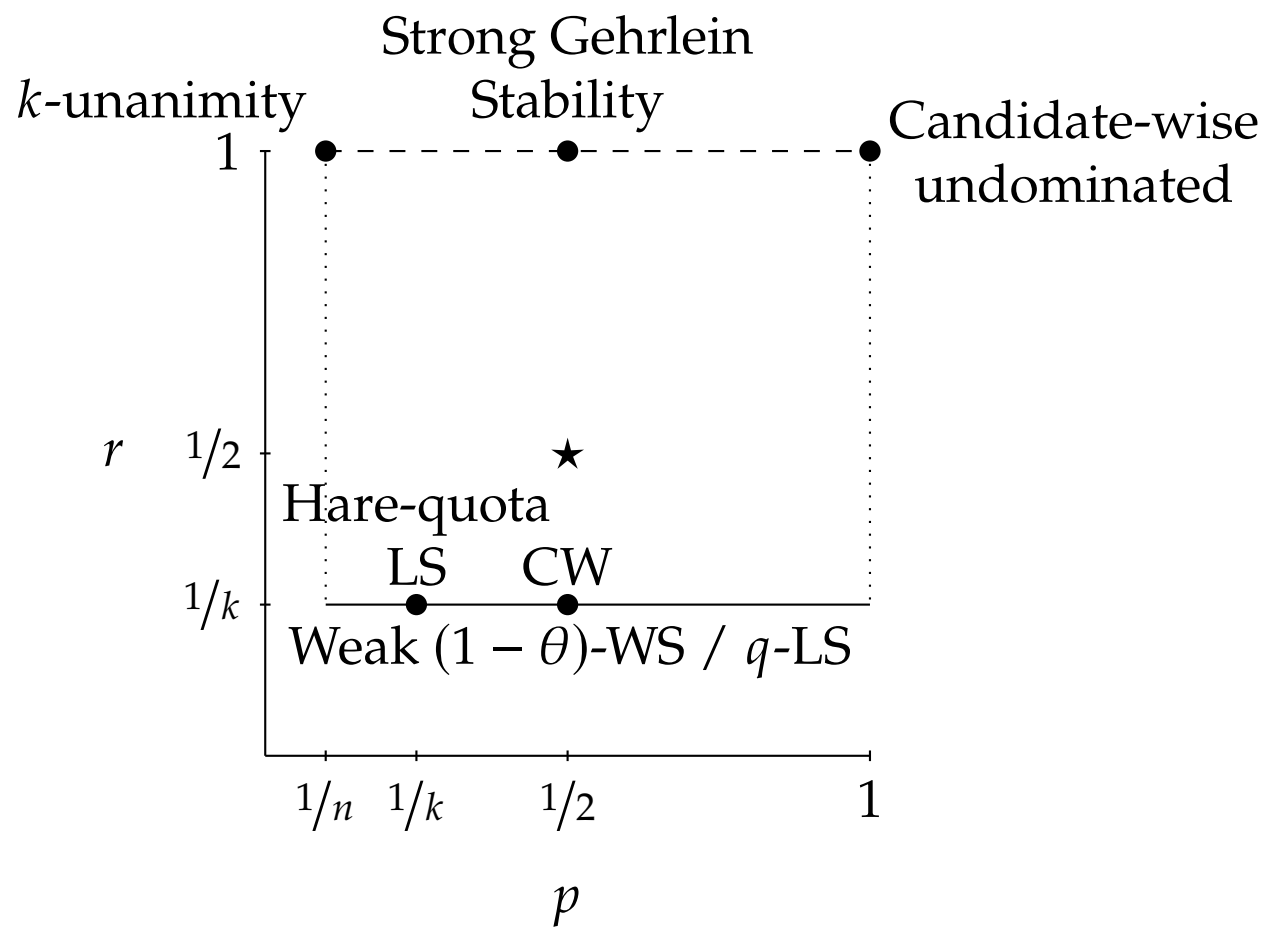

Similar to the coalitions framework, we present a visual representation of the $(p,r)$-stability framework in Figure \ref{fig:prstability}. Propositions \ref{prop:WGS-LS}, \ref{prop:theta1k}, and \ref{prop:candidate_undom_11_stable} identify existing criteria as points on this figure, as does Corollary \ref{cor:top_left_unanimity} (presented in Section \ref{subsubsec:homog_pref_stability}).

\begin{prop}\label{prop:WGS-LS}
    $(\nicefrac{1}{2},1)$-stability is equivalent to strong Gehrlein stability, and $(\nicefrac{q}{n},\nicefrac{1}{k})$-stability is equivalent to $q$-local stability. Consequently, $q$-local stability with the Hare quota $q = \nicefrac{n}{k}$ corresponds to $(\nicefrac{1}{k},\nicefrac{1}{k})$-stability.
\end{prop}

\citet{elkind2015condorcet} introduce the concept of Condorcet winning sets and $\theta$-winning sets, which is closely related to local stability. Local stability, however, as noted by \citet{aziz2017condorcet}, differs from Condorcet winning sets (and $\theta$-winning sets more broadly) in terms of strong and weak inequalities. For this reason, we define a modification of $\theta$ winning sets, which we call \emph{weak $\theta$-winning sets}:
\begin{defn}
    $S$ is a weak $\theta$-winning set of an election $E=(C,V,\succ)$ for $\theta \in [0,1)$ if for every $c\in C\setminus S$ \[\abs{\set{v\in V:\exists s\in S \ s\succ_v c}}\ge \theta n\] 
    For the special case when $\theta = \nicefrac{1}{2}$, we refer to a weak $\theta$-winning set as a weak Condorcet winning set.
\end{defn}

\begin{prop}\label{prop:theta1k}
    $S$ is a weak \(\theta\)-winning set if and only if it is \((1-\theta, \nicefrac{1}{k})\)-stable. A weak Condorcet winning set is thus \((\nicefrac{1}{2},\nicefrac{1}{k})\)-stable. 
\end{prop}

Another point of interest is $(1,1)$-stability, which ensures that unanimously preferred candidates are not excluded in favor of other candidates. We define the following notions to capture this behavior.
\begin{defn}
    We say candidate $c$ \emph{dominates} candidate $d$ if $c \succ_v d$ for all $v \in V$. We say a candidate $d$ is \emph{dominated} if some candidate $c$ dominates $d$. We say a committee $S \subseteq C$ is \emph{candidate-wise undominated} if, whenever $c$ dominates $d$ and $d \in S$, we also have $c \in S$.\footnote{We note that this is related in spirit to the concept of Pareto efficiency. However, as noted in \citet{laine2025pareto}, when defining Pareto efficiency for ordinal multiwinner elections, there is a substantive difference between Pareto efficiency for candidates and Pareto efficiency for committees.} We say a rule $\Rule$ is \emph{candidate-wise undominated} if it produces a candidate-wise undominated committee whenever one exists.
\end{defn}
\begin{prop}
\label{prop:candidate_undom_11_stable}
    A committee or rule is candidate-wise undominated if and only if it satisfies $(1,1)$-stability.
\end{prop}

\subsubsection{Comparison with Coalitions} Stability and coalitions are defined in distinct ways; however, they are mathematically connected and share certain structural properties. First, we show that stability is a stronger condition than coalitions. 
\begin{prop}\label{prop:stab_coal}
    If a set satisfies \((p,r)\)-stability, it also satisfies the \((p,r)\)-coalitions property.
\end{prop}

In particular, Proposition \ref{prop:stab_coal} allows us to relate proportional stability notions to proportional coalitions notions. That is, we immediately deduce the well-known result of \citet{aziz2017condorcet} that $q$-local stability implies solid $q$-coalitions, as $(p,r) = (\nicefrac{q}{n}, \nicefrac{1}{k})$ for both conditions. We can similarly relate majoritarian stability notions to majoritarian coalition notions, as formalized in the following corollary.
\begin{cor}
\label{cor:sgs_implies_fm}
    Strong Gehrlein stability implies Fixed Majority.
\end{cor}

Second, we show that stability exhibits a stronger lifting-style result than the corresponding result for coalitions, which we refer to as \emph{movement}. Whereas coalitions is guaranteed to be preserved whenever $p$ does not decrease and $r$ is fixed (Proposition \ref{prop:p-coal-lifting}), stability is preserved whenever $p$ does not decrease and $r$ does not increase.

\begin{prop}[Movement for stability]
\label{prop:lifting} 
    If a committee is \((p,r)\)-stable, and \(p^*\ge p\) and \(r^* \le r\), then it is also \((p^*, r^*)\)-stable.
\end{prop}

The contrapositive of Proposition \ref{prop:lifting} yields the following.
\begin{cor}
    If \(S\) is not \((p,r)\)-stable, then \(S\) is also not \((p',r')\)-stable if \(p'\leq p\) and \(r'\geq r\).
\end{cor}

\subsection{Realizability for Stability}

We next turn our attention to the realizability of $(p,r)$-stability in our framework. Compared to coalitions, the analysis and characterization of stability is more nuanced, and requires a more detailed examination of specific parameter ranges for $p$ and $r$.

\subsubsection{Universal Realizability}
To start, recall that by Theorem \ref{thm:coalitions_ur}, \((p,r)\)-coalitions is universally realizable if and only if \(\ceil{rk}\floor{\nicefrac{1}{p}} \le k\). Combining this with the contrapositive of Proposition \ref{prop:stab_coal}, we deduce the following necessary (but not sufficient) condition regarding the universal realizability of stability.
\begin{cor}\label{cor:stab_ur}
    If $\ceil{rk}\floor{\nicefrac{1}{p}} > k$, then \((p,r)\)-stability is not universally realizable.
\end{cor}
The following result gives a sufficient condition for universal realizability in a restricted parameter space.

\begin{algorithm}
\caption{An algorithm for producing a \((1,1)\)-stable set of size $k$ for an election $E$. By Proposition \ref{prop:lifting}, this also produces a \((1,r)\)-stable set for any $r \in (0,1]$.}\label{alg:lexicosort}
    \begin{algorithmic}[1]
    \State{Let $M$ be the ${n\times m}$ matrix where $M_{v,i}$ is voter $v$'s $i^{th}$ ranked candidate. }
    \Procedure{PrefLexicographicSort}{$M,k$}
    \State{$\Phi \in \R^{m\times m}, \text{ where } \Phi_{i,j} = \abs{\set{v \in [n]:M_{v,i}=c_j}} $}\label{alg:lexicosort:iposmat}
    \State{$\bm{g}=\paren{(mn)^{-1},\dots, (mn)^{-m}} $}\label{alg:lexicosort:geo}
    \State{$\bm{f}=\bm{g} \Phi$}\label{alg:lexicosort:fscore}
    \State{Let $\bm{a}$ denote the argsort of $\bm{f}$ in decreasing order, breaking ties arbitrarily.}\label{alg:lexicosort:argsort}
    \State{$S=\set{c_{\bm{a}_i}:i\in[k]}$}\label{alg:lexicosort:picktopk}\\
    \Return{$S$}
    \EndProcedure
    \end{algorithmic}
\end{algorithm}

\begin{thm}\label{thm:1r-univreal}
    \((1,r)\)-stability is universally realizable for all $r \in (0,1]$. Further, Algorithm \ref{alg:lexicosort} produces such a set in $\bigO(m^2 n)$ time.
\end{thm}


\subsubsection{Regions of stability realizable only for homogeneous preferences}
\label{subsubsec:homog_pref_stability}

Similar to coalitions, we characterize the class of restricted preferences that yield $(\nicefrac{1}{n}, r)$-stability. To do so, we prove a stronger fact: namely, when $p = \nicefrac{1}{n}$, the notions of $(p,r)$-stability and $(p,r)$-coalitions are equivalent.

\begin{prop}
\label{prop:stab_coal_equiv_1n}
      For any $r \in (0,1]$, a committee is $(\nicefrac{1}{n}, r)$-stable if and only if it satisfies $(\nicefrac{1}{n}, r)$-coalitions.
\end{prop}

Applying Theorem \ref{thm:coal_hom} and Proposition \ref{prop:stab_coal_equiv_1n} immediately yields the following two results.
\begin{thm}    \label{thm:left_vertical_line_characterization}
    $(\nicefrac{1}{n}, r)$-stability is realizable from $E=(C,V,\succ)$ if and only if $\succ$ is $(\ceil{rk},k)$-\emph{homogeneous}.
\end{thm}

\begin{cor}
\label{cor:top_left_unanimity}
    $(\nicefrac{1}{n}, 1)$-stability is realizable for an election $E=(C,V,\succ)$ if and only if $\succ$ is $k$-unanimous. Therefore, a rule $\Rule$ is  $(\nicefrac{1}{n}, 1)$-stable if and only if it is unanimous.
\end{cor}


\subsubsection{Regions of stability realizable contingent on preference graphs}

\citet{aziz2017condorcet} show that the realizability of strong Gehrlein stability is dependent on the structure of the majority graph of the election. We extend this result to other values on the $(p,1)$-line of Figure \ref{fig:prstability} with the following generalization of the majority graph.

\begin{defn}\label{def:wgraph}
    The $p$-preference graph $G^{(p)}=(C,L_p)$ of an election $E=(C,V,\succ)$ consists of set of nodes $C$ and a set links $L_p$ (also referred to as edges), defined by $(a,b) \in L_p \iff |\{v \in V \ : \ a \succ_v b\}| \ge np$.\footnote{We use $L$ for \emph{links} instead of $E$ for \emph{edges} to avoid notationally clashing with $E$ denoting an election.}
\end{defn}


Recall that we can partition a graph into strongly connected components; that is, $a$ and $b$ are in the same strongly connected component if and only if $a$ is reachable from $b$ and $b$ is reachable from $a$. The strongly connected components form equivalence classes, and hence partition $C$. Denote the strongly connected components of $G^{(p)}$ as $\mathcal{G}^{(p)}_1,...,\mathcal{G}^{(p)}_z$ for some positive integer $z$.

\begin{thm}
    \label{thm:bottom_horizontal_line}
    $(p, 1)$-stability is realizable from $E$ if and only if there exists a sub-collection of strongly connected components, indexed by $J \subseteq \{1,...,z\}$, such that
     \begin{enumerate}
         \item $\sum_{j \in J} |\mathcal{G}^{(p)}_j| = k$, and
         \item For all $i \notin J$, there is no edge from a candidate in $\mathcal{G}^{(p)}_i$ to any candidate in $\cup_{j \in J} \mathcal{G}^{(p)}_j$.
     \end{enumerate}
\end{thm}

Theorem \ref{thm:bottom_horizontal_line} characterizes the necessary and sufficient conditions to achieve $(p,1)$-stability. We next remark on an efficient strategy to produce a $(p,r)$-stable set of size $k$, using a (potentially smaller) $(p,1)$-stable set as a building block.

\begin{prop}\label{prop:ginc}
        Suppose \(G\) is \((p,1)\)-stable with \(\abs{G}\in[rk,k]\) and $r \in (0,1)$. If \(G\subseteq S\) with \(\abs{S}=k\), then \(S\) is \((p,r)\)-stable.
\end{prop}

This indicates that we can view $(p,1)$-stable sets as building blocks for $(p,r)$ stable sets in general, provided the $(p,1)$-stable set is of size at least $rk$. For instance, one could construct a $(\nicefrac{1}{2}, \nicefrac{1}{2})$-stable set by finding a strong Gehrlein stable set of size at least $\nicefrac{k}{2}$, and filling the rest of the committee arbitrarily. 

Lastly, we conclude by generalizing the result of \citet{aziz2017condorcet} -- which states that strongly Gehrlein stable sets are unique -- to the uniqueness of all $(p,1)$-stable sets when $p \le \nicefrac{1}{2}$.

\begin{prop}\label{prop:bot_left_unique}
    Suppose \(S\) is a \((p,1)\)-stable set with \(p\le \nicefrac{1}{2}\). If \(T\) is \((p^*,1)\)-stable for \(p^*\le \nicefrac{1}{2}\), then \(T=S\).
\end{prop}

This result does not hold in general when $p > \nicefrac{1}{2}$, as the following example demonstrates.

\begin{ex}\label{ex:non-unique}
    Consider the following election, with $k=2$.
    \begin{center}
    \begin{tabular}{c|c}
        \# Voters & Preferences \\\hline
        1 & \(c_1\succ c_2\succ c_3\succ c_4\succ c_5\succ c_6\succ c_7\) \\
        1 & \(c_3\succ c_4\succ c_5\succ c_6\succ c_1\succ c_2\succ c_7\)\\
        1& \(c_5\succ c_6\succ c_1\succ c_2\succ c_3\succ c_4\succ c_7\)\\
    \end{tabular}
    \end{center}
    The $\nicefrac{4}{5}$-preference graph is displayed in Figure \ref{fig:ex-non-unique}.
    Consider the $(\nicefrac{4}{5}, \nicefrac{1}{2})$-stable sets.
    \begin{figure}
        \centering
        \begin{tikzpicture}
        \fill (-3, 2) circle[radius=2pt] node[anchor = east]  {$c_{1}$};
        \fill (0, 2) circle[radius=2pt] node[anchor = east]  {$c_{3}$};
        \fill (3, 2) circle[radius=2pt] node[anchor = west]  {$c_{5}$};
        \fill (-3, 0) circle[radius=2pt] node[anchor = east]  {$c_{2}$};
        \fill (0, 0) circle[radius=2pt] node[anchor = east]  {$c_{4}$};
        \fill (3, 0) circle[radius=2pt] node[anchor = west]  {$c_{6}$};
        \fill (0, -2) circle[radius=2pt] node[anchor = north]  {$c_{7}$};
        \draw[-latex] (-3, 2) to (-3, 0);
        \draw[-latex] (0, 2) to (0, 0);
        \draw[-latex] (3, 2) to (3, 0);
        \draw[-latex] (-3, 0) to (0, -2);
        \draw[-latex] (0, 0) to (0, -2);
        \draw[-latex] (3, 0) to (0, -2);
    \end{tikzpicture}
        \caption{The $\nicefrac{4}{5}$-preference graph of the election in Example \ref{ex:non-unique}. Edges implied by transitivity (e.g.\ $c_1 \to c_7$ via $c_1 \to c_2 \to c_7$) are omitted for clarity.}
        \label{fig:ex-non-unique}
    \end{figure}
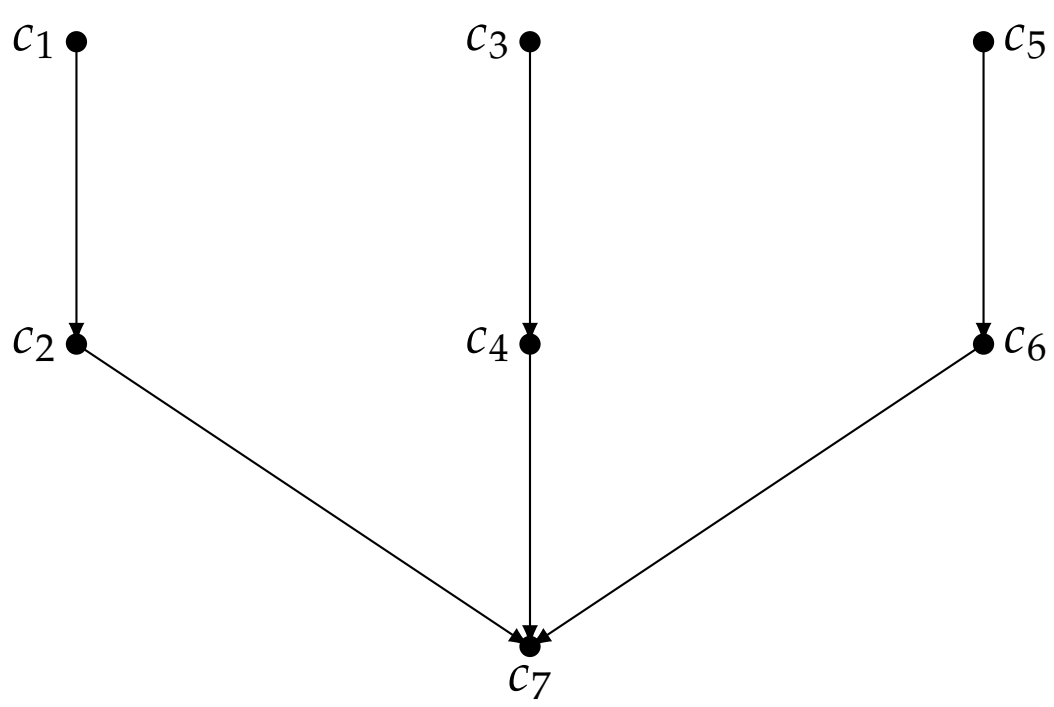
    By Theorem \ref{thm:bottom_horizontal_line}, if there is a set of two candidates which are not pointed to by any other candidates in this graph, those two will form a $(\nicefrac{4}{5}, 1)$-stable set, and hence, by movement (Proposition \ref{prop:lifting}), a $(\nicefrac{4}{5}, \nicefrac{1}{2})$-stable set as well. Take, then, $\set{c_1,c_3}$, $\set{c_1,c_5}$, and $\set{c_3,c_5}$, which all satisfy this property, so there is no unique winning set of this type.
\end{ex}
\subsection{Computational Considerations for Stability}

Stability exhibits markedly different computational behavior depending on the values of $p$ and $r$. For instance, \citet{aziz2017condorcet} show that deciding whether a locally stable committee exists is NP-complete for the Droop quota $q=\lfloor n/(k+1)\rfloor+1$, which corresponds to the point $\left((\lfloor n/(k+1)\rfloor+1)/n,\ \nicefrac1k\right)$ on the $r=\nicefrac1k$ line of Figure \ref{fig:prstability} (via Proposition \ref{prop:WGS-LS}). Notably, a different point on this same line, $p=\nicefrac1n$ (corresponding to the smallest possible quota, $q=1$), is efficiently decidable via homogeneity, as we show in Theorem \ref{thm:left_vertical_line_characterization} below.

More broadly, our analysis in this section identifies several slices of the $(p,r)$-plane that are efficiently computable: the $p=1$ line, for any $r$; the $p=\nicefrac1n$ line, for any $r$; and the $r=1$ line, for $p\le\nicefrac12$. Together with the hardness result of \citet{aziz2017condorcet} above, this indicates that tractability along the $(p,r)$-plane depends delicately on the specific point in question, rather than on any simple monotonic pattern in $p$ or $r$ alone.

\paragraph{Tractability for $(1,r)$ for any $r \le 1$.} Theorem \ref{thm:1r-univreal} establishes universal realizability for $(1,r)$-stability, and further shows that a stable set can be found in $\bigO(m^2n)$ time via Algorithm \ref{alg:lexicosort}. Hence, a stable set can always be found for any point on this segment efficiently.

\paragraph{Tractability for $(\nicefrac1n,r)$ for any $r \le 1$.} Proposition \ref{prop:stab_coal_equiv_1n} establishes, for $r \le 1$, that a set $S$ satisfies $(\nicefrac1n,r)$-stability if and only if $S$ satisfies $(\nicefrac1n,r)$-coalitions. Theorem \ref{thm:coal_pnp} establishes that $(\nicefrac1n,r)$-coalitions is decidable in $\bigO(\max(n\ceil{rk},k))$ time whenever $r \le 1$. Hence, $(\nicefrac1n,r)$-stability is decidable in $\bigO(\max(n\ceil{rk},k))$ time  for any $r \le 1$.

\paragraph{Tractability for $(p,1)$ for any $p\le\nicefrac12$.} \citet{aziz2017condorcet} provide a polynomial time algorithm for finding a strong Gehrlein stable set, or determining one does not exist. In the language of our stability framework, this corresponds to $(\nicefrac{1}{2},1)$-stability. Proposition \ref{prop:bottom_line_pnp_1} extends this result to $(p,1)$-stability for $p\le\nicefrac{1}{2}$ by utilizing Theorem \ref{thm:bottom_horizontal_line} and the proof of Theorem 1 from \citet{aziz2017condorcet}.

\begin{prop}\label{prop:bottom_line_pnp_1}
    There exists an algorithm that, given an election $(C,V,\succ)$ and a value of $p\le\nicefrac{1}{2}$ and $k$, in $\bigO(m^2n)$ time produces a unique $(p,1)$-stable committee, or decides that one does not exist.
\end{prop}

In fact, since values of $p$ change their behavior only in multiples of $\nicefrac{1}{n}$, we can efficiently find the smallest value of $p \le \nicefrac{1}{2}$ (represented as $p=\nicefrac{w}{n}$) for which a stable subset exists for a given election, or determine if one does not exist.

\begin{cor}\label{cor:bottom_line_pnp_2}
    Given an election $E$, there is an $\bigO(m^2n^2)$ time algorithm for finding the lowest value of $p$ where $p=\nicefrac{w}{n}$ for $w\in \Z \cap [1,\nicefrac{n}{2}]$ for which a $(p,1)$-stable set exists, or deciding that one does not exist on this election.
\end{cor}

\section{Applications to Party List Preferences}\label{sec:partylists}
We next demonstrate how our coalitions and stability frameworks can be used to analyze party list elections. In a party list election, candidates are organized into ordered lists (typically corresponding to political parties) and voters cast their ballot for a list rather than for individual candidates directly; when a party wins seats, the seats are filled by candidates from its list in the order the party specified. 

Party list electoral systems are widely used in practice, and have been analyzed as a restricted domain of elections by \citet{aziz2017condorcet}, where voters are all divided into blocks who share preferences up to at least their top $k$. We will analyze stability notions within party list preferences, and consider a further refinement we call \emph{complete} party list preferences which assume that voters can rank party lists among each other, which allows us to get stronger results.

\begin{defn}[Adapted from Definition 8 of \citet{aziz2017condorcet}] \label{def:party_list}
    An election $E=(C,V,\succ)$ is a party list election for committee size $k$ if there exists a positive integer $h$ where $C$ is a disjoint union of ordered party lists $\set{C_i: i\in [h]}$, letting $C_i=(c_{i,1},\dots, c_{i,\abs{C_i}})$, where:
    \begin{enumerate}
        \item $\abs{C_i}= k$ for all $i\in[h]$;\footnote{Existing work such as that of \citet{aziz2017condorcet} only requires that $\abs{C_i}\ge k$. We require $\abs{C_i}=k$, which corresponds to a party naming exactly as many candidates as there are seats to fill.
        }
        \item For each voter, there is some list $C_i$ such that they prefer every candidate in $C_i$ to every candidate outside of $C_i$ (and we call the set of such voters $V_i$); and
        \item If $\topp(v)\in C_i$, then the top $k$ positions of $\succ_v$ are $C_i$ in the order listed; that is, every voter ranks the candidates in their top list in the same order.\footnote{When $h=1$, all voters rank all candidates in the same order, and thus it is a unanimous preference ordering.}
    \end{enumerate}
\end{defn}
We will write $\mathcal{E}_{\mathrm{list},k}$ to denote all finite party list elections for list size $k$. One popular method for party lists elections is the D'Hondt method, also called the Jefferson method \cite{dhondt,balinski1978,medzihorsky2019}, which allocates seats proportionally to each list's vote share by repeatedly awarding the next seat to whichever list currently has the highest votes-per-seat ratio.

Additionally, certain real world systems (particularly the majoritarian systems we will discuss in Section \ref{sec:majorityjackpot}) can be viewed as assuming that voters not only have their top list organized in order, but that they also have ordinal preferences over the party lists themselves. We introduce the following restriction of party list preferences which we call \emph{complete} party list preferences. Intuitively, complete party list preferences are party list preferences where every voter, rather than merely favoring one list over all others, ranks \emph{every} list from most to least preferred, and ranks the candidates within each list in that list's fixed order (and not just their first list). 

\begin{defn}\label{def:cpl}
    An election $E=(C,V,\succ)$ is a \emph{complete} party list election for committee size $k$ if $C$ is a disjoint union of ordered party lists $\set{C_i: i\in [h]}$, letting $C_i=(c_{i,1},\dots, c_{i,\abs{C_i}})$ where:
    \begin{enumerate}
        \item $\abs{C_i}=k$ for all $i\in[h]$;
        \item For each voter $v \in V$, there is some permutation $(i_1,\dots,i_h)$ of $[h]$ where $C_{i_1}\succ_v C_{i_2}\succ_v\dots\succ_v C_{i_h}$; and
        \item For each voter $v \in V$, $i\in [h]$ and $j\in [\abs{C_i}]$, $c_{i,j}\succ_v c_{i,j+1}$; that is, every voter ranks the candidates in \emph{each} list in the same order.
    \end{enumerate}
\end{defn}
We will at times write $\mathcal{E}_{\mathrm{CPL},k}$ for the set of all finite complete party list elections of size $k$. We note that preferences like this are similar to those of \citet{fishburn1981}, where voters express preferences over different possible committees.

The additional structure of complete party lists yields new results for stability. Since $(p,r)$-stability implies $(p,r)$-coalitions (Proposition \ref{prop:stab_coal}), these results also give sufficient conditions for the corresponding coalitions properties to be realizable. The remainder of this section develops specific results for $(p,1)$-stability and $(\nicefrac{1}{2}, r)$-stability.


\subsection{Stability for Complete Party Lists when $r=1$}

Under the additional structure of complete party list elections, we can refine Theorem \ref{thm:bottom_horizontal_line} by determining whether there exists a special candidate that we call a \textit{$p$-winner}.
\begin{defn}\label{def:pwinner}
A candidate $c \in C$ is a $p$-winner of an election $E=(C,V,\succ)$ if for all $ d\in C\setminus \set{c}$,  $\abs{\set{v\in V : d\succ_v c}}< np$.
\end{defn}
An immediate consequence of this definition in terms of the $p$-preference graph is that a candidate $c \in C$ is a $p$-winner if and only if it has an in-degree of 0. Note that when $p=\nicefrac{1}{2}$, the $p$-winner is a (strong) Condorcet winner. Also, when $p =1$, the $p$-winner is a candidate that is not unanimously defeated by any other candidate,\footnote{Substituting $p=1$ into Definition \ref{def:pwinner}, $c$ is not a $p$-winner if and only if there is a $d\ne c$ such that $\abs{\set{v\in V : d\succ_v c}}\ge n$, which is true if and only if there is a $d\ne c$ which unanimously defeats $c$.} and hence always exists in any election. Towards a result for $(p,1)$-stable sets in general, we employ the following lemmas.

\begin{lem}\label{lem:cpl_cc_1}
    In a complete party list election, if there are elements from at least two different lists in the same connected component of the \(p\)-preference graph (as in Definition \ref{def:wgraph}), then the entirety of all those lists must also be in that component.
\end{lem}

\begin{lem}\label{lem:cpl_cc_2}
    In a complete party list election, all connected components of the $p$-preference graph are either singletons or of size at least \(2k\). 
\end{lem}

\begin{thm}
\label{thm:cpl_p1-stability}
    In a complete party list election, there is a \((p,1)\)-stable set if and only if there is a \(c\in C\) that is a \(p\)-winner.
\end{thm}

This allows us to make the following observation about strong Gehrlein stable sets.
\begin{cor}\label{cor:cpl_gehrlein}
    In a complete party list election \(E\in \mathcal{E}_{\mathrm{CPL},k}\), there is a strong Gehrlein stable set if and only if there is a Condorcet winner in the corresponding single winner election for \(E\).
\end{cor}

As a point of comparison with the party list setting, \citet{aziz2017condorcet} note that no guarantees can be made about the party list setting given strong Gehrlein stability, other than if there is a strict majority of voters who rank a particular party first then their top $k$ candidates will comprise the committee. On the other hand, Corollary \ref{cor:cpl_gehrlein} demonstrates that the additional structure of the \emph{complete} party list setting provides a provable guarantee about strong Gehrlein stability, namely, that the existence of a Condorcet winner also guarantees the existence of a Gehrlein stable set. 


\subsection{Other Majority-based Stability Notions} 
In the event that there is no Condorcet winner in the single winner election, the additional flexibility of the $(p,r)$-stability framework allows us to consider finding values of $r$ for which we can still find $(\nicefrac{1}{2}, r)$-stable sets. To do this, we introduce the following definitions.

\begin{defn}\label{def:under_p}
    Consider a complete party list election \(E\in\mathcal{E}_{\mathrm{CPL},k} \) with party lists \(C_1,\dots,C_h\). For \(D\in\set{C_1,\dots,C_h} \) define 
    $$
    \mathrm{under}_p(D) = \set{i\in[h] : \left|\set{v\in V : D \succ_v C_i}\right| \ge np}
    $$
    That is, $\mathrm{under}_p(D)$ is the set of indices of the party lists that at least $np$ voters rank entirely below $D$.
\end{defn}
\begin{defn}
    Given $S \subseteq C$, we call $c\in C\setminus S$ a \((p,r)\)-destabilizer of an election \(E=(C,V, \succ)\) if there is a set of at least \(np\) voters who prefer \(c\) to some subset of \(S\) of size greater than \(k(1-r)\). 
\end{defn}

The following equivalence is immediate by Definition \ref{def:prstabind}.

\begin{lem}\label{lem:stableiffnodestab}
    \(S \subseteq C\) is \((p,r)\)-stable if and only if $S$ has no \((p,r)\)-destabilizers.
\end{lem}

We use the following lemmas and definitions towards a result about the realizability of $(\nicefrac{1}{2},r)$-stability.
\begin{lem}\label{lem:cpl_nodestab}
    In a complete party list election, for $D \in \set{C_1,\dots,C_h}$, \(c\in D\setminus S\) cannot be a \((p,r)\)-destabilizer of \(S\) if \[\sum_{i\in \mathrm{under}_p(D)} \abs{C_i\cap S} \le (1-r)k\]
\end{lem}

\begin{defn}
\label{defn:top_cc_preference_graph_candidates}
    Let $p\le \nicefrac{1}{2}$, $G$ the $p$-preference graph, and $G'$ its condensation with vertices representing the strongly connected components of $G$. Since $p\le\nicefrac12$, $G'$ is a transitive tournament (as shown in the proof of Proposition \ref{prop:bottom_line_pnp_1}), and hence has a unique component with in-degree zero in its total ordering. We define $\mathcal{G}^*_p$ to be this component.
\end{defn}

\begin{lem}\label{lem:topcc_cpl}
    Consider a complete party list election $E$ where each party $i$ is assigned a non-negative integer $k_i$ (interpreted as the number of seats in the committee they will receive). Let $I^*_p = \set{i \in [h] : C_i \subseteq \mathcal{G}^*_p}$ be the indices of the lists comprising $\mathcal{G}^*_p$. If the system of equations below is satisfiable, then \((p,r)\)-stability is realizable for $E$.
    \begin{align}
        \sum_{i\in \mathrm{under}_p(C_j)} k_i &\le (1-r)k \ \text{  for all $j \in I^*_p$}\label{eqn:topcc_cpl_1}\\
        \sum_{i\in I^*_p} k_i &= k \label{eqn:topcc_cpl_2}
    \end{align}
    In particular, \(S=\bigcup_{i\in I^*_p} \topp_{k_i}(C_i)\) is a \((p,r)\)-stable set.
\end{lem}

With all of these lemmas (as well as additional technical results on linear systems presented in Appendix \ref{app:partylists}), we can now prove our main result for $(\nicefrac{1}{2},r)$-stability for complete party list elections.

\begin{thm}\label{thm:cpp-r-bound}
    Let $\ell$ be the number of parties in the top connected component of a complete party list election $E=(C,V,\succ)$ with party lists $\set{C_i}$. If \(p=\nicefrac{1}{2}\) and \(n\) is odd, then \((\nicefrac{1}{2},r)\)-stability is realizable on this preference profile for all $r\le \frac{1}{2}-\frac{\ell}{2k}$.
\end{thm}

\section{Applications to Majority Jackpot Systems}\label{sec:majorityjackpot}
We next turn our attention to majority jackpot systems, which aim to increase the likelihood of a single party having a majority by giving extra representation to a ``winning'' party. This is typically done by assigning some fraction $r$ of the elected committee to a winning party, often the largest party when it receives more than some fraction $p$ of the vote. Such systems are currently used in various places, including Italy's provincial elections \cite{britannica-italy} and the national elections of San Marino \cite{sanmarino} and Armenia \cite{armenia}. Because this resembles the form of our $(p,r)$ parameterization, we analyze majority jackpot systems as a case study from the perspective of our coalitions and stability frameworks.





\subsection{Existing Systems and Criteria}
We begin with two illustrative examples of majority jackpot systems, those of Armenia and San Marino, and discuss their fundamental mathematical properties.

\begin{ex}[Armenia's electoral system \cite{armenia}]\label{ex:armenia}
    Armenia's electoral system (as per the 2015 constitutional provision) assumes a party list election $E\in \mathcal{E}_{\mathrm{list},k}$, with candidates organized in lists \(C_1,C_2,\dots,C_h\) which we assume have length \(k=101\) (matching the size of the parliament). Voters choose one list, and seats are distributed proportionally, unless a list is assigned at least $51$ seats, but less than $55$, in which case it is assigned $55$ seats.
\end{ex}

Armenia's bonus is small, and only applies in cases of a majority party. The particular importance of this is that it can be analyzed simply with respect to voters' top preferences.
\begin{ex}[San Marino's electoral system \cite{sanmarino}]\label{ex:sanmarino}
    San Marino's electoral system assumes a complete party list election \(E\in \mathcal{E}_{\mathrm{CPL}, k}\), with candidates organized in lists \(C_1,C_2,\dots,C_h\) which we assume have length \(k=60\) (matching the size of the parliament).  First, voters choose one list.
    \begin{enumerate}
        \item If the list with the greatest number of votes has over \(50\%\) of the vote, then the legislature is chosen proportionally.
        \item Otherwise, hold a runoff between the top two lists. The winner receives \(\floor{\nicefrac{k}{2}}+5\) seats, and the remainder are distributed proportionally.\footnote{In this system, the proportional distribution is by the d'Hondt method \cite{dhondt}.}
    \end{enumerate}
\end{ex}

Italy's provincial elections use a very similar system to San Marino, with $k=32$ and the winner receiving 60\% \cite{britannica-italy}. 

We note that Armenia's electoral system is modeled using party list preferences, whereas San Marino's requires complete party list preferences. This distinction arises because which two lists reach the runoff in San Marino's system is only determined by the vote itself. For this reason, modeling the rule as a function of a single profile $E$ requires knowing each voter's preference between \emph{every} pair of lists, not just their top choice. Armenia's rule, by contrast, only ever depends on first-place vote totals, so party list preferences suffice. 

With this in mind, we introduce a formal definition of a majority jackpot system over (complete) party list preferences for our analysis. 

A key tool of our definition is a \emph{winning determination function} $\nu: \mathcal{E}_{\mathrm{list}, k} \to [h] \cup \set{\emptyset}$ that outputs either the index of a party list or the empty set.\footnote{Since $\mathcal{E}_{\mathrm{CPL}, k} \subseteq \mathcal{E}_{\mathrm{list}, k}$, the winning determination function also applies to complete party lists.} Conceptually, we can interpret a party list $i \in [h]$ to be the \emph{winning party} if $\nu(E) = i$; in the case where $\nu(E) = \emptyset$, we will say there is no winning party. When there is a winning party, a majority jackpot system will have this winning party assigned a majority of the seats in the committee. 

In principle, $\nu$ can implement any logic for selecting a winning party. For example, Armenia's system (Example \ref{ex:armenia}) determines $\nu$ by the party with the most first-preference votes, but the winning determination function could select a winning party by other rules (e.g., the Condorcet winner among parties when it exists, or via a runoff between the top two parties when a single party doesn't achieve a majority of top votes, as in San Marino's system).

\begin{defn}[Majority jackpot for (complete) party list preferences]\label{def:mj_cpl}
    Let $\tilde{\mathcal{E}} \in \set{\mathcal{E}_{\mathrm{list},k}, \mathcal{E}_{\mathrm{CPL},k}}$. A rule $\Rule$ (defined on  elections $E=(C,V,\succ) \in \tilde{\mathcal{E}} $ with party lists $C_1,\dots,C_h$) is a majority jackpot rule (over $\tilde{\mathcal{E}} $) if there is a winning determination function $\nu$, not identically $\emptyset$ on $\tilde{\mathcal{E}} $, such that for every election $E\in\tilde{\mathcal{E}} $, if $\nu(E)\ne \emptyset$ then $|C_{\nu(E)} \cap \Rule(E,k)| > \nicefrac{k}{2}$. Additionally,
    \begin{itemize}
        \item If $\nu(E)\ne \emptyset$ for all $E\in \tilde{\mathcal{E}} $, then we call $\Rule$ an \emph{unconditional} majority jackpot rule.
        \item If for all $E\in \tilde{\mathcal{E}} $, $\nu(E)\ne \emptyset$ if and only if some condition $\chi$ is satisfied for the election $E$, then we call $\Rule$ a majority jackpot rule \emph{conditional on $\chi$}. Under this definition, a conditional majority jackpot rule has exactly one such condition $\chi$.
    \end{itemize}
\end{defn}
Armenia's election system, for instance, is a majority jackpot system \emph{conditional} on the existence of a party with a majority of the party vote. The majority party, in this case, is the output of the winning determination function. On the other hand, the electoral systems of San Marino and the Italian provinces are unconditional majority jackpot methods.


\subsection{Coalitions Analysis}
\label{sec:mjs_coalitions}

As noted above, a majority jackpot rule's winning determination function $\nu$ does not need to select the party with the most first-preference votes. This modeling flexibility allows us to characterize when various electoral phenomena can (or cannot) occur using our coalitions and stability frameworks.

One such phenomenon is an \emph{electoral inversion} -- a situation where a party gains a weak majority of the first-preference votes but not a weak majority of the seats. Electoral inversions are a phenomenon mostly discussed in the context of parliamentary and electoral college elections (see \citet{may1948} and \citet{miller2011} for analysis), but they are also present in proportional systems (such as the single transferable vote in Cincinnati city council elections until 1955 \cite{barber1995,hoag1926,santucci2021}). The following example provides a simple case of an electoral inversion in a complete party list election with three parties, when $\nu$ selects a ``compromise party'' instead of the party favored by a majority of first-preference votes.

\begin{ex}
Consider the following complete party list profile with parties $C_1, C_2, C_3$:
\begin{center}
\begin{tabular}{c|c}
   \# Voters  & Preferences \\
   \hline
    51 & $C_1\succ C_2\succ C_3$\\
    49 & $C_3\succ C_2\succ C_1$
\end{tabular}
\end{center}
Depending on the system in place, a winning determination function $\nu$ could select party 2 as a compromise between the two factions, and assign party 2 more than $\nicefrac{k}{2}$ of the seats to the committee. This would still satisfy the definition of a majority jackpot rule, despite $\nu(E)\ne 1$.    
\end{ex}

In a complete party list election, $(\nicefrac{1}{2},\nicefrac{1}{2})$-coalitions is closely tied to avoiding electoral inversions. Since a majority of voters favoring the same party automatically agree on that party's entire top-$k$ block (Definition \ref{def:party_list}), $(\nicefrac{1}{2},\nicefrac{1}{2})$-coalitions guarantees such a party at least $\ceil{\nicefrac{k}{2}} \ge \nicefrac{k}{2}$ seats. When $k$ is even, this is only half of the seats, so avoiding an electoral inversion is a weaker goal than what majority jackpot systems are designed to achieve (namely, guaranteeing the winning party a \emph{strict} majority of seats). When $k$ is odd, $\ceil{\nicefrac{k}{2}}>\nicefrac{k}{2}$, so the two goals coincide.

The idea that any party getting half the vote gets a little over half the seats can accordingly be expressed as $(\nicefrac{1}{2}, r)$-coalitions, with $r>\nicefrac{1}{2}$ -- which, for even $k$, is a strictly stronger property than $(\nicefrac{1}{2},\nicefrac{1}{2})$-coalitions. For example, Armenia's electoral system satisfies $(\nicefrac{1}{2},\nicefrac{55}{101})$-coalitions and San Marino's satisfies $(\nicefrac{1}{2},\nicefrac{35}{60})$-coalitions. This suggests that $(\nicefrac{1}{2},r)$-coalitions (for $r > \nicefrac{1}{2}$) is a natural point of interest in the framework.

The following result characterizes the relationship between $(\nicefrac12,r)$-coalitions (for $r>\nicefrac12$) and majority jackpot rules more broadly, under a few technical conditions.

\begin{thm}\label{thm:mj_majoritycoal}
    Let condition $\psi$ hold for $E$ if and only if there is a group of candidates $C_\mathrm{maj} \subseteq C$ where $ \abs{C_\mathrm{maj}}= \floor{\nicefrac{k}{2}}+1$ and 
    \[\abs{\set{v\in V: {\forall} c\in C_\mathrm{maj}, \ {\forall}  d\in C\setminus C_\mathrm{maj} , \ c\succ_v d}}>\nicefrac{n}{2}\]
    In other words, $\psi$ holds if there is a group of candidates who a majority of voters prefer over all other candidates.

    \begin{enumerate}
        \item Consider the setting of complete party list elections. Suppose $n$ is odd. If $\Rule$ is candidate-wise undominated and a majority jackpot rule conditional on $\chi$ where $\psi\implies \chi$, and $C_\mathrm{maj} \subseteq C_{\nu(E)}$ when $\chi$ holds, then $\Rule$ satisfies $(\nicefrac{1}{2}, r)$-coalitions for some $r>\nicefrac{1}{2}$.\footnote{We note the similarity of this to the mutual majority criterion, which states that a single winner election should elect a member of such a $C_\mathrm{maj}$ as described by \citet[p. 153]{tideman2017}.} 
        \item Consider the setting of party list elections. If a rule satisfies $(\nicefrac{1}{2}, r)$-coalitions with $r > \nicefrac{1}{2}$, then it is a majority jackpot rule conditional on a condition at least as strong as $\psi$.
    \end{enumerate}
\end{thm}


\subsection{Stability Analysis}

Next, we examine three additional voting phenomena, studied by \citet{vandeemen1993theo} in the context of party preferences, and connect them to majority jackpot systems using our stability framework. Since Van Deemen's paradoxes are phrased in terms of voters' preferences over entire party lists rather than individual candidates, we work within the setting of complete party list preferences. 

\medskip

\noindent \underline{\textsc{\citet{vandeemen1993theo}'s Three Paradoxes:}}

\begin{enumerate}
    \item The \emph{More Preferred Less Seats} paradox: a party $A$ preferred by a majority of voters to party $B$ may nonetheless receive fewer seats than $B$.
    \item The \emph{Condorcet Winner Turns Loser} paradox: a party $A$ that beats every other party head-to-head may still fail to win the most seats. This is a strengthening of the first paradox, since it requires $A$ to beat \emph{every} rival, not just one.
    \item The \emph{Majority Reversal} paradox: the seat assignments may exactly reverse the majority preference for every pair of parties -- that is, for every pair of parties $A$ and $B$, a majority of voters prefers $A$ to $B$ if and only if $B$ receives more seats than $A$. This is a strengthening of the second paradox, since it requires the reversal to hold consistently for every pair of parties, whereas the second paradox holds for the top-ranked party. Within the literature, the electoral inversions from Section \ref{sec:mjs_coalitions} are also sometimes called majority reversals \cite{may1948,miller2011}; to distinguish between the two, we refer to this stronger, all-pairs version specifically as a \emph{full} majority reversal.
\end{enumerate}

As noted by \citet{vandeemen1993theo}, Paradox 3 implies Paradox 2, which implies Paradox 1. Below we provide an example of a full majority reversal in a complete party list election.

\begin{ex} [Full Majority Reversal]
Consider the following complete party list election.
    \begin{center}
    \begin{tabular}{c|c}
       \# Voters  & Preferences \\
       \hline
        40 & $C_1\succ C_3\succ C_2$\\
        35 & $C_2\succ C_3\succ C_1$\\
        25 & $C_3\succ C_2\succ C_1$\\
    \end{tabular}
    \end{center}
Suppose our committee distributes the seats proportionally according to first preference vote. Then, the seats belonging to $C_1$ will be greater than $C_2$ which are greater than $C_3$. However, 60 voters prefer both $C_3$ and $C_2$ to $C_1$, and 65 prefer $C_3$ over $C_2$, so the majority preference order is $C_3\succ C_2\succ C_1$, exactly the opposite of the seat assignments.
\end{ex}

Each of these paradoxes can give rise to a new desideratum characterized by their avoidance. We mildly deviate from \citet{vandeemen1993theo}'s characterization and use weak majorities (at least $\nicefrac{n}{2}$ voters) instead of strict majorities (more than $\nicefrac{n}{2}$ voters), so that we may reason about the single, fixed value of $p = \nicefrac{1}{2}$ in $(p, r)$-stability. Using strict majorities instead enables analogous results to be proven for $(p,r)$-stability with some $p > \nicefrac{1}{2}$.

\medskip

\noindent \underline{\textsc{Desiderata Motivated by \citet{vandeemen1993theo}'s Three Paradoxes:}} We label these with the subscript $\mathrm{CPL}$ to indicate that they are applicable only in complete party list preferences.

\begin{enumerate}
    \item \emph{More Preferred More Seats} ($\mathrm{MPMS}_\mathrm{CPL}$), satisfied if whenever a weak majority of voters prefer $A$ to $B$, then $A$ gets at least as many seats as $B$ in the committee.
    \item \emph{Condorcet Winner Plurality} ($\mathrm{CWP}_\mathrm{CPL}$), satisfied if a weak Condorcet winner among parties (if one exists) wins at least as many seats as every other party, though not necessarily a majority. Mathematically, $C_w$ is a weak Condorcet winner among parties in a complete party list election $E\in \mathcal{E}_{\mathrm{CPL},k}$ with party lists $\set{C_i:i\in [h]}$ if, for every $i\in [h]$ with $C_i\ne C_w$, $\abs{\set{v\in V:\forall c\in C_w \ \forall d\in C_i, \ c\succ_v d}}\ge\nicefrac{n}{2}$.
    \item \emph{No Full Majority Reversal} ($\mathrm{NFMR}_\mathrm{CPL}$), satisfied when there is at least one instance where a weak majority of voters prefer party $A$ over party $B$, and $A$ has at least as many seats as $B$ in the committee. (Note: since full majority reversal requires this pattern to \emph{fail} for every pair, avoiding it only requires one pair where it holds.)
\end{enumerate}

We make two observations. First, these three desiderata exhibit the following implication chain.

\begin{prop}
\label{prop:mpms_cwp_nfmr_chain} $\mathrm{MPMS}_\mathrm{CPL}\implies\mathrm{CWP}_\mathrm{CPL}\implies\mathrm{NFMR}_\mathrm{CPL}$
\end{prop}

This chain of implications is expected, as each desideratum (up to the weak-inequality convention) is the negation of the corresponding paradox; hence, negating \citet{vandeemen1993theo}'s chain of implications (Full Majority Reversal Paradox $\implies$ Condorcet Winner Turns Loser Paradox $\implies$ More Preferred Less Seats Paradox) reverses its direction, yielding exactly the chain above.\footnote{This formal chain of logic also has implications for the occurrence of such paradoxes in practice. Since the More Preferred Less Seats paradox is the weakest (and hence easiest to trigger) of the three and the Majority Reversal paradox the strongest (and hence hardest to trigger), one would expect More Preferred Less Seats to be violated most often in practice and Majority Reversal least often. This is exactly what \citet{vandeemen1998emp} find in their empirical investigation of Dutch elections between 1982 and 1994: More Preferred Less Seats is violated much more often than Condorcet Winner Turns Loser, while the Full Majority Reversal was never observed in their sample.}

To facilitate our analysis, we introduce a preference graph over party lists, which records majority preferences between entire lists rather than between individual candidates.
\begin{defn}\label{def:cpl_graph}
    The $p$-complete party list preference graph $G^{(p)}_\mathrm{CPL}=(P,L_{\mathrm{CPL},p})$ of a complete party list election $E=(C,V,\succ)$ consists of nodes identified with the set of party lists $P=\set{C_i:i\in [h]}$ and a set of links $L_{\mathrm{CPL}, p}$ (also referred to as edges), defined by 
    $$(C_i,C_j) \in L_{\mathrm{CPL},p} \iff |\{v \in V  :  \forall c\in C_i , \ \forall d\in C_j, \  c\succ_v d\}| \ge np$$ 
    We will mostly discuss the graph for $p=\nicefrac{1}{2}$, and will denote $G^{(\nicefrac{1}{2})}_\mathrm{CPL}$ as $G_\mathrm{CPL}$ for brevity. We call this the \emph{majority party preference graph} for $E$.
\end{defn}
Our second observation uses this graph: when a complete party list election's majority party preference graph $G_\mathrm{CPL}$ is a tournament, our weak majority desiderata coincide exactly with \citet{vandeemen1993theo}'s original (strict) formulations. This follows as a tournament guarantees exactly one edge exists between every pair of parties, so every weak majority is already a strict one.

\subsubsection{A General Connection to Stability for Complete Party List Elections}

Next, we relate these desiderata to points in our stability framework. Our first result shows that $\mathrm{CWP}_\mathrm{CPL}$ and $\mathrm{NFMR}_\mathrm{CPL}$ are implied by $(\nicefrac{1}{2},1-\nicefrac{1}{h})$-stability for complete party list elections.

\begin{prop}
\label{prop:cwp_stability}
    Consider an election $E\in \mathcal{E}_{\mathrm{CPL},k}$ with $h$ party lists. If a committee $S$ satisfies $(\nicefrac{1}{2},1-\nicefrac{1}{h})$-stability, then it also satisfies $\mathrm{CWP}_\mathrm{CPL}$. Consequently, such an $S$ also satisfies $\mathrm{NFMR}_\mathrm{CPL}$.
\end{prop}

Hence, any committee satisfying $(\nicefrac{1}{2},1-\nicefrac{1}{h})$-stability automatically satisfies both $\mathrm{CWP}_\mathrm{CPL}$ and $\mathrm{NFMR}_\mathrm{CPL}$. As such, it avoids both the Condorcet-Winner-Turns-Loser and Full Majority Reversal paradoxes (in their weak majority forms).

$\mathrm{MPMS}_\mathrm{CPL}$, on the other hand, does not exhibit a clean relationship to $(\nicefrac{1}{2},r)$-stability. As the following example shows, there is an election where two different committee allocations can result in $(\nicefrac{1}{2},r)$-stability, yet one committee meets $\mathrm{MPMS}_\mathrm{CPL}$ while the other does not.

\begin{ex}\label{ex:mpms_fail}
    Consider the following election featuring three parties, electing a committee of size $9$:
    \begin{center}
        \begin{tabular}{c|c}
            \# Voters & Preferences \\\hline
            51 & $C_1\succ C_2\succ C_3$ \\
            49 & $C_3\succ C_1\succ C_2$
        \end{tabular}
    \end{center}
    Consider the seat allocation allotting $8$ seats to $C_1$, and 1 seat to $C_3$. By Proposition \ref{prop:ginc}, this satisfies $(\nicefrac{1}{2},\nicefrac{8}{9})$-stability. However, the \emph{majority} preference ordering of parties is $C_1\succ C_2\succ C_3$, so the allocation violates $\mathrm{MPMS}$. Instead, allocating $8$ seats to $C_1$ and $1$ seat to $C_2$ still satisfies $(\nicefrac{1}{2},\nicefrac{8}{9})$-stability, and also satisfies $\mathrm{MPMS}$, despite leaving the large minority of voters completely without representation. 
    \begin{center}
        \begin{tabular}{c|c|c|c|c}
              $C_1$ seats & $C_2$ seats & $C_3$ seats & Stability satisfied? & MPMS satisfied? \\\hline
             8& 1 & 0 &$(\nicefrac{1}{2},\nicefrac{8}{9})$ & Yes\\
             8& 0 & 1 &$(\nicefrac{1}{2},\nicefrac{8}{9})$ & No\\
             5& 0 & 4& $(\nicefrac{1}{2},\nicefrac{1}{2})$ & No\\
             5& 4 & 0 &$(\nicefrac{1}{2},\nicefrac{1}{2})$ & Yes\\
        \end{tabular}
    \end{center}
\end{ex}

The allocations in Example \ref{ex:mpms_fail} further illustrate that $\mathrm{MPMS}_\mathrm{CPL}$ and $(\nicefrac{1}{2},r)$-stability place different constraints on how seats are distributed. Allocating $5$ seats to $C_1$ and $4$ to $C_3$ satisfies $(\nicefrac{1}{2},\nicefrac{1}{2})$-stability but not $\mathrm{MPMS}_\mathrm{CPL}$, while allocating $5$ seats to $C_1$ and $4$ to $C_2$ satisfies both. This suggests that $\mathrm{MPMS}_\mathrm{CPL}$ and $(\nicefrac{1}{2},r)$-stability are different concepts in general.

\subsubsection{A Sharper Connection to Majority Jackpot Rules}
\label{sec:sharper_connection_mjs}

We next show how we can sharpen the stability results when we restrict attention to committees produced by majority jackpot rules. Proposition \ref{prop:cwp_stability} established $(\nicefrac{1}{2},1-\nicefrac{1}{h})$-stability as \emph{sufficient} for $\mathrm{CWP}_\mathrm{CPL}$ and $\mathrm{NFMR}_\mathrm{CPL}$ for an arbitrary committee. In Theorem \ref{thm:cwp_majjackpot_selection_function} below, we obtain a sharper, two-sided characterization: $(\nicefrac{1}{2},\nicefrac{1}{2})$-stability exactly determines which party a majority jackpot rule's winning determination function $\nu$ may select, under technical conditions.

To do so, we will restrict attention to complete party list elections $E\in \mathcal{E}_{\mathrm{CPL},k}$ whose majority party preference graph $G_\mathrm{CPL}$ is a tournament. This will allow us to leverage strong Gehrlein stable sets as part of our analysis, which require a strict majority (in Definition \ref{defn:strong_g_stability}) to establish that one candidate beats another. Since our desiderata are otherwise stated using weak majorities, the tournament assumption guarantees no pairwise comparison between parties results in a tie, so a weak majority for one party over another is automatically a strict one.

To ease exposition, we will let $\mathcal{T}_{\mathrm{CPL},k}$ be the subset of elections in $\mathcal{E}_{\mathrm{CPL},k}$ whose majority party preference graph $G_\mathrm{CPL}$ is a tournament. Note that for any $E  \in \mathcal{T}_{\mathrm{CPL},k}$, whenever a weak Condorcet winning party exists it must be unique due to the tournament structure.

We first establish that a Condorcet winning party's own top $\ell$ candidates already form a strongly Gehrlein stable committee.

\begin{lem}\label{lem:clist_gehr}
    Let $E \in \mathcal{T}_{\mathrm{CPL},k}$. Suppose $C_w$ is the weak Condorcet winner among parties $\set{C_i:i\in [h]}$. Then, for any $\ell\le k$, the top $\ell$ candidates of $C_w$ form a strongly Gehrlein stable set of size $\ell$.
\end{lem}

We now characterize exactly which party a majority jackpot rule's winning determination function may select in order to satisfy $(\nicefrac{1}{2},\nicefrac{1}{2})$-stability.

\begin{thm}
\label{thm:cwp_majjackpot_selection_function}
     Let $E \in \mathcal{T}_{\mathrm{CPL},k}$ and $\Rule$ be a majority jackpot rule.
    \begin{enumerate}
        \item If $\Rule$ satisfies $(\nicefrac{1}{2},\nicefrac{1}{2})$-stability, then $\nu$ selects the weak Condorcet winning party when it exists, and outputs $\nu(E)=\emptyset$ otherwise.
        \item If a weak Condorcet winning party exists, $\nu$ selects it, and $\Rule$ is candidate-wise undominated, then $\Rule$ satisfies $(\nicefrac{1}{2},\nicefrac{1}{2})$-stability.
    \end{enumerate}
\end{thm}

We can translate this characterization of $\nu$ into a guarantee on $\mathrm{CWP}_\mathrm{CPL}$ and $\mathrm{NFMR}_\mathrm{CPL}$.

\begin{cor}
\label{cor:nfmr_mjs}
      Let $E \in \mathcal{T}_{\mathrm{CPL},k}$. If a weak Condorcet winning party exists, any majority jackpot rule $\Rule$ satisfying $(\nicefrac{1}{2},\nicefrac{1}{2})$-stability satisfies $\mathrm{CWP}_\mathrm{CPL}$, and hence also satisfies $\mathrm{NFMR}_\mathrm{CPL}$.
\end{cor}

Hence, for any election $E\in\mathcal{T}_{\mathrm{CPL},k}$, any majority jackpot rule satisfying $(\nicefrac{1}{2},\nicefrac{1}{2})$-stability automatically satisfies both $\mathrm{CWP}_\mathrm{CPL}$ and $\mathrm{NFMR}_\mathrm{CPL}$ whenever a weak Condorcet winning party exists. As such, it avoids both the Condorcet-Winner-Turns-Loser and Full Majority Reversal paradoxes in \citet{vandeemen1993theo}'s original (strict) sense, since the tournament assumption makes weak majorities and strong majorities coincide.

\section{Discussion}
\label{sec:discussion}
We introduced two frameworks of committee criteria for multiwinner elections: $(p,r)$-coalitions and $(p,r)$-stability. For coalitions, we characterized the pairs $(p,r)$ for which a satisfying committee exists in every election, and, for a given election, characterized when one exists at $p=\nicefrac{1}{n}$ through homogeneity of preferences. We also gave an efficient algorithm that, for any $p$ and $r$, either finds a committee satisfying $(p,r)$-coalitions or determines that none exists. For stability, computational difficulty depends on location in the $(p,r)$-plane. Deciding whether a locally stable committee exists is NP-complete for the Droop quota \citep{aziz2017condorcet}, while we identified efficiently solvable slices: $p=1$, $p=\nicefrac{1}{n}$, and $r=1$ with $p\le\nicefrac{1}{2}$.

We then applied our frameworks to two settings motivated by real electoral systems: complete party list elections and majority jackpot systems. For complete party list elections, we showed that a $(p,1)$-stable committee exists if and only if some candidate is a $p$-winner, and identified values of $r$ for which $(\nicefrac{1}{2},r)$-stability is guaranteed to be realizable. For majority jackpot rules, under technical assumptions, we related $(\nicefrac{1}{2},r)$-coalitions with $r>\nicefrac{1}{2}$ to the rule's winning determination function, and showed that $(\nicefrac{1}{2},\nicefrac{1}{2})$-stability constrains which party that function may select. We also connected these notions to the paradoxes of \citet{vandeemen1993theo}: for complete party list elections, $(\nicefrac{1}{2},1-\nicefrac{1}{h})$-stability is sufficient for $\mathrm{CWP}_\mathrm{CPL}$ and $\mathrm{NFMR}_\mathrm{CPL}$.

Our results suggest two natural directions for future work. First, we identified several efficiently solvable slices of the stability landscape. It would be interesting to determine the complexity at other points of the $(p,r)$-plane, in particular around $(\nicefrac{1}{2},\nicefrac{1}{2})$, which features in our analysis of complete party lists and majority jackpot rules. Second, our frameworks are defined for unrestricted preferences, but our sharpest results, such as the characterization of $(p,1)$-stability through $p$-winners, hold in the restricted domain of complete party lists. It would be interesting to identify other restricted domains, such as single-peaked preferences, in which one can characterize more precisely which points of the $(p,r)$-plane admit a stable committee, or the complexity of finding one.


\newpage

\bibliographystyle{apalike}
\bibliography{citations}

\newpage
\appendix

\begin{appendices}

\section{Proofs for Section \ref{sec:coalitions}}

\subsection*{Proof of Proposition \ref{prop:sc-prcoal}}

\begin{proof}
    From the definition, a committee \(S\) satisfies the solid \((\nicefrac{q}{n},\nicefrac{1}{k})\)-coalitions property if, for all groups of voters \(V^*\) with \(\abs{V^*}\ge q\)  and a candidate \(c^*\) where \(\topp V^* = c^*\), then \(c^* \in S\), which is just the \(q\)-solid coalitions property. 
\end{proof}


\subsection*{Proof of Proposition \ref{prop:fixmaj}}

\begin{proof}
    By definition, a committee \(S\) satisfies the \((\nicefrac{1}{2}+\nicefrac{1}{2n},1)\)-coalitions property if, for all groups of voters \(V^*\) with  \(\abs{V^*}\ge n(\nicefrac{1}{2} + \nicefrac{1}{2n}) = \nicefrac{(n+1)}{2} \) (which is equivalent to \(\abs{V^*}>\nicefrac{n}{2}\) since $|V^*|$ is an integer and exactly one of $\nicefrac{n}{2}$ and $\nicefrac{n+1}{2}$ is an integer) where \(\abs{\topp_{k} V^*} = k\), then \(\topp_{k} V^* \subseteq S\). Since \(\abs{\topp_{k} V^*} = k\), we have \(\topp_{k} V^* = S\), which is equivalent to the fixed majority criterion.
\end{proof}


\subsection*{Proof of Proposition \ref{prop:prdummett}}

\begin{proof}
    By definition, a committee $S$ satisfies $(\nicefrac{\ell q}{n},\nicefrac{\ell}{k})$-coalitions if and only if, for all groups of voters $V^*$ with $\abs{V^*}\ge \nicefrac{\ell q}{n} \cdot n = \ell q$, and $\abs{\topp_{\ceil{\nicefrac{\ell}{k}\cdot k}} V^*}=\abs{\topp_{\ell} V^*}=\ell$, then $\topp_{\ell} V^* \subseteq S$. This is precisely the definition of Dummett's proportionality criterion with quota $q$.
\end{proof}


\subsection*{Proof of Proposition \ref{prop:p-coal-lifting}}

\begin{proof}
Consider a \(V'\) where \(\abs{V'}\ge p'n\) with \(\abs{\topp_{\ceil{rk}} V'} = \ceil{rk}\). Since $S$ satisfies $(p,r)$-coalitions, for all $V^*$ of size at least $pn$ for which $|\topp_{\ceil{rk}} V^*| = \ceil{rk}$, we have $\topp_{\ceil{rk}}V^* \subseteq S$. Since $p' \ge p$, we have $|V'|\ge p'n \ge pn$, so $\topp_{\ceil{rk}}V' \subseteq S$.
\end{proof}


\subsection*{Proof of Lemma \ref{lem:coalitions_numunique}}

\begin{proof}
     There are two cases to consider: either $\mathcal{V}$ is empty or not. When $\mathcal{V} = \emptyset$, we have $\zeta = |\mathcal{V}| = 0$. Additionally, for all $p \in (0,1]$, we have $1 \le \floor{\nicefrac{1}{p}}$. Therefore $\zeta \le \floor{\nicefrac{1}{p}}$. 
     
     Next, we consider the case when  $\mathcal{V} \ne \emptyset$. By construction, the elements of $\mathcal{V}$ are disjoint. We first note that $\zeta \le \nicefrac{1}{p}$; otherwise if $\zeta > \nicefrac{1}{p}$, we reach the following contradiction:
$$
n = |V_1\cup \dots \cup V_\kappa| \ge \left|\cup_{V_j \in \mathcal{V}} V_j \right| = \sum_{V_j \in \mathcal{V}} |V_j| \ge \sum_{V_j \in \mathcal{V}} pn = \zeta pn > n
$$
 Since $\mathcal{V} \ne \emptyset$, $\zeta$ is a positive integer and it cannot exceed $\nicefrac{1}{p}$, we also have $\zeta\le \floor{\nicefrac{1}{p}}$.
\end{proof}


\subsection*{Proof of Theorem \ref{thm:coalitions_ur}}

\begin{proof}

    We first prove the bidirectional statement of the theorem. 

    \medskip

     \noindent \((\impliedby)\) We proceed by the contrapositive and suppose \((p,r)\)-coalitions is not universally realizable for some $n$ and $m$. Then there exists a collection of preferences $\succ$ for $n$ voters over $m$ candidates for which $(p,r)$-coalitions is not realizable. 
    
    Consider the set $\mathcal{V} = \set{V_j : j \in [\kappa] \text{ and } |V_j| \ge pn }$. We claim that $\mathcal{V} \ne \emptyset$; otherwise, if $\mathcal{V} = \emptyset$, then $\left|\cup_{V_j \in \mathcal{V}} \topp_{\ceil{rk}}V_j\right| = 0 \le k$, so $(p,r)$-coalitions would be realizable by Corollary \ref{cor:coalitions_checker}, yielding a contradiction.

    By Lemma \ref{lem:coalitions_numunique}, $\zeta \le \floor{\nicefrac{1}{p}}$. Since $(p,r)$-coalitions is not realizable on this preference profile, by Corollary \ref{cor:coalitions_checker} and the union bound we must have 
    $$
    k < \left|\cup_{V_j \in \mathcal{V}} \topp_{\ceil{rk}}V_j\right| \le \sum_{V_j \in \mathcal{V}}\left|\topp_{\ceil{rk}}V_j\right| = \zeta \ceil{rk} \le \ceil{rk}\floor{\nicefrac{1}{p}}
    $$
    Hence $k < \ceil{rk}\floor{\nicefrac{1}{p}}$, completing the contrapositive. So, if $\ceil{rk}\floor{\nicefrac{1}{p}} \le k$, $(p,r)$-coalitions is universally realizable.

    \medskip

    \noindent \((\implies)\) Suppose $(p,r)$-coalitions is universally realizable. Since $p \in (0,1]$, $\floor{\nicefrac{1}{p}} \ge 1$. If $\floor{\nicefrac{1}{p}} = 1$, then since $r \le 1$ we have $\ceil{rk}\le k$. Hence, $\ceil{rk} \floor{\nicefrac{1}{p}} = \ceil{rk}\le k$ so the inequality automatically holds. It remains to consider the case when $\floor{\nicefrac1p} \ge 2$.
    
    The universal realizability of $(p,r)$-coalitions implies that for all possible finite elections $E = (C,V, \succ)$ of $\mathcal{E}$, $(p,r)$-coalitions is realizable for $E$. Since $(p,r)$-coalitions holds for all finite elections, it must also hold for the specific election $E'$ we construct below. The election $E'$ will consist of $n$ voters and $m$ candidates (whose specific values are determined below) with $\floor{\nicefrac{1}{p}}$ groups of voters ($V_1,\dots,V_{\floor{\nicefrac{1}{p}}})$ of size exactly $\ceil{np}$, each group agreeing on a distinct top-$\ceil{rk}$ block of preferences.
 
    To construct such an $n$, we need $\lceil np \rceil$ to equal a specific target value $\omega$ exactly, so that $\floor{\nicefrac{1}{p}}$ disjoint groups of size $\omega$ partition $n$ with no leftover voters. Since $\floor{\nicefrac1p} \ge 2$ and $p \in (0,1]$ by definition, we must have $0 < p \le \nicefrac{1}{2}$. 
    
    Every such $p$ falls into one of two cases: either $p = \nicefrac{1}{z}$ for some integer $z$, or $\nicefrac{1}{(z+1)} < p < \nicefrac{1}{z}$ for some integer $z$. We consider both cases in turn below.
    \begin{enumerate}
        \item When $p=\nicefrac{1}{z}$ for some integer $z$, let $n = \omega z$ for a positive integer $\omega$. Then $\abs{V_i}=\ceil{np}=\ceil{\omega z \cdot \nicefrac{1}{z}}=\ceil{\omega}=\omega$.
        \item When $\nicefrac{1}{(z+1)}<p<\nicefrac{1}{z}$, let $\omega = z$ and $n = \omega z = z^2$. Then $\abs{V_i}=\ceil{np}=\ceil{\omega z p}$. Since $\omega z \cdot \frac{1}{z+1} < \omega zp < \omega z \cdot \frac1z = \omega$, and $\omega z \cdot \frac{1}{z+1} = \omega\left(1 - \frac{1}{z+1}\right) = \omega - \frac{\omega}{z+1} > \omega - 1$, we have $\omega - 1 < \omega zp < \omega$, so $|V_i| = \ceil{\omega zp} = \omega$.
    \end{enumerate}
    Hence, in either case we can write $n = \omega z$ for some positive integers $\omega$ and $z$, such that $|V_i| = \omega$ for every group of voters $V_i$.
    
    Next, we also note that in either case, we also have $z = \floor{\nicefrac{1}{p}}$ (as in the first case, if $p=\nicefrac{1}{z}$ then $z=\nicefrac{1}{p}=\floor{\nicefrac{1}{p}}$ since $z$ is an integer by definition; in the second case, if  $\nicefrac{1}{(z+1)} < p < \nicefrac{1}{z}$ then $z < \nicefrac{1}{p} < z+1$ so $z = \floor{\nicefrac{1}{p}}$ as well).
 
    We construct the preferences of the voters as below, using $m = \ceil{rk} z$ distinct candidates $c_{i,j}$ (one set of $\ceil{rk}$ per group, all pairwise disjoint across groups):
    \begin{center}
        \begin{tabular}{c|c}
        Group of voters (of size $\omega$) & Preferences\\\hline
             \(V_1\) & \(c_{1,1}\succ c_{1,2}\succ \cdots \succ c_{1,\ceil{rk}}\succ \cdots\) \\
             \(V_2\) & \(c_{2,1}\succ c_{2,2}\succ \cdots \succ c_{2,\ceil{rk}}\succ \cdots\) \\
             \(\vdots\) & \(\vdots\)\\
             \(V_{z}\) & \(c_{z,1}\succ c_{z,2}\succ \cdots \succ c_{z,\ceil{rk}}\succ \cdots\)
        \end{tabular}
    \end{center}
    Note that since $\floor{\nicefrac{1}{p}} \ge 2$, we have $m > 1$. Additionally, $n = \omega z$ in both cases, and since $z = \floor{\nicefrac{1}{p}} \ge 2$ and $\omega$ is a positive integer, we also have $n > 1$. Hence, this election $E' \in \mathcal{E}$ (as both $m,n > 1$). 
    
    Since $n = \omega z$ exactly in both cases, these $z$ groups exhaust all $n$ voters (as $|V_i| = \omega$ for all groups $V_i$). Since each group's top-$\ceil{rk}$ block is distinct, they form exactly $\kappa = z$ top-$\ceil{rk}$-agreement classes. As each group has size $\omega = \ceil{np} \ge pn$, all $z$ classes meet quota, so $\mathcal{V} = \{V_1, \dots, V_z\}$ and hence $\zeta = z = \floor{\nicefrac{1}{p}}$.
 
    By definition, any set satisfying \((p,r)\)-coalitions must include every \(c_{i,j}\) for \(i\in [\zeta]\) and \(j\in[\ceil{rk}]\), since $c_{i,j}$ is among the top \(\ceil{rk}\) candidates of $V_i$. Since the $c_{i,j}$ are pairwise disjoint across groups, this amounts to $\ceil{rk}\zeta$ distinct required candidates, so our committee needs to contain at least \(\ceil{rk}\zeta\) candidates, which must be at most \(k\). Therefore,
    $$k \ge \ceil{rk}\zeta  \implies k \ge \ceil{rk}\floor{\nicefrac{1}{p}}$$
    yielding the claimed inequality.

    This concludes the proof of the bidirectional statement of the theorem. To prove the last part of the theorem when $r$ is an integer multiple of $\nicefrac{1}{k}$, write $r = \nicefrac{w}{k}$ for some positive integer $w$. Then
    $$
    k \ge \ceil{rk}\floor{\nicefrac{1}{p}} = w\floor{\nicefrac{1}{p}}  \implies \frac{w}{k} \le \floor{p^{-1}}^{-1} \implies r \le \floor{p^{-1}}^{-1}  
    $$
    as advertised.
\end{proof}


\subsection*{Proof of Theorem \ref{thm:coal_hom}}

\begin{proof}

    By construction, we can partition $V$ into $\kappa$ subsets of voters $V_1,\dots, V_\kappa$, where every voter in the same $V_i$ agrees on their top-$\ceil{rk}$ choices (in any order), so $\bigcup_{j\in[\kappa]}V_j = V$. Since $\kappa$ counts actual blocks that appear amongst the voters' preferences, each $V_j$ is nonempty by construction. Hence $|V_j| \ge 1$ for all $j \in [\kappa]$. When $p=\nicefrac{1}{n}$, the quota is $pn=1$, so every $V_j$ meets this quota and qualifies for $\mathcal V$. Thus $\mathcal{V} = \set{V_1,\dots,V_\kappa}$, implying $\mathcal{V}$ is a partition of $V$.
     
    By Corollary \ref{cor:coalitions_checker}, $(\nicefrac{1}{n},r)$-coalitions is realizable on $E$ if and only if $\left|\bigcup_{V_j \in \mathcal{V}} \topp_{\ceil{rk}}(V_j)\right| \le k$. Since $\topp_{\ceil{rk}}(V_j) = \bigcup_{v \in V_j}\topp_{\ceil{rk}}(v)$ by definition, and $\bigcup_{V_j \in \mathcal{V}} V_j = V$ (because $\mathcal{V}$ is a partition of $V$), we have
    $$
    \bigcup_{V_j \in \mathcal{V}} \topp_{\ceil{rk}}(V_j) = \bigcup_{v \in V} \topp_{\ceil{rk}}(v)
    $$
    Hence, $(\nicefrac{1}{n},r)$-coalitions is realizable on $E$ if and only if $|\cup_{v \in V} \topp_{\ceil{rk}}(v)| \le k$, which is the condition of $(\ceil{rk},k)$-homogeneity.

\end{proof}


\subsection*{Proof of Theorem \ref{thm:coal_pnp}}

\begin{proof}
We bound the cost of each step of Algorithm \ref{alg:cocomhash} in turn, then sum these costs to obtain the overall runtime.
 
\medskip
 
\noindent \textbf{Constructing $H$ by looping over $[n]$.} Algorithm \ref{alg:cocomhash} takes the preference matrix $M$ of the election $E$ directly as input. We create an empty hashmap $H$, initially with no keys and no values. For each $v \in [n]$, we extract $T_v = M[v, 1:\ceil{rk}]$ as a set (Line \ref{alg:cocomhash:extract}) and check whether it is a key in $H$ (Line \ref{alg:cocomhash:inkeys}), updating its associated value accordingly. As discussed in the main text, this extraction and lookup take $\bigO(\ceil{rk})$ time per voter under an order-independent hash and equality check. Since there are $n$ voters, constructing $H$ in full takes $\bigO(n\ceil{rk})$ time. Upon completion, $H$ contains a collection of $(key,value)$ pairs, where the $value$ for a given $key$ tells us how many voters had that $key$ as their top-$\ceil{rk}$ preferences.
 
\medskip
 
\noindent \textbf{Filtering to $H^*$.} Since each of the $n$ voters contributes at most one key to $H$, $H$ has at most $n$ keys. Checking each key's value against the quota $pn$ (Line \ref{alg:cocomhash:keycheck}) costs $\bigO(1)$ per key, so constructing $H^* = \{K \in \text{Keys}(H) : H(K) \ge pn\}$ takes $\bigO(n)$ time in total. 
 
\medskip
 
\noindent \textbf{Constructing $U$.} Line \ref{alg:cocomhash:unionconstruct} then computes $U = \bigcup_{K \in H^*} K$. Since $H^*$ contains at most $n$ sets (being a subset of $H$'s at most $n$ keys), each of size $\ceil{rk}$, this union can be computed by inserting all $\ceil{rk}$ elements of each of the at most $n$ sets into a running result, at a cost of $\bigO(\ceil{rk})$ per set; over at most $n$ sets, this totals $\bigO(n\ceil{rk})$ as above.
 
\medskip
 
\noindent \textbf{Resolving the committee.} By construction, $U = \cup_{V_j \in \mathcal{V}} \topp_{\ceil{rk}}(V_j)$, so by Corollary \ref{cor:coalitions_checker}, the \textsc{Coalitions} problem can be resolved by comparing $|U|$ to $k$. This is what the remainder of the algorithm does, via three cases: when $|U| = k$ (Line \ref{alg:cocomhash:exact}), $U$ itself is a $(p,r)$-coalitions set, and this check takes $\bigO(1)$ time. When $|U| < k$ (Lines \ref{alg:cocomhash:select} - \ref{alg:cocomhash:tie}), we select any $(k-|U|)$-element subset $W$ of $C\setminus U$ (e.g.\ the first $k-|U|$ candidates under any fixed enumeration of $C$) and return $U \cup W$; since $k - |U| \le k$, this selection takes $\bigO(k)$ time in the worst case. When $|U| > k$, since any set satisfying $(p,r)$-coalitions must contain all of $U$, no size-$k$ set can, and Line \ref{alg:cocomhash:fail} returns $\bot$ in $\bigO(1)$ time. Hence this step costs $\bigO(k)$ in the worst case.
 
\medskip
 
\noindent \textbf{Total.} Summing the costs above of $\bigO(n\ceil{rk})$ to construct $H$, $\bigO(n)$ to filter to $H^*$, $\bigO(n\ceil{rk})$ to construct $U$, and at most $\bigO(k)$ to resolve the committee, the overall runtime is
$$
\bigO(n\ceil{rk}) + \bigO(n) + \bigO(n\ceil{rk}) + \bigO(k) = \bigO(\max(n\ceil{rk},k))
$$
since $\ceil{rk}\ge1$ for all $r \in (0,1]$ and $k\ge2$, yielding $n\ceil{rk} \ge n$.
\end{proof}


\section{Proofs for Section \ref{sec:stability}}
\label{app:stability}

\subsection*{Proof of Proposition \ref{prop:SGS-quantifiers}}

\begin{proof}

    \noindent $(\implies)$ Let $S$ be a strong Gehrlein stable set. For $s \in S$ and $c \in C\setminus S$, define $V(s,c) = \set{v \in V :s\succ_v c}$. Since $S$ is strongly Gehrlein, $|V(s,c)| > n/2$. 

    Next, consider any  $V^* \subseteq V$ with $|V^*| \ge n/2$. Then $V^* \cap V(s,c)$ is non-empty; otherwise, if these two sets are disjoint we reach the following contradiction:
    $$
    n = |V| \ge |V(s,c) \cup V^*| = |V(s,c)| + |V^*| > \frac{n}{2} + \frac{n}{2} = n
    $$
    Hence, $V^* \cap V(s,c)$ is non-empty, so there is some voter in $V^*$ who prefers $s$ to $c$.

    \smallskip

    \noindent $(\impliedby)$ We proceed by the contrapositive and suppose that $S$ is not strongly Gehrlein stable. So there is a $c \in C\setminus S$ and $s \in S$ where $\abs{\set{v\in V: s\succ_v c}}\le \nicefrac{n}{2}$, which implies $\abs{\set{v\in V: c\succ_v s}}\ge \nicefrac{n}{2}$. But then this set $\set{v\in V: c\succ_v s}$ is a subset of voters $V^*$ that does not contain any voter who prefers $s$ to $c$, completing the contrapositive.
\end{proof}


\subsection*{Proof of Proposition \ref{prop:WGS-LS}}

\begin{proof}

    We first prove the portion of the claim regarding strong Gehrlein stability. Substituting $p=\nicefrac{1}{2}$ and $r=1$ into Definition \ref{def:prstabind}, we get that $S$ violates $(\nicefrac{1}{2},1)$-stability if there is a candidate $c$ outside $S$, a set $V^*$ of at least $\nicefrac{n}{2}$ voters, and a subset $S^*$ of $S$ where $\abs{S^*}>k(1-1)=0$ (and since $\abs{S^*}$ is an integer, this is equivalent to $\abs{S^*}\ge 1$) and all voters in $V^*$ prefer $c$ to all members of $S^*$. There exists such an $S^*$ if and only if there is a candidate $s\in S$ where every voter in $V^*$ prefers $c$ to $s$, which is equivalent to strong Gehrlein stability by Proposition \ref{prop:SGS-quantifiers}.

    Next, we prove the remaining portion of the claim regarding local stability. Substituting $p = \nicefrac{q}{n}$ and $r = \nicefrac{1}{k}$ into Definition \ref{def:prstabind}, we have that $S$ violates $(\nicefrac{q}{n}, \nicefrac{1}{k})$-stability if there exists $c \in C\setminus S$, a set of voters $V^*$ of size at least $q$, and a subset $S^*$ of the committee $S$ of size more than $k-1$ such that all voters in $V^*$ prefer $c$ to all $s \in S^*$. Since there is only one subset $S^*$ of $S$ for which $\abs{S^*}>k-1$, we deduce that $S^* = S$. Hence, $c$ is preferred by all $v \in V^*$ to all members of $s \in S$, meeting the definition of $q$-local stability.
\end{proof}


\subsection*{Proof of Proposition \ref{prop:theta1k}}

\begin{proof}
    By Proposition \ref{prop:WGS-LS}, it is sufficient to show that a weak \(\theta\)-winning set is $(n\cdot(1-\theta))$-locally stable. This is shown in Remark 1 of \citet{aziz2017condorcet}.
\end{proof}


\subsection*{Proof of Proposition \ref{prop:candidate_undom_11_stable}}

\begin{proof}
    We first prove the claim for committees. Plugging in $(p,r) = (1,1)$ into the definition of stability, we have
    $$
    (\forall c\in C\setminus S) \  (\forall V^*\subseteq V :\abs{V^*}\ge n)  \ (\forall S^*\subseteq S  : \abs{S^*}> 0) \quad (\exists v\in V^* \ \exists s\in S^* \text{ such that } s\succ_v c)
    $$
    Since there is only one such $V^* \subseteq V$ for which $|V^*| \ge n$ (namely $V$ itself), $(1,1)$-stability is equivalent to  
    $$
    (\forall c\in C\setminus S)   \ (\forall S^*\subseteq S  : \abs{S^*}> 0) \quad (\exists v\in V \ \exists s\in S^* \text{ such that } s\succ_v c)
    $$

    \smallskip
    
    \noindent $(\impliedby)$ Suppose $S$ is $(1,1)$-stable. By way of contradiction, suppose $S$ is not candidate-wise undominated. Then there exists a dominated $b \in S$ where its dominator $a \notin S$. Since $S$ is $(1,1)$-stable, we can consider specific $S^* = \{b\}$. Then stability implies
    $$
    (\forall c\in C\setminus S)   \quad (\exists v\in V  \text{ such that } b\succ_v c)
    $$
    Letting $c = a$, we arrive at a contradiction, as every voter $v \in V$ prefers $a$ to $b$ (since $a$ dominates $b$).

    \smallskip

    \noindent $(\implies)$ We proceed by the contrapositive and show that if a committee $S$ is not $(1,1)$-stable, then it is not candidate-wise undominated. Taking the negation of $(1,1)$-stability above, 
    $$
    (\exists c\in C\setminus S)   \ (\exists S^*\subseteq S  : \abs{S^*}> 0) \quad (\forall v\in V \ \forall s\in S^* \text{ we have } c\succ_v s)
    $$
    This means that there is a $s\in S^*\subseteq S$ and a $c\notin S$ where all voters prefer $c\succ_v s$. But this means that $c$ dominates $s$ candidate-wise, and $s$'s dominator is not included in $S$, so $S$ is not candidate-wise undominated.

    This completes the proof of the proposition for committees. The proof of the proposition for rules follows immediately by extrapolation from committees to rules.
\end{proof}


\subsection*{Proof of Proposition \ref{prop:stab_coal}}

\begin{proof}
    We proceed by the contrapositive. Suppose that \(S\) fails the \((p,r)\)-coalitions property. Then, there exists a group of voters \(V^*\) with \(\abs{V^*}\geq np\) who all prefer, within their top \(\ceil{rk}\) choices some set of candidates \(C^*\) where \(\abs{C^*}= \ceil{rk}\), and \(C^*\not\subseteq S\). Therefore, $\abs{S\cap C^*}<\ceil{rk}$. Because $\abs{S\cap C^*}\in \Z$, we also get $\abs{S\cap C^*}<rk$. 
    
    Next, pick \(c\in C^*\setminus S\) and let $S^*=S\setminus C^*$. Then, there exists a group of voters (namely \(V^*\)) of size at least \(np\) that prefer this \(c\) over \(S^*\), and since \( \abs{S^*} = \abs{S\setminus C^*}=\abs{S}-\abs{S\cap C^*}> (1-r)k \), so \(S\) fails \((p,r)\)-stability.
\end{proof}


\subsection*{Proof of Corollary \ref{cor:sgs_implies_fm}}

\begin{proof}
    Proposition \ref{prop:stab_coal} provides that Strong Gehrlein stability ($(\nicefrac{1}{2}, 1)$-stability) implies the $(\nicefrac{1}{2},1)$-coalitions property. By $p$-lifting for coalitions (Proposition \ref{prop:p-coal-lifting}), it also implies $(\nicefrac{1}{2}+\nicefrac{1}{2n},1)$-coalitions, which is the fixed majority criterion by Proposition \ref{prop:fixmaj}.
\end{proof}


\subsection*{Proof of Proposition \ref{prop:lifting}}

\begin{proof}
    Suppose $S$ is a $(p,r)$-stable committee. Consider arbitrary $c \in C\setminus S$, $S^* \subseteq S$ of size greater than $(1-r^*)k$, and $V^* \subseteq V$ of size at least $p^*n$. Since $p^* \ge p$ and $r^* \le r$, we have $p^*n \ge pn$ and $(1-r^*)k \ge (1-r)k$. Since the set $S^*$ is of size greater than $(1-r)k$, and the set $V^*$ is of size at least $pn$, the $(p,r)$-stability of $S$ implies that there exists $v \in V^*$ and $s \in S^*$ such that $s \succ_v c$. Hence, $S$ is also a $(p^*,r^*)$-stable committee.
\end{proof}


\subsection*{Proof of Theorem \ref{thm:1r-univreal}}

\begin{proof}
    For $i \in [m]$, we let \(\phi_i(c_j)\) denote the total number of \(i^\text{th}\) preferences for a candidate \(c_j \in C\). This corresponds to $\Phi_{i,j}$ in Line \ref{alg:lexicosort:iposmat} of the algorithm.
    
    We will construct our committee $S$ by choosing the top candidates in terms of their \(\phi_1\) scores, breaking ties by \(\phi_2\) and so on. If there are two candidates $x$ and $y$ for which \(\phi_i(x)=\phi_i(y)\) for all \(i \in [m]\), then break ties arbitrarily. In other words, we lexicographically sort the candidates by the number of voters having them in their top preferences. This is essentially the same as the single non-transferable vote, but with tie breaking occurring based on lower choices.

    Given an election $E=(C,V,\succ)$ in matrix form $M$, we can compute $\phi_i(x)$ for all $x \in C$ and all $i \in [m]$ in $\bigO\paren{m^2 n}$ time. This follows because, for each $\Phi_{i,j}$ we require $\bigO(n)$ time to examine the $i^{\text{th}}$ position of all $n$ voters' preferences to check for the presence or absence of $c_j$. Hence, we can produce the $m\times m$ array $\Phi$ in Line \ref{alg:lexicosort:iposmat} in $\bigO\paren{m^2 n}$ time.

    Constructing $\bm{g}$ in Line \ref{alg:lexicosort:geo} can be done in $\bigO(m)$. Given the values of $\phi_i(x)$, Line \ref{alg:lexicosort:fscore} constructs a composite score $\bm{f}_j $ for each candidate $c_j \in C$. To ease notation later in the proof, we will denote $\bm{f}_j$ as $ f(c_j)$. By construction,
    $$
    f(c_j) = \bm{f}_j = (\bm{g}\Phi)_j = \sum_{i=1}^m \frac{1}{{(mn)}^i} \phi_i(c_j)
    $$
    For each $j$, this is accomplished in Line \ref{alg:lexicosort:fscore} to produce a vector $\bm{f}$ of these scores. Computing $\bm{f}$ via vector-matrix multiplication takes $\bigO(m^2)$ time.

     Then, in line \ref{alg:lexicosort:argsort}, we argsort this array in non-increasing order, breaking ties arbitrarily, which we can accomplish in $\bigO(m\log(m))$ time. Then, in line \ref{alg:lexicosort:picktopk} we pick the $k$ candidates whose values correspond to the top $k$ positions in $\bm{a}$, giving us the final time complexity of \[\bigO(m^2 n+m^2+m+m\log m) = \bigO(m^2 n)\]

  Next, we show that $S$ is $(1,r)$-stable for $r > 0$. By movement (Proposition \ref{prop:lifting}), since decreasing $r$ preserves stability, it suffices to show that $S$ is $(1,1)$-stable. When $r=1$ and $p=1$,
    $$
    (\forall c\in C\setminus S) \  (\forall V^*\subseteq V :\abs{V^*}\ge n)  \ (\forall S^*\subseteq S  : \abs{S^*}> 0) \quad (\exists v\in V^* \ \exists s\in S^* \text{ such that } s\succ_v c)
    $$
    Since there is exactly one such $V^* \subseteq V$ of size at least $n$ (namely, $V^* = V$), this is equivalent to
    $$
    (\forall c\in C\setminus S) \  (\forall S^*\subseteq S  : \abs{S^*}\ge 1) \quad (\exists v\in V \ \exists s\in S^* \text{ such that } s\succ_v c)
    $$
    This condition holding for every nonempty $S^*\subseteq S$ is equivalent to it holding for every singleton $S^* = \{s\}$: if some $v \in V$ satisfies $s \succ_v c$ for a particular $s$, then that same pair $(v,s)$ witnesses the condition for any $S^* \supseteq \{s\}$ as well, so checking singletons is both necessary and sufficient.\footnote{This conclusion can also be arrived at by invoking Lemma \ref{lem:r_one_line_equiv} and setting $p = 1$.} Hence, \((1,1)\)-stability is equivalent to
    $$
    (\forall c\in C\setminus S) \  (\forall s \in S) \quad (\exists v\in V \text{ such that } s\succ_v c)
    $$
    Hence, we suppose \(s\in S\) and \(c\in C\setminus S\) and show that there is some $v \in V$ such that $s \succ_v c$. We consider two cases: either $s$ is ranked first by someone, or $s$ is not.
    
    \begin{enumerate}
        \item If \(\phi_1(s)\ge 1\), then there exists some voter who prefers \(s\) to everyone else in \(C\), including \(c\).
        \item  Alternatively, if $\phi_1(s) = 0$, let $j>1$ be the first index for which $\phi_j(s)\ge 1$. Then \(\phi_i(s)=0\) for all \(i<j\). 
        
        We must also have \(\phi_i(c)=0\) for all \(i<j\); otherwise, if \(\phi_i(c)\ge 1\) for any \(i<j\), then 
        \begin{align*}
            f(c)-f(s)&= \sum_{l=1}^m (mn)^{-l}(\phi_l(c)-\phi_l(s))\\
            &= \left(\sum_{l<j} (mn)^{-l}(\phi_l(c)-\phi_l(s)) \right)+ \left(\sum_{l\ge j} (mn)^{-l}(\phi_l(c)-\phi_l(s))\right)\\ 
            &\ge(mn)^{-i} \phi_i(c)+\sum_{l= j}^m (mn)^{-l}(\underbrace{\phi_l(c)}_{\ge 0})-\underbrace{\phi_l(s)}_{\le n})\\
            &\ge (mn)^{-i} - n\sum_{l\ge j} (mn)^{-l}\\
            &> (mn)^{-i} - n\sum_{l=j}^\infty (mn)^{-l} \\
            &= (mn)^{-i} - n\frac{(mn)^{-j}}{1-(mn)^{-1}} \\
            &= (mn)^{-i} - n\frac{(mn)^{-j}}{\frac{mn-1}{mn}} \\
            &=(mn)^{-i} - \frac{n}{mn-1}(mn)^{-j+1} \\
            &> (mn)^{-i} - (mn)^{-j+1}\\
            &= (mn)^{-i}(1-(mn)^{(i+1)-j}) \\
            & \ge 0
        \end{align*}
        where the second to last inequality follows because $m > 1$ and the last inequality follows because $ i < j$. But then, our algorithm would pick \(c\) to include in $S$ before \(s\), which contradicts our hypothesis that $s\in S$ and $c \notin S$. 
        
        Since \(\phi_j(s)\ge 1\), then there is a voter \(v\) who ranked \(s\) in the \(j^\text{th}\) position, and since \(\phi_i(c)=0\) for all \(i<j\), then \(v\) cannot have ranked \(c\) higher than \(j\), which means they are ranked lower than the \(j^\text{th}\) position and thus \(s\succ_v c\). 
    \end{enumerate}
    Therefore in either case, we conclude there is some $v \in V$ such that $s \succ_v c$. Hence, \(S\) is \((1,1)\)-stable.
\end{proof}


\subsection*{Proof of Proposition \ref{prop:stab_coal_equiv_1n}}

\begin{proof}
    By Proposition \ref{prop:stab_coal}, $(\nicefrac{1}{n}, r)$-stability implies $(\nicefrac{1}{n}, r)$-coalitions. To prove the reverse direction, consider a size $k$ committee $S \subseteq C$ that satisfies $(\nicefrac{1}{n}, r)$-coalitions for $E$. We let $w=\ceil{kr}$. We want to show that $S$ is $(\nicefrac{1}{n}, r)$-stable, which by Definition \ref{def:prstabind} requires
    $$
    \text{$(\forall c \in C \setminus S) \ (\forall \ V^* \subseteq V$ of size at least $1)$ $ \ (\forall S^* \subseteq S$ of size more than $k-kr)$, $ \ \exists v \in V^*$ $ \ \exists s \in S^*$  : $s \succ_v c$}
    $$
    When $kr \in \Z$, $k - kr = k-w$; alternatively, when $kr \notin \Z$, a set $S^*$ of size more than $k-kr$ must also be of size more than $k-w$ (and vice-versa), because the sizes of sets are integers.\footnote{The details of this equivalence are as follows. Since $w = \ceil{kr}$, we have $k - kr \ge k - \ceil{kr} = k - w$, so if $\abs{S^*} > k-kr$, then $\abs{S^*} > k-w$. Conversely, recall the general fact that for an integer $k$ and a real number $x$, $\floor{k-x} = k - \ceil{x}$. Applying this with $x = kr$, we get $k - w = k - \ceil{kr} = \floor{k - kr}$. So if $\abs{S^*} > k - w = \floor{k-kr}$, then since $\abs{S^*}$ is an integer strictly greater than $\floor{k-kr}$, we have $\abs{S^*} \ge \floor{k-kr} + 1 > k - kr$, where the last inequality holds because $\floor{k-kr} > (k-kr)-1$ by definition of the floor function. Hence $\abs{S^*} > k-kr$.} Hence, the above condition is equivalent to 
    $$
    \text{$(\forall c \in C \setminus S) \ (\forall \ V^* \subseteq V$ of size at least $1)$ $ \ (\forall S^* \subseteq S$ of size more than $k-w)$, $ \ \exists v \in V^*$ $ \ \exists s \in S^*$  : $s \succ_v c$}
    $$
    which is also equivalent to
    $$
    \text{$(\forall c \in C \setminus S) $ \ $(\forall S^* \subseteq S$ of size more than $k-w) \ \forall  v \in V$ $ \ \exists s \in S^*$  : $s \succ_v c$}
    $$
    Since $S$ satisfies $(\nicefrac{1}{n}, r)$-coalitions,
    we have for any $V^*\subseteq V$ where $\abs{\topp_{w} V^*}=w$ and $\abs{V^*}\ge 1$, 
    $\topp_{w} V^*\subseteq S$. So in particular, we can consider sets $V^*$ that are singletons; in this case, $\abs{\topp_{w} V^*}=w$ trivially since $V^*$ is a singleton, and hence the coalitions requirement implies that for every $v \in V$, we have $\topp_{w}(\succ_v) \subseteq S$.
    
    We claim that $S$ is $(\nicefrac{1}{n}, r)$-stable. Take $c \in C \setminus S$, $S^* \subseteq S$ of size more than $k-w$, and $v \in V$ arbitrary. Since $\abs{\topp_{w}(\succ_v)} = w$ and $\abs{S^*} > k-w$, we must have $\topp_{w}(\succ_v) \cap S^* \ne \emptyset$; otherwise, if we had $\topp_{w}(\succ_v) \cap S^* = \emptyset$ we get a contradiction, as 
    $$
    k = |S| = |\topp_{w}(\succ_v) \cup S| \ge |\topp_{w}(\succ_v) \cup S^*| =  |\topp_{w}(\succ_v)| + |S^*| > w+ (k-w) = k
    $$
    Hence, there exists some $s \in \topp_{w}(\succ_v) \cap S^*$. By definition of $\topp_{w}(\succ_v)$, the position of $s$ in $\succ_v$ is at most $w$. But since $c \notin S$ and $\topp_{w}(\succ_v) \subseteq S$, $c \notin \topp_{w}(\succ_v)$, so the position of $c$ is after $w$, implying $s \succ_v c$.
\end{proof}


\subsection*{Proof of Theorem \ref{thm:bottom_horizontal_line}}

Before we present the proof of Theorem \ref{thm:bottom_horizontal_line}, we introduce helpful lemmas about $(p,1)$-stability and graph-theoretic results.

\begin{lem}
    \label{lem:r_one_line_equiv}
    $S$ is $(p,1)$-stable if and only if
    $$
    \textrm{$(\forall c \in C \setminus S) \ (\forall \ V^* \subseteq V$ of size at least $np)$ $ \ (\forall s \in S)$, $ \ \exists v \in V^*    : s \succ_v c$}
    $$
\end{lem}

\begin{proof}
    By definition, $S$ is $(p,1)$-stable if and only if
    $$
    \text{$(\forall c \in C \setminus S) \ (\forall \ V^* \subseteq V$ of size at least $np)$ $ \ (\forall S^* \subseteq S$ of size more than $k(1-r))$, $ \ \exists v \in V^*$ $ \ \exists s \in S^*$  : $s \succ_v c$}
    $$
    When $r=1$, we get
    $$
    \text{$(\forall c \in C \setminus S) \ (\forall \ V^* \subseteq V$ of size at least $np)$ $ \ (\forall S^* \subseteq S$ of size more than $0)$, $ \ \exists v \in V^*$ $ \ \exists s \in S^*$  : $s \succ_v c$}
    $$
    Note that the above condition holds for all nonempty subsets $S^*$ of $S$ if and only if it holds for all singletons, or equivalently all elements $s\in S$.
    This allows us to rewrite the statement as follows:
    $$
    \text{$(\forall c \in C \setminus S) \ (\forall \ V^* \subseteq V$ of size at least $np)$ $ \ (\forall s \in S)$, $ \ \exists v \in V^*    : s \succ_v c$}
    $$
\end{proof}

\begin{lem}
    \label{lem:bottom_line_counting}
    $S$ is $(p, 1)$-stable if and only if  $(\forall c \in C \setminus S) \ (\forall s \in S) \ |\{v \in V \ : \ c \succ_v s\}| < np$.
\end{lem}

\begin{proof}
    First, we show $(\implies)$ by the contrapositive. Suppose there exists $c \in C \setminus S$ and $s \in S$ arbitrary such that $|\{v \in V \ : \ c \succ_v s\}| \ge np$. Let $V^* = \{v \in V \ : \ c \succ_v s\}$. Then for all $v \in V^*$ we have $c  \succ_v s$, so $S$ is not $(p, 1)$-stable.
    
    Next, we show $(\impliedby)$. Take $c \in C \setminus S$, $s \in S$, and $V^* \subseteq V$ of size at least $np$ arbitrary. Let $V(c,s) = \{v \in V \ : \ c \succ_v s\}$ and $V(s,c) = \{v \in V \ : \ s \succ_v c\}$. Since $|V(c,s)| \le np-1$ by assumption, we have $|V(s,c)| \ge n-np+1$. 

    We must have that $V^* \cap V(s,c) \ne \emptyset$, because otherwise if $V^* \cap V(s,c) = \emptyset$ then
    $$
    n = |V| \ge |V^* \cup V(s,c)| = |V^*| + |V(s,c)| \ge np + (n-np+1) = n+1
    $$
    yielding a contradiction. Hence, there is some $v \in V^* \cap V(s,c)$, so there is some voter $v \in V^*$ such that $s \succ_v c$, establishing the $(p, 1)$-stability of $S$.
\end{proof}

\begin{lem}
    \label{lem:s_closure_single}
    Suppose $S$ is $(p,1)$-stable and $s \in S$. If $(t,s) \in L_p$ then $t \in S$. 
\end{lem}

\begin{proof}
    By way of contradiction, suppose $(t,s) \in L_p$ but $t \notin S$. Then we contradict Lemma \ref{lem:bottom_line_counting} since $(t,s) \in L_p$ is equivalent to $|\{v \in V \ : \ t \succ_v s\}| \ge np$.
\end{proof}

Since strongly connected components are equivalence classes, let $[[s]]$ denote the equivalence class containing $s$.

\begin{lem}
    \label{lem:s_closure_class}
    Suppose $S$ is $(p, 1)$-stable and $s \in S$. Then $[[s]] \subseteq S$.
\end{lem}

\begin{proof}
    Let $t \in [[s]]$. For $t = s$ the claim follows immediately. For $t \ne s$, let $d$ be the edge distance from $t$ to $s$. We proceed by cases. If $d = 1$, then $(t,s) \in L_p$ by Lemma \ref{lem:s_closure_single}, so $t \in S$. On the other hand, if $d > 1$ then there exist candidates $c_1,...,c_{d-1} \in [[s]]$ such that $(t,c_1),...,(c_{d-1},s) \in L_p$. Since $s \in S$, we can apply Lemma \ref{lem:s_closure_single} repeatedly and deduce $c_{d-1},...,c_1, t \in S$. Thus, $t \in S$ in either case, completing the proof. 
\end{proof}

With all of these results in hand, we now present the proof of Theorem \ref{thm:bottom_horizontal_line}.

\begin{proof} (Of Theorem \ref{thm:bottom_horizontal_line}) The $(\impliedby)$ direction of the theorem follows immediately by letting $S = \cup_{j \in J} \mathcal{G}^{(p)}_j$ and invoking Lemma \ref{lem:bottom_line_counting}. To show $(\implies)$, suppose $S$ is a set of $k$ candidates that satisfies $(p, 1)$-stability. By Lemma \ref{lem:s_closure_class}, we know that for every $s \in S$, $\{s\} \subseteq [[s]] \subseteq S$. Hence, we can rewrite $S$ as
    $$
    S = \bigcup_{s \in S} \{s\} = \bigcup_{s \in S} [[s]] 
    $$
    Let $J$ be the set of indices for which $[[s]] = \mathcal{G}^{(p)}_j$ for some $j$ for $s \in S$. Then,
    $$
    S = \bigcup_{s \in S} [[s]]  = \bigcup_{j \in J} \mathcal{G}^{(p)}_j  
    $$
    Since strongly connected components are disjoint, we have
    $$
    k = |S| = \left| \bigcup_{j \in J} \mathcal{G}^{(p)}_j  \right| = \sum_{j \in J} |\mathcal{G}^{(p)}_j|
    $$
    This proves the first part of the claim. To prove second part, note that if there exists an edge from some $\mathcal{G}^{(p)}_i$ with $i \notin J$ to some candidate in $\cup_{j \in J} \mathcal{G}^{(p)}_j$ we contradict Lemma \ref{lem:bottom_line_counting}. 
\end{proof}


\subsection*{Proof of Proposition \ref{prop:ginc}}

\begin{proof} 
    Suppose for contradiction that $G \subseteq S$ yet $S$ is not $(p,r)$-stable for $r \in (0,1)$. Then there exists some $c \in C \setminus S$, $V^* \subseteq V$ of size at least $np$, and $S^* \subseteq S$ of size more than $(1-r)k$ for which $c \succ_v s$ for all $v \in V^*$ and $s \in S^*$. 

    There are two possible cases: either $S^* \cap G = \emptyset$ or not. We will show both cases lead to a contradiction. First, consider the case where $S^* \cap G = \emptyset$. Since $S^*, G \subseteq S$, we have $k = |S| \ge |S^* \cup G| = |S^*| + |G| > (1-r)k + rk = k$, yielding a contradiction. Next, consider the case where there exists some $g \in S^* \cap G$. By Lemma \ref{lem:bottom_line_counting} the $(p,1)$-stability of $G$  implies $|\{v \in V \ : \ c \succ_v g\}| < np$. Thus no subset of voters can ever have at least $np$ voters who prefer $c$ to $g$. But $V^*$ is a set of at least $np$ voters who prefer $c$ to $g$, again yielding a contradiction.
\end{proof}


\subsection*{Proof of Proposition \ref{prop:bot_left_unique}}

\begin{proof}
    If $S$ is $(p,1)$-stable and $T$ is $(p^*,1)$-stable, then by movement, $S$ and $T$ are both $(\nicefrac{1}{2},1)$-stable. From \citet{aziz2017condorcet}, we get that strong Gehrlein stable sets, i.e. $(\nicefrac{1}{2},1)$-stable sets, are unique, so $S=T$. 
\end{proof}


\subsection*{Proof of Proposition \ref{prop:bottom_line_pnp_1}}

\begin{proof}
    We begin by constructing a preference matrix \(P^{(v)} \in \{0,1\}^{m \times m}\) for each voter \(v\), where \(P^{(v)}_{ij}=1\) if \(c_i\succ_v c_j\) and \(0\) otherwise. This construction can be done in \(\bigO(m^2)\) time by simply iterating through the preference list. So in total it is $\bigO(m^2n)$ to complete this process for all $n$ of the voters. Then, we construct the adjacency matrix for $G^{(p)}$, denoted as $\mathrm{Adj}(G^{(p)})$ by the following formula:
    \[
    \mathrm{Adj}(G^{(p)})_{i,j}=1\iff \sum_{v\in V} P^{(v)}_{ij} \ge np
    \]
    Computing a single entry this way takes $\bigO(n)$ time, and since $\mathrm{Adj}(G^{(p)})$ has $m^2$ entries, constructing it in full takes $\bigO(m^2 n)$ time.
 
    We then produce the condensation of \(G^{(p)}\), \(G'\), which can be constructed in \(\bigO(m^2)\) time (such as by using the methods of \citet{tarjan1972} and \citet{sharir1981}). Consider the following facts:
    \begin{enumerate}
        \item \(G'\) is acyclic, since it is the condensation of a directed graph.
        \item If \(p\le\nicefrac{1}{2}\), then there is at least one directed edge between any two candidates. Therefore, between any two connected components \(\mathcal{G}_i\) and \(\mathcal{G}_j\), there exists at least one edge from one to the other.
    \end{enumerate}
 
    The combination of these two facts means that $G'$ is a tournament on the strongly connected components, as every pair of distinct components has exactly one edge between them. The acyclic nature of $G'$ additionally implies that $G'$ is transitive (as (i) a tournament contains a directed cycle if and only if it contains a directed 3-cycle, (ii) a tournament is transitive if and only if it contains no directed 3-cycles; since $G'$ is acyclic, there are no directed cycles by definition, and hence by (i) no directed 3-cycles, and hence by (ii) $G'$ is transitive). Hence, $G'$ is a transitive tournament.
 
    The remainder of the proof follows by a similar argument in Theorem 1 of \citet{aziz2017condorcet}, which we include with additional details to establish the runtime of $\bigO(m^2n)$ (and not just a polynomial time runtime as established in \citet{aziz2017condorcet} for the corresponding case of $p=\nicefrac{1}{2}$).
 
    A transitive tournament admits a unique total ordering, which we denote by $\mathcal{G}_i < \mathcal{G}_j \iff (\mathcal{G}_i, \mathcal{G}_j) \in G'$. We then relabel so that \(i<j \iff \mathcal{G}_i <\mathcal{G}_j\), which is the unique topological ordering of \(G'\). Let $z$ denote the number of vertices of $G'$. Since $G'$ is a tournament with $z\le m$ vertices, it has  $\binom{z}{2}=\bigO(z^2)$ edges; we can construct this topological ordering via topological sort in $\bigO(z + z^2) = \bigO(z^2)$ time, which is in the worst-case $\bigO(m^2)$.
 
    Next, we make a structural observation of this topological ordering that will be helpful in providing a stable set $S$ or determining if one does not exist using the criteria from Theorem \ref{thm:bottom_horizontal_line}.
 
    Since the relabeling $i<j \iff \mathcal{G}_i<\mathcal{G}_j$ is the topological ordering of $G'$, by definition every edge of $G'$ points from a lower index to a higher one: if $(\mathcal{G}_i,\mathcal{G}_j)\in G'$, then $i<j$. In the language of Theorem \ref{thm:bottom_horizontal_line}, this forces the index set $J$ of any $(p,1)$-stable committee to be exactly an initial segment $\{1,\dots,\ell\}$ of the ordered list $1,\dots,z$. This is because for $j\in J$ and $i<j$, the topological property gives $(\mathcal{G}_i,\mathcal{G}_j)\in G'$ (an edge cannot instead run from $\mathcal{G}_j$ to $\mathcal{G}_i$, since that would require $j<i$). So if $i\notin J$, this is an edge from outside $J$ into $J$, contradicting Condition 2 of Theorem \ref{thm:bottom_horizontal_line}. Hence $i\in J$ as well, so every $(p,1)$-stable committee is of the form $\cup_{i\le\ell}\mathcal{G}_i$ for some $\ell$, and it suffices to check only the unions of $\mathcal{G}_1,\dots,\mathcal{G}_z$ in this order. Conversely, every prefix $\cup_{i\le\ell}\mathcal{G}_i$ automatically satisfies Condition 2, as an edge from outside the prefix into it would be some $(\mathcal{G}_{i'},\mathcal{G}_j)\in G'$ with $i'>\ell\ge j$, but the topological property requires $i'<j$ for any such edge, yielding a contradiction. Hence, by Theorem \ref{thm:bottom_horizontal_line}, a prefix $\cup_{i\le\ell}\mathcal{G}_i$ is $(p,1)$-stable whenever it has size exactly $k$.
 
    Hence, we can set $S \leftarrow \emptyset$ and iteratively run through $\ell \in [z]$ with $S \leftarrow S \cup \mathcal{G}_\ell$, stopping whenever $|S| \ge k$ and returning $S$. Since $k \le m$ in our multiwinner election setup, this algorithm is guaranteed to halt. Upon halting, the resulting $S$ will either be of size $k$ or size greater than $k$.

    If $|S| = k$, then by Theorem \ref{thm:bottom_horizontal_line} this $S$ is $(p,1)$-stable. Furthermore, this algorithm will find it in at most $z\le m$ unions and checks for whether one has size exactly $k$, which takes $\bigO(m)$ time in the worst-case. 

    Alternatively, if $|S| > k$, no stable set exists. This follows because if the set $S$ produced at the time of halting has size greater than $k$, its previous iteration must have had a set $S$ of size less than $k$ (as if this prior $S$ has $|S|\ge k$, the algorithm would have halted earlier). Hence, by Theorem \ref{thm:bottom_horizontal_line}, there is no $\ell$ where $\abs{\cup_{i\le\ell}\mathcal{G}_i}=k$, which is $\bigO(m)$ time in the worst-case.
 
    So, the whole algorithm runs in $\bigO(m^2n)$ time, dominated by constructing the $P^{(v)}$ matrices and $G^{(p)}$. Furthermore, by Proposition \ref{prop:bot_left_unique}, $S$ is unique whenever it exists.
\end{proof}


\subsection*{Proof of Corollary \ref{cor:bottom_line_pnp_2}}

\begin{proof}
    Let $W$ be the ordered list of integers $1,\dots,\floor{\nicefrac{n}{2}}$. We can iterate through $w$ in $1,\dots,\floor{\nicefrac{n}{2}}$, and use the algorithm described in Proposition \ref{prop:bottom_line_pnp_1} to find a $(\nicefrac{w}{n},1)$-stable set or determine if one does not exist in $\bigO(m^2n)$ time. If a stable set $S$ exists for the given $w$, return both $S$ and $w$ and halt. Otherwise, continue looping through the remaining values in the ordered list following this process. 

    Upon completion, there are two options: either a pair $(S,w)$ was returned or not. If so, the algorithm terminated after searching through at most $\floor{\nicefrac{n}{2}}$ values of $w$. If not, the algorithm searched through exactly $\floor{\nicefrac{n}{2}}$ values of $w$. Hence,  this process runs in at most $\bigO(m^2n^2)$ time.
\end{proof}

\section{Proofs for Section \ref{sec:partylists}}
\label{app:partylists}

\subsection*{Proof of Lemma \ref{lem:cpl_cc_1}}

\begin{proof}
    Suppose $c\in C_i$ and $d\in C_j$ are in the same connected component with $i \ne j$. We will show that all of $C_i$ and $C_j$ are in this connected component. 

    First, note that in a complete party list election, if $c_{i,\ell}\succ_v c_{j,\ell'}$, then $c_{i,q}\succ_v c_{j,q'}$ for any $\ell,\ell',q,q'$. Thus, $\set{v\in V: c_{i,\ell}\succ c_{j,\ell'}}=\set{v\in V: c_{i,q}\succ c_{j,q'}}$.

    Since $c$ and $d$ are in the same connected component, there is a path from $c$ to $d$ in the $p$-preference graph of $E$. Consider $f$ to be the first element on this path that is not in $C_i$. From the above, there is an edge from $c_{i,k}$ to $f$. If $f\in C_j$ we are done. Otherwise, consider $g$ the last element of the path that is not in $C_j$. Since there must be an edge from $g$ to something in $C_j$, there must also be an edge from $g$ to $c_{j,1}$, which in turn has an edge to $c_{j,k}$. Likewise, following the path from $d$ to $c$, we get a path from $d$ to $c_{i,1}$ to $c_{i,k}$. 

    Hence, $c_{i,1}, c_{i,k},c_{j,1},c_{j,k}$ are all in the same connected component. Finally, since every voter prefers $c_{i,\eta}$ to $c_{i,\zeta}$ if $\zeta > \eta$, all other elements of $C_i$ and $C_j$ are in this connected component.
\end{proof}


\subsection*{Proof of Lemma \ref{lem:cpl_cc_2}}

\begin{proof}
    Any connected component will fall into three categories: it is either a singleton, it contains elements of multiple lists, or it contains more than one element from only one list.
    
    Suppose for contradiction we are in the last case. In a complete party list election, let \(c\) be above \(d\) in the same list with both in the same connected component. Then, all voters must prefer \(c\) over \(d\) by definition, and \((c,d)\) is an edge in \(G\), but there is no edge from any element later in the list to one higher in the list. If there must be some path from \(d\) to \(c\), then \(G\) would have to contain an edge from lower in the list to higher in the list, otherwise the path in question could only go to a candidate further down.
    
    Thus, if the component is not a singleton, there must be candidates from at least two different lists, which, by Lemma \ref{lem:cpl_cc_1}, implies that any connected component has either a single element, or at least two entire lists, which means that it has at least \(2k\) elements. 
\end{proof}


\subsection*{Proof of Theorem \ref{thm:cpl_p1-stability}}

\begin{proof}
    First, note that when $p=1$, we get $(1,1)$-stability which is universally realizable by Theorem \ref{thm:1r-univreal}, and since $1$-winners always exist in any election, the claim holds. Hence, it remains for us to prove the claim when $p \in (0,1)$.

    From Theorem \ref{thm:bottom_horizontal_line}, a set \(S\) is \((p,1)\)-stable if and only if, for the \(p\)-preference graph \({G}\) of \(E=(C,V, \succ)\) with its condensation \(G'=\set{\mathcal{G}_1,\dots, \mathcal{G}_z}\) containing \(z\) elements, there is a set  \(K\subseteq [z]\) where \(S=\cup_{i\in K} \mathcal{G}_i\), and for any \(i\in K,\ j\in [z]\setminus K\), the edge \((\mathcal{G}_i,\mathcal{G}_j)\notin G'\).

    \smallskip
    
    \noindent $(\impliedby)$ Suppose \(c\) is a \(p\)-winner. Let $C'$ be the list containing $c$. First, \(c\) must be at the top of \(C'\); otherwise for \(d\) be the top element of the list \(c\) is on, \(\abs{\set{v\in V : d\succ_v c}}=n>np\) for any \(p<1\), contradiction to $c$ being a $p$-winner. 

    We claim that $C'$ is a $(p,1)$-stable set. To prove this, we proceed by contradiction and suppose there is an edge from \(d\) outside of \(C'\) to some element in \(C'\). That means that at least \(np\) voters prefer \(d\) to some element of \(C'\). However, since it is a complete party list election, all those voters must also prefer \(d\) to \(c\), contradicting the fact that $c$ is a $p$-winner. So, there is no edge from outside \(C'\) to any element of \(C'\).

    Consequently, no strongly connected component of the $p$-preference graph can contain elements both inside and outside $C'$ (as any such component would require a path -- hence at least one edge -- entering $C'$ from outside, which we have just ruled out). Hence every connected component is either entirely contained in $C'$ or entirely disjoint from it, so $C'$ is exactly the union of those components lying inside it, with no edge entering it from any component outside. Thus $C'$ satisfies the hypotheses of Theorem \ref{thm:bottom_horizontal_line}, and so $C'$ is \((p,1)\)-stable.

    \smallskip

    \noindent $(\implies)$ We proceed by contradiction of the contrapositive. That is, suppose there is no \(p\)-winner yet there is a committee $W$ that is a $(p,1)$-stable set.
    
    If our committee $W$ contains any non-singleton connected component, it must contain the entire component, which from Lemma \ref{lem:cpl_cc_2} will be of size at least \(2k\), a contradiction with our set being of size \(k\). Thus, $W$ must contain only singleton connected components. 

    Let $s_1\in W$. If $s_1$ has an in-degree of $0$, then it is a $p$-winner, contradiction -- so assume that $s_1$ has a positive in-degree. Since $\{s_1\}$'s component lies in $W$, by Theorem \ref{thm:bottom_horizontal_line} the source component of any edge into $\{s_1\}$ must also lie entirely in $W$. Since every component meeting $W$ is a singleton (as shown above), this source component is itself a singleton $\{s_2\}$ for some $s_2 \ne s_1$, giving an edge $(\{s_2\},\{s_1\}) \in G'$. Repeating this argument, there exist $s_3,\dots,s_k \in W$ such that for $i>1$, $(\{s_i\},\{s_{i-1}\})$ is an edge in $G'$ (if instead $s_{i-1}$ had in-degree $0$, it would itself be a $p$-winner, a contradiction).
    
    Since $G'$ is a condensation, it is acyclic, and $s_i\ne s_j$ for any $j \in [k]$. Since there is no $p$-winner, none of these can have a zero in-degree. In particular, $s_k$ has a non-zero in-degree; however, we know that none of $s_1,\dots,s_{k-1}$ can point to $s_k$ by construction. Hence, there has to be some $c\notin W$ where $(c,s_k)$ is an edge of $G$, which contradicts $W$ being a $(p,1)$-stable set.    
\end{proof}


\subsection*{Proof of Corollary \ref{cor:cpl_gehrlein}}

\begin{proof}
    Strong Gehrlein stability is $(\nicefrac12,1)$-stability (Proposition \ref{prop:WGS-LS}), and a $\nicefrac12$-winner is a Condorcet winner. The claim then follows by substituting $p=\nicefrac12$ into Theorem \ref{thm:cpl_p1-stability}.
\end{proof}


\subsection*{Proof of Lemma \ref{lem:cpl_nodestab}}

\begin{proof}
    We proceed by the contrapositive. If \(c\) is a \((p,r)\)-destabilizer of \(S\), then there is a set \(S^*\subseteq S\) where \(\abs{S^*}> (1-r)k\) and \(\abs{\set{v:c\succ_v S^*}}\ge np\).

    Note that \(S^*\subseteq \cup_{i\in \text{under}_p(D)}C_i \). Indeed, if there is some $s\in S^*$ with $s \in C_i$, then $\set{v:c\succ_v s}\supseteq \set{v:c\succ_v S^*}$, so $\abs{\set{v:c\succ_v s}}\ge np$. Since $c\in D$ and we are in a complete party list election, every voter who prefers $c$ to $s$ prefers all of $D$ to all of $C_i$. Hence every such voter has $D \succ_v C_i$, so $\abs{\set{v:D\succ_v C_i}}\ge np$, giving $i\in \mathrm{under}_p(D)$ by Definition \ref{def:under_p}.

    Combining all of these pieces, we have
    \begin{align*}
        \sum_{i\in \mathrm{under}_p(D)} \abs{C_i\cap S} &= \abs{ \cup_{i\in \text{under}_p(D)}C_i \cap S}\\
        & = \abs{S \cap \cup_{i\in \text{under}_p(D)}C_i} \\
        &\ge \abs{S^* \cap \cup_{i\in \text{under}_p(D)}C_i} \\
        &= \abs{S^*} \\
        &>(1-r)k
    \end{align*}
    where the first equality follows because all $C_i$ are disjoint, and the last equality follows since $S^\ast \subseteq \cup_{i\in\text{under}_p(D)} C_i$.
\end{proof}


\subsection*{Proof of Lemma \ref{lem:topcc_cpl}}

\begin{proof}
    Suppose the system of Equations (\ref{eqn:topcc_cpl_1}) and (\ref{eqn:topcc_cpl_2}) is satisfiable. By Equation (\ref{eqn:topcc_cpl_2}), $S=\bigcup_{i\in I^*_p}\topp_{k_i}(C_i)$ is a set of size $k$. To complete the proof, we show $S$ is $(p,r)$-stable. By Lemma \ref{lem:stableiffnodestab}, it is equivalent to show that $S$ has no $(p,r)$-destabilizers.
 
    By way of contradiction, suppose $c$ is a $(p,r)$-destabilizer for $S$. Since $c$ is a destabilizer, there is $S^*\subseteq S$ with $\abs{S^*}>(1-r)k$ and a set $V^*\subseteq V$ of at least $np$ voters such that every voter in $V^*$ prefers $c$ to every candidate in $S^*$. 
    
    Take any $s\in S^*$. Since $S=\bigcup_{i\in I^*_p}\topp_{k_i}(C_i)\subseteq \mathcal{G}^*_p$, we have $s\in\mathcal{G}^*_p$, and $c$ beats $s$ for at least $np$ voters, implying an edge $(c,s)$ in $G$. If $c\notin\mathcal{G}^*_p$, this is an edge from outside $\mathcal{G}^*_p$ into it, contradicting $\mathcal{G}^*_p$ having an in-degree of zero in $G'$. Hence $c\in \mathcal{G}^*_p$.

    By construction, $\abs{C_i\cap S}=k_i$ for $i\in I^*_p$ and $\abs{C_i\cap S}=0\le k_i$ otherwise, so Equation (\ref{eqn:topcc_cpl_1}) gives $\sum_{i\in\mathrm{under}_p(C_j)}\abs{C_i\cap S}\le(1-r)k$ for all $j\in I^*_p$. By Lemma \ref{lem:cpl_nodestab}, none of the lists $C_j$ with $j\in I^*_p$ can contain a $(p,r)$-destabilizer. Since $c \in \mathcal{G}^*_p$ and $\mathcal{G}^*_p$ is exactly the union of the lists $C_j$, $j \in I^*_p$, this contradicts $c$ being a $(p,r)$-destabilizer. Thus, there are no $(p,r)$-destabilizers at all, and so $S$ is $(p,r)$-stable.
\end{proof}


\subsection*{Proof of Theorem \ref{thm:cpp-r-bound}}

Before we present the proof of Theorem \ref{thm:cpp-r-bound}, we introduce additional helpful notation and technical results about linear systems. 

We denote the relation $\lel$ for vectors $\bm{x}$ and $\bm{y}$ as $\bm{x} \lel \bm{y} \iff x_i < y_i$ for all $i$. We similarly write $\bm{x}\leqdot \bm{y} \iff x_i \le y_i$ for all $i$. 

The first helpful result we use is (an equivalent reformulation of) Gordan's Theorem \cite{gordan1873}.

\begin{lem}[Gordan's Theorem, \cite{gordan1873}; We use an equivalent version courtesy of \citet{mangasarian1994} and \citet{perng2017}]\label{thm:Gordan}
     Let $A$ be a real matrix. Exactly one of the following is true:
    \begin{enumerate}
        \item The system $A\bm{x}\leqdot \bm{0}, A\bm{x}\neq \bm{0} , \bm{x}\geqdot \bm{0}$ has a solution  $\bm{x}$. 
        \item The system $A^\top \bm{y}\geqdot \bm{0}, \bm{y}\gel  \bm{0}$ has a solution  $\bm{y}$.
    \end{enumerate}
\end{lem}

Using this equivalent version of Gordan's Theorem \cite{gordan1873} allows us to prove the following result.

\begin{prop}\label{prop:wd_graph}
    Let $\mathcal{G}$ be a strongly connected tournament containing $\ell$ elements, and a set of edges $H$. Then, the following system of equations has a solution:
    \begin{align}
        \sum_{j:(i,j)\in H} r_j &\le \nicefrac{1}{2}  \ \text{ for all } i\in [\ell] \label{eqn:wd_graph_1}\\
        \sum_{i\in [\ell]} r_i &= 1 \label{eqn:wd_graph_2}
    \end{align} 
\end{prop}

\begin{proof}
    Let $A$ be the adjacency matrix of the tournament. We can denote the length $\ell$ vector whose components are all $\nicefrac{1}{2}$ as $\frac{1}{2}\bm{1}_{\ell\times\ell }\bm{r}$ by Equation (\ref{eqn:wd_graph_2}).

    Then the system of equations in (\ref{eqn:wd_graph_1}) is equivalent to $A\bm{r}\leqdot \frac{1}{2}\bm{1}_{\ell\times\ell }\bm{r}$, which can further be rewritten as $(A-\frac{1}{2}\bm{1}_{\ell\times \ell})\bm{r}\leqdot \bm{0}$. Let $M=(A-\frac{1}{2}\bm{1}_{\ell\times \ell})$. By Gordan's Theorem (Lemma \ref{thm:Gordan}), one of the following is true: either (i) there is a vector $\bm{x}\geqdot \bm{0}$ for which $M\bm{x}\leqdot \bm{0}$ and $M\bm{x}\ne \bm{0}$, or (ii) there is a vector $\bm{y}\gel \bm{0}$ where $M^\top \bm{y}\geqdot \bm{0}$. 

    We will show that (i) must be true. By way of contradiction, suppose (ii) is true. Since $A$ is an adjacency matrix of a tournament, when $A_{ij}=1$, $A_{ji}=0$, and $A_{ii}=0$, so $M_{ij}+M_{ji}=(1-\nicefrac{1}{2})+(-\nicefrac{1}{2})=0$, so we can write $M=S-\nicefrac{1}{2} I$ where $S$ is skew-symmetric. By assumption, $M^\top \bm{y}\geqdot \bm{0}$, so we have 
    \begin{align*}
        & (S-\nicefrac{1}{2} I)^\top \bm{y} \geqdot \bm{0} \\
        \implies & 2S^T\bm{y} \geqdot \bm{y} \\
        \implies & 2\bm{y}^\top S^\top \bm{y} \geqdot  \bm{y}^\top \bm{y}  \\
        \implies &  -2\bm{y}^\top S\bm{y} > 0
    \end{align*}
where the last implication uses the fact that $S$ is skew-symmetric and that $\bm{y} \gel 0$. But, $\bm{y}^\top S\bm{y}$ is a scalar, and in particular $\bm{y}^\top S\bm{y}=(\bm{y}^\top S\bm{y})^\top =\bm{y}^\top S^T\bm{y}=-\bm{y}^\top S\bm{y}=0$, so we have a contradiction.

Hence (i) is true, so we can take an arbitrary solution $\bm{x}\geqdot 0$ to $M\bm{x}\leqdot \bm{0}$ with $M\bm{x}\neq \bm{0}$. Since $M\bm{x}\neq \bm{0}$, $\bm{x}\ne \bm{0}$, and since $\bm x \geqdot \bm 0$, this gives $\bm 1_\ell \cdot \bm x > 0$. Hence, we can let $\bm{r}=\frac{1}{\bm{1}_\ell \cdot\bm{x}}\bm{x}$. Then $M\bm{r}\leqdot \bm{0}$ will still hold (since $M\bm r = \frac{1}{\bm 1_\ell\cdot\bm x}M\bm x \leqdot \bm 0$ because $\frac{1}{\bm 1_\ell\cdot\bm x}>0$) with $\bm{1}\cdot\bm{r}=1$, so we have found our vector $\bm{r}$.
\end{proof}

Next, we derive two additional technical lemmas.

\begin{lem}
\label{lem:adj_lip}
    Let $A\in \set{0,1}^{\ell\times \ell}$. Then the linear map of $A$ from metric space $(\R^\ell, d_{1})$ to $(\R^\ell, d_{\infty})$ is Lipschitz continuous with constant 1, where $d_1$ and $d_\infty$ are the metrics induced by the 1- and $\infty$-norms respectively.
\end{lem}
\begin{proof}
    Let $A_i$ be the $i^\text{th}$ row of $A$.
    For any $\bm{x},\bm{y}\in \R^\ell$,
    \begin{align*}
        \norm{A\bm{x}-A\bm{y}}_{\infty} &= \norm{A\paren{\bm{x}-\bm{y}}}_{\infty} \\
        &= \max_{i \in [\ell]}\left \{|A_i\paren{\bm{x}-\bm{y}}| \right\} \\
        &\le \max_{i \in [\ell]}\left \{\norm{A_i}_{\infty}\norm{\bm{x}-\bm{y}}_{1} \right\}\\
        &\le \norm{\bm{x}-\bm{y}}_{1}
    \end{align*}
    where the first inequality follows by H\"{o}lder's inequality, and the last inequality follows because $\norm{A_i}_{\infty} \le 1$ for all $i \in [\ell]$.
\end{proof}
\begin{lem}\label{lem:vector_rounding}
    Let $\bm{x}\in [0,1)^\ell$ with $\norm{\bm{x}}_1 = t\in \Z_{>0}$. Let, then, $\bm{z}\in \set{0,1}^\ell$ where $z_i=1$ if $x_i$ is one of the $t$-highest values (so $\norm{\bm{z}}_1=t$). Then $\norm{\bm{x}-\bm{z}}_1\le\nicefrac{\ell}{2}$.
\end{lem}
\begin{proof}
    Let $\bm{x}^*$ be the vector $\bm{x}$ but with its values arranged in decreasing order, and $\bm{z}^*$ the result of performing this rounding operation on $\bm{x}^*$. Clearly, $\norm{\bm{x}^*-\bm{z}^*}_1=\norm{\bm{x}-\bm{z}}_1$. Then,

    \begin{align*}
        \norm{\bm{x}^* - \bm{z}^*}_1 &= \sum_{i\le t}(1-x_i) + \sum_{i>t} x_i \\
        &= t-\sum_{i\le t} x_i + \sum_{i>t}x_i \\
        &= t-\paren{t-\sum_{i>t} x_i}+\sum_{i>t} x_i \\
        &= 2 \sum_{i>t}x_i
    \end{align*}

 Now, suppose for contradiction that the average value of $x_i$ for $i>t$ exceeds $\nicefrac{t}{\ell}$. Since $\bm x^*$ is sorted in decreasing order, every $x_i$ with $i \le t$ is at least as large as every $x_j$ with $j > t$, so the average of $x_i$ for $i \le t$ is also greater than $\nicefrac{t}{\ell}$. But then the overall average of all $\ell$ entries -- a weighted combination of these two averages -- would also exceed $\nicefrac{t}{\ell}$, contradiction to the fact that the overall average is exactly $\nicefrac{t}{\ell}$ (since $\norm{\bm x}_1 = t$). 
 
 Hence the average of $x_i$ for $i>t$ is at most $\nicefrac{t}{\ell}$, so $\sum_{i>t} x_i \le (\ell-t)\cdot\nicefrac{t}{\ell}$. So,
    \begin{gather*}
         \norm{\bm{x}^* - \bm{z}^*}_1
         = 2 \sum_{i>t}x_i \le 2(\ell - t)\nicefrac{t}{\ell} = 2t-2\nicefrac{t^2}{\ell}\le \nicefrac{\ell}{2}
    \end{gather*}
    where the last inequality follows because the maximum value of the quadratic is $\nicefrac{\ell}{2}$.
\end{proof}


With these technical tools in hand, we now prove Theorem \ref{thm:cpp-r-bound}.

\begin{proof} (Of Theorem \ref{thm:cpp-r-bound}) Consider the majority graph of $E$, $G^{(1/2)}$, and let $\mathcal{G}^*$ be the (unique) top connected component of $G^{(1/2)}$. By Lemma \ref{lem:cpl_cc_1}, $\mathcal{G}^*$ decomposes into $\ell$ entire party lists $C_{i_1},\dots,C_{i_\ell}$. Since $p=\nicefrac12$, every pair of distinct candidates has an edge in $G^{(1/2)}$, and since $E$ is a complete party list election, every pair of candidates drawn from two different lists among $C_{i_1},\dots,C_{i_\ell}$ shares the same edge direction (as in the proof of Lemma \ref{lem:cpl_cc_1}). Hence these $\ell$ lists induce a well-defined tournament $\mathcal{T}$ with $\ell$ vertices, one per list, where $(C_{i_a}, C_{i_b}) \in \mathcal{T}$ if and only if every candidate of $C_{i_a}$ beats every candidate of $C_{i_b}$. Moreover $\mathcal{T}$ is strongly connected, since $\mathcal{G}^*$ is. By Proposition \ref{prop:wd_graph} applied to $\mathcal{T}$, there is a solution $\bm{r}=(r_1,\dots,r_\ell)$ to $A\bm{r}\leqdot \frac{1}{2}$ with $\bm{1}_\ell \cdot \bm{r} = 1$, where $A$ is the adjacency matrix of $\mathcal{T}$.

    Our aim will be to produce a vector of non-negative integers $\bm{k}=(k_1,\dots, k_\ell)$ where $\frac{1}{k}A\bm{k}\leqdot \nicefrac{1}{2}+\nicefrac{\ell}{2k}$ and $\sum_i k_i = k$; our committee will consist of the top $k_i$ members of each $C_{i_a}$. By Lemma \ref{lem:topcc_cpl}, this set is $(\nicefrac{1}{2},\nicefrac{1}{2}-\nicefrac{\ell}{2k} )$-stable, since $1-(\nicefrac{1}{2}+\nicefrac{\ell}{2k})=\nicefrac{1}{2}-\nicefrac{\ell}{2k}$.

    Let $\bm{r}=\bm{f}+\bm{e}$ where $f_i=\frac{1}{k}\floor{kr_i}$, and $e_i=r_i-f_i$. We will then produce a vector $\bm{b}$ where $kb_i\in \Z_{\ge 0}$ and $\bm{e}\cdot \bm{1}=\bm{b}\cdot \bm{1}$; we will let our vector $\bm{k}=k(\bm{f}+\bm{b})$. If we can show that there is a vector $\bm{b}$ where $\norm{A(\bm{f}+\bm{b})}_\infty\le\nicefrac{1}{2}+\nicefrac{\ell}{2k}$, we're done.
    
    By the triangle inequality and Lemma \ref{lem:adj_lip}, we have 
    $$
    \norm{A(\bm{f}+\bm{b})}_\infty= \norm{A(\bm{r}-\bm{e}+\bm{b})}_\infty\le \norm{A\bm{r}}_\infty+ \norm{A(\bm{b}-\bm{e})}_\infty\le \frac{1}{2}+ \norm{\bm{b}-\bm{e}}_1 
    $$ 
    
    Thus, we need to find a $\bm{b}$ where $\norm{\bm{b}-\bm{e}}_1\le \nicefrac{\ell}{2k}$, or where $\norm{k\bm{b}-k\bm{e}}_1\le \nicefrac{\ell}{2}$. Let $t=\norm{k\bm{e}}_1$. We produce $k\bm{b}$ by mapping the highest $t$ values of $k\bm{e}$ to $1$, and mapping the rest to $0$, making $\bm{e}\cdot \bm{1}=\bm{b}\cdot \bm{1}$ and $\bm{b}\in\frac{1}{k} \Z^\ell_{\ge 0}$. By Lemma \ref{lem:vector_rounding}, $\norm{k\bm{b}-k\bm{e}}_1\le \frac{\ell}{2} $, as desired.
 
    Hence, we have shown that $(\nicefrac{1}{2}, r)$-stability is realizable for $r = \nicefrac{1}{2}-\nicefrac{\ell}{2k}$. By movement, it follows that $(\nicefrac{1}{2}, r)$-stability is realizable for all $r \le \nicefrac{1}{2}-\nicefrac{\ell}{2k} $.
\end{proof}

\section{Proofs for Section \ref{sec:majorityjackpot}}


\subsection*{Proof of Theorem \ref{thm:mj_majoritycoal}}

\begin{proof} We prove each part in turn.

\medskip

\noindent \textbf{(Proof of Part 1.)} We proceed by the contrapositive and assume that $\Rule$ does not satisfy $(\nicefrac{1}{2},r)$-coalitions for any $r>\nicefrac{1}{2}$. Then $\Rule$ must not satisfy  $(\nicefrac{1}{2},r)$-coalitions for the specific value of 
    $$
    r = \frac{\floor{\frac{k}{2}}+1}{k}
    $$
    Note that this specific value of $r$ is indeed greater than $\nicefrac{1}{2}$. By algebraic manipulation, $rk = \floor{\frac{k}{2}}+1$. Since the right-hand side is an integer, so too is the left-hand side, so $rk = \ceil{rk}$. Hence,
    $$
    \ceil{rk} = \floor{\frac{k}{2}}+1
    $$
    Since $\Rule$ fails $(\nicefrac{1}{2},r)$-coalitions, there is a complete party list election $E$ for which a group of voters $V^*$ with $\abs{V^*}\ge \nicefrac{n}{2}$ and  (denoting $\topp_{\ceil{rk}} V^*$ as $S^*$) $\abs{S^*}=\ceil{rk}$, but $S^*\not\subseteq \Rule(E,k)$. Since $n$ is odd by assumption, $|V^*| \ge \nicefrac{n}{2}$ implies $|V^*| > \nicefrac{n}{2}$.

    Since $E$ is a complete party list election, any two members of $S^*$ appear in the top $k$ positions of some voter's ranking, so they belong to the same list $C_i$. Moreover, since every voter in $V^*$ prefers all of $S^*$ over everything outside $S^*$, and complete party lists compare entire lists as wholes, each such voter must in fact prefer all of $C_i$ over every other list. Hence, each such voter has $C_i$ entirely at the top of their ranking, in $C_i$'s fixed canonical order. It follows that $S^*$ is exactly the first $\floor{\nicefrac{k}{2}}+1$ candidates of $C_i$'s canonical order.

    Let $C_\mathrm{maj} = S^\ast$. Since $|V^*|>\nicefrac{n}{2}$ voters prefer all of $C_\mathrm{maj}$ over every candidate outside it, and $|C_\mathrm{maj}|=\floor{\nicefrac{k}{2}}+1$, $\psi$ holds for $E$ with this witness. Since $\psi\implies\chi$, $\chi$ holds for $E$ as well. As $\Rule$ is a majority jackpot rule conditional on $\chi$, $\nu(E)\ne\emptyset$, and by hypothesis, $C_\mathrm{maj}\subseteq C_{\nu(E)}$. Since $C_\mathrm{maj}\subseteq C_i$ and party lists are pairwise disjoint, $C_{\nu(E)}$ (itself an entire list) must equal $C_i$. That is, $\nu(E)=i$. By definition of majority jackpot, since $\nu(E)\ne\emptyset$, $\abs{C_i\cap\Rule(E,k)}>\nicefrac{k}{2}$. 
    
    Towards a contradiction, suppose $\Rule$ is candidate-wise undominated. Since every voter ranks $C_i$'s members in the same fixed canonical order, for $c_{i,j}, c_{i,j'}\in C_i$ with $j<j'$, every voter prefers $c_{i,j}$ to $c_{i,j'}$. Hence, $c_{i,j}$ dominates $c_{i,j'}$. Candidate-wise undominatedness then forces $C_i\cap\Rule(E,k)$ to be downward-closed under canonical order. That is, if $c_{i,j}\in\Rule(E,k)$, every $c_{i,j'}$ with $j'<j$ (each of which dominates $c_{i,j}$) must also be in $\Rule(E,k)$. So $C_i\cap\Rule(E,k)$ is exactly the top-$t$ values of $C_i$ in canonical order, where $t=\abs{C_i\cap\Rule(E,k)}>\nicefrac{k}{2}$. Since $t$ is an integer, $t\ge\floor{\nicefrac{k}{2}}+1$.

     Hence $C_i\cap\Rule(E,k)$ must contain (at least) the top $\floor{\nicefrac{k}{2}}+1$ candidates of $C_i$, which includes $C_\mathrm{maj}$ itself. But $C_\mathrm{maj}=S^*$, so we then have $S^*\subseteq\Rule(E,k)$, contradicting our assumption that $S^*\not\subseteq\Rule(E,k)$. 

    \medskip
    
    \noindent \textbf{(Proof of Part 2.)} Let $E$ be an arbitrary party list election. Suppose $\Rule$ satisfies $(\nicefrac{1}{2},r)$-coalitions for some $r>\nicefrac{1}{2}$. It is sufficient to show that when $\psi$ holds, we have $\abs{C_\mathrm{maj} \cap \Rule(E,k)}> \nicefrac{k}{2}$. This is because all the elements of $C_\mathrm{maj}$ are on the same list $C_i$ (as we show below), so if $\nu(E)=i$ we will have half of the committee consisting of a single list, fulfilling Definition \ref{def:mj_cpl}.

    First we show that all the elements of $C_\mathrm{maj}$ need to be on the same list $C_i$. Let $v$ be any voter witnessing $\psi$ (i.e., $c\succ_v d$ for all $c\in C_\mathrm{maj}$ and $d\in C\setminus C_\mathrm{maj}$). Note that such a $v$ exists, since there are more than $\nicefrac{n}{2}$ of such voters witnessing the property of $\psi$. We claim that $\topp(v) \in C_\mathrm{maj}$ -- otherwise, if $\topp(v) \notin C_\mathrm{maj}$, then since $C_\mathrm{maj}$ is nonempty, some $c\in C_\mathrm{maj}$ would satisfy $c \succ_v \topp(v)$, contradicting that $\topp(v)$ is $v$'s most preferred candidate. Let $C_i$ be the list containing $\topp(v)$. Since $E$ is a party list election and $\topp(v) \in C_i$, $v$'s entire top-$k$ block is $C_i$ in canonical order.
    
    We next show $C_\mathrm{maj} \subseteq C_i$. Suppose for contradiction some $c^* \in C_\mathrm{maj}$ satisfies $c^* \notin C_i$. Since $c^*$ is a member of $C_\mathrm{maj}$ lying outside $C_i$, $\abs{C_\mathrm{maj} \cap C_i} \le \abs{C_\mathrm{maj}} - 1 = \floor{\nicefrac{k}{2}}$. Since $\abs{C_i} = k$, it follows that $\abs{C_i \setminus C_\mathrm{maj}} = \abs{C_i} - \abs{C_i \cap C_\mathrm{maj}} \ge k - \floor{\nicefrac{k}{2}} > 0$, so some $b \in C_i \setminus C_\mathrm{maj}$ exists. Since $v$'s entire top-$k$ block is $C_i$ and $c^* \notin C_i$, $b$ must be ranked by $v$ above $c^*$, i.e., $b \succ_v c^*$. But this yields a contradiction, as $b \notin C_\mathrm{maj}$ and $c^* \in C_\mathrm{maj}$, and the defining property of $\psi$ gives $c^* \succ_v b$. Therefore we must have $C_\mathrm{maj} \subseteq C_i$.

    Since $C_\mathrm{maj} \subseteq C_i$ and every $d \in C_i \setminus C_\mathrm{maj}$ satisfies $c \succ_v d$ for all $c \in C_\mathrm{maj}$ (by the defining property of $\psi$, as $d \notin C_\mathrm{maj}$), every element of $C_\mathrm{maj}$ is ranked by $v$ above every element of $C_i \setminus C_\mathrm{maj}$. Since $v$ ranks $C_i$ in its fixed canonical order, the candidates in $C_\mathrm{maj}$ therefore occupy exactly the top $\floor{\nicefrac{k}{2}}+1$ spots of $C_i$'s canonical order.

    Note that the set $C_i$ is associated to the voter $v$ selected above. We now show every voter witnessing $\psi$ has this same $C_i$ as their entire top-$k$ block. Since $C_\mathrm{maj}$ is a fixed set of candidates (not depending on any particular voter) and we established $C_\mathrm{maj} \subseteq C_i$ above, the same argument as above applies verbatim to any other voter $\hat v$ witnessing $\psi$ to conclude $\topp(\hat v) \in C_\mathrm{maj} \subseteq C_i$. Because this is a party list election, $\hat v$'s entire top-$k$ block is therefore $C_i$ as well, in $C_i$'s canonical order. In particular, every voter witnessing $\psi$ agrees on their top-$\ceil{rk}$ block, namely the first $\ceil{rk}$ candidates of $C_i$ in canonical order.

    Let $V^*$ denote the (more than $\nicefrac{n}{2}$) voters witnessing $\psi$. By the previous paragraph, they share the same top-$\ceil{rk}$ candidates, namely the top $\ceil{rk}$ candidates of $C_i$, implying $V^* \in \mathcal{V}$. Since all members of $\mathcal{V}$ are disjoint by construction and of size at least $\nicefrac{n}{2}$, $\mathcal{V} = \set{V^*}$ -- otherwise, if there exists some other $W \in \mathcal{V}$ then $|W| \ge \nicefrac{n}{2}$ and  $V^* \cup W \subseteq V$, yet $n = |V| \ge |V^*| + |W| > \nicefrac{n}{2} + \nicefrac{n}{2} = n$, yielding a contradiction.

    Hence, $(\nicefrac{1}{2}, r)$-coalitions is realizable for this election $E$ by Corollary \ref{cor:coalitions_checker}. Since $\Rule$ satisfies $(\nicefrac{1}{2},r)$-coalitions and this property is realizable on $E$, this shared top-$\ceil{rk}$ block must be included in the committee. Since $r>\nicefrac{1}{2}$, $\ceil{rk} \ge rk > \nicefrac{k}{2}$, and since $\ceil{rk}$ is an integer strictly larger than $\nicefrac{k}{2}$, it must be at least the smallest integer exceeding $\nicefrac{k}{2}$. Hence, $\ceil{rk} \ge \floor{\nicefrac{k}{2}}+1$. Thus, the shared top-$\ceil{rk}$ block has size at least $\floor{\nicefrac{k}{2}}+1$, and by the first part of the argument its first $\floor{\nicefrac{k}{2}}+1$ candidates are exactly $C_\mathrm{maj}$. So the intersection $C_\mathrm{maj}\cap\Rule(E,k)$ has size at least $\floor{\nicefrac{k}{2}}+1 > \nicefrac{k}{2}$.
\end{proof}


\subsection*{Proof of Proposition \ref{prop:mpms_cwp_nfmr_chain}}

\begin{proof}

($\mathrm{MPMS}_\mathrm{CPL} \implies \mathrm{CWP}_\mathrm{CPL}$) Suppose $\mathrm{MPMS}_\mathrm{CPL}$ holds for committee $S$. If no weak Condorcet winner exists among the party lists, then the implication $\mathrm{CWP}_\mathrm{CPL}$ holds trivially. Otherwise, let $C_w$ be a weak Condorcet winner and let $C_i \ne C_w$ be arbitrary. By definition of a weak Condorcet winner, at least $\nicefrac{n}{2}$ voters prefer $C_w$ to $C_i$. By $\mathrm{MPMS}_\mathrm{CPL}$, we then have $\abs{C_w \cap S} \ge \abs{C_i \cap S}$,  so $\mathrm{CWP}_\mathrm{CPL}$ holds.

\medskip

\noindent ($\mathrm{CWP}_\mathrm{CPL} \implies \mathrm{NFMR}_\mathrm{CPL}$) Suppose $\mathrm{CWP}_\mathrm{CPL}$ holds for committee $S$. We consider two cases: either a weak Condorcet winner exists, or not. We will show that in either case, $\mathrm{NFMR}_\mathrm{CPL}$ holds whenever $\mathrm{CWP}_\mathrm{CPL}$ holds. 

\medskip

\noindent \textbf{Case 1:} Suppose a weak Condorcet winner $C_w$ exists among the party lists. Let $C_i \ne C_w$ be any other party list. Since $\mathrm{CWP}_\mathrm{CPL}$ holds, $\abs{C_w \cap S} \ge \abs{C_i \cap S}$. And by definition of weak Condorcet winner, at least $\nicefrac{n}{2}$ voters prefer $C_w$ to $C_i$. Hence the pair $(C_w, C_i)$ witnesses the criteria of $\mathrm{NFMR}_\mathrm{CPL}$.

\medskip

\noindent \textbf{Case 2:} Suppose no weak Condorcet winner exists. Then $h \ge 3$. To see why, note that when $h = 1$ there is only one party, and this party is a weak Condorcet winner, contradicting the assumption of Case 2. Additionally when $h = 2$, the two vote counts between the two parties must sum to exactly $n$, so at least one party must have at least $\nicefrac{n}{2}$ of the votes in the majority preference graph. But then that party is a weak Condorcet winner, also contradicting the assumption of Case 2.

Since no weak Condorcet winner exists among the $h\ge3$ party lists, the majority party preference graph $G_\mathrm{CPL}$ among them must contain a directed cycle $D_1 \to \cdots \to D_b \to D_1$ for some $b \ge 2$ (with distinct $D_1, \dots, D_b \in \set{C_1,\dots,C_h}$). This follows because if no such cycle existed, a topological sort of $G_\mathrm{CPL}$ would place some party $D$ ahead of all others. But this party $D$ would then be a weak Condorcet winner, contradicting the assumption of Case 2. 

Suppose for contradiction that $\mathrm{NFMR}_\mathrm{CPL}$ is not witnessed by any consecutive pair in this cycle. Then for each $t \in [b]$ (which we use as indices mod $b$), since a weak majority prefers $D_t$ to $D_{t+1}$ but this pair does not witness $\mathrm{NFMR}_\mathrm{CPL}$, we must have $\abs{D_t \cap S} < \abs{D_{t+1} \cap S}$. Chaining these strict inequalities around the cycle gives
    $$
    \abs{D_1 \cap S} < \cdots < \abs{D_b \cap S} < \abs{D_1 \cap S}
    $$
    yielding a contradiction. Hence some consecutive pair $(D_t, D_{t+1})$ in the cycle satisfies $\abs{D_t \cap S} \ge \abs{D_{t+1} \cap S}$. Coupled with the fact that a weak majority prefers $D_t$ to $D_{t+1}$, this pair witnesses $\mathrm{NFMR}_\mathrm{CPL}$.
\end{proof}


\subsection*{Proof of Proposition \ref{prop:cwp_stability}}

To facilitate the proof of Proposition \ref{prop:cwp_stability}, we introduce two concepts. Given a committee $S$ and $h$ parties $C_1,\dots,C_h$, consider $|C_1\cap S|, \dots, |C_h \cap S|$. We say the \emph{maximal size of a party in $S$} is $P_{\max} = \max_{i \in [h]} |C_i \cap S|$ and the \emph{minimal size of a party in $S$} is $P_{\min} = \min_{i \in [h]} |C_i \cap S|$. The following combinatorial lemma places constraints on $P_{\min}$ and $P_{\max}$.

\begin{lem}\label{lem:mpms}
    $P_{\min} \le \frac{k}{h} \le P_{\max}$ 
\end{lem}

\begin{proof}
    By way of contradiction, suppose $P_{\max} < \nicefrac{k}{h}$. By definition of $P_{\max} $, all $|C_i \cap S| < \nicefrac{k}{h}$ for $i \in [h]$. Since all party lists are disjoint, we have
    $$
    k = |S| =  \sum_{i \in [h]} |C_i \cap S| < h \times \frac{k}{h} = k
    $$
    yielding a contradiction. Similarly, to show the upper bound on $P_{\min}$, we proceed by contradiction again and suppose $P_{\min} > \nicefrac{k}{h}$. By definition of $P_{\min}$, all $|C_i \cap S| > \nicefrac{k}{h}$ for $i \in [h]$. But then
    $$
    k = |S| = \sum_{i \in [h]} |C_i \cap S| > h \times \frac{k}{h} = k 
    $$
    also yielding a contradiction.  
\end{proof}

\noindent We now prove Proposition \ref{prop:cwp_stability}.

\begin{proof} (Of Proposition \ref{prop:cwp_stability})
    We proceed by the contrapositive, and will show that if a committee $S$ violates $\mathrm{CWP}_\mathrm{CPL}$, then $S$ also violates $(\nicefrac{1}{2},1-\nicefrac{1}{h})$-stability.

    If $S$ violates $\mathrm{CWP}_\mathrm{CPL}$, there is a weak Condorcet winner $C_w$ among the party lists of $E$ and a party list index $\ell\ne w$ for which $\abs{C_\ell \cap S} > \abs{C_w\cap S}$. Fix such $C_w$ and $\ell$. Let $i \in \argmax_{j\in[h]} \abs{C_j\cap S}$, so $\abs{C_i\cap S} = P_{\max} \ge \abs{C_\ell \cap S} > \abs{C_w \cap S}$. In particular, this implies $C_i \ne C_w$.

    We claim $P_{\max} > \nicefrac{k}{h}$. By Lemma \ref{lem:mpms}, $P_{\max}\ge\nicefrac{k}{h}$, so assume for contradiction that $P_{\max}=\nicefrac{k}{h}$. Since $P_{\min}\le\nicefrac{k}{h}\le P_{\max} = \nicefrac{k}{h}$, every party's seat count lies in $[P_{\min}, \nicefrac{k}{h}]$. However, the sum of the seats across the $h$ parties must sum to $k = h\cdot\nicefrac{k}{h}$, so none of the parties' seats can fall strictly below $\nicefrac{k}{h}$ (or the total would fall short of $k$). Thus, $P_{\min}=\nicefrac{k}{h}$ as well. But then $\abs{C_w\cap S}=P_{\max}\ge\abs{C_i\cap S}$, contradicting $\abs{C_i\cap S}>\abs{C_w\cap S}$. Hence $P_{\max}>\nicefrac{k}{h}$, which means $\abs{C_i\cap S}>\nicefrac{k}{h}$.

    Let $c\in C_w \setminus S$. Since $C_w$ is a weak Condorcet winner among the party lists, and $E$ is a complete party list election, there is a weak majority of voters who prefer $c$ to all candidates in $C_i$. We then pick our $V^*$ to be this set of voters, and our $S^*$ to be $C_i\cap S$. Since $|S^*| = |C_i \cap S| > \frac{k}{h} = k\left(\frac{1}{h}\right)$, when $r=1-\nicefrac{1}{h}$ we then have $|S^*| > k(1-r)$. Thus $S$ violates $(\nicefrac{1}{2},1-\nicefrac{1}{h})$-stability.

    This completes the contrapositive, establishing that if $S$ satisfies $(\nicefrac{1}{2},1-\nicefrac{1}{h})$-stability, then it also satisfies $\mathrm{CWP}_\mathrm{CPL}$. The last part of the proposition's claim follows as $\mathrm{CWP}_\mathrm{CPL}$ implies $\mathrm{NFMR}_\mathrm{CPL}$.
\end{proof}


\subsection*{Proof of Lemma \ref{lem:clist_gehr}}

\begin{proof}

Let $G_\ell$ be the top $\ell$ candidates of $C_w$ in the canonical order of the party list. We will show that $G_\ell$ meets the criteria of strong Gehrlein stability (as specified in Definition \ref{defn:strong_g_stability}). That is, we will show for every $g \in G_\ell$ and $g' \in C \setminus G_\ell$, $g$ beats $g'$ in the pairwise election: $\abs{\set{v\in V: g\succ_v g'}}>\nicefrac{n}{2}$. 

By definition, $G_\ell \subseteq C_w$, so  $C \setminus G_\ell = (C_w \setminus G_\ell) \cup (C \setminus C_w)$. Hence, we proceed in two cases: when $g' \in C_w \setminus G_\ell$ and when $g' \notin C_w$.

For the first case, suppose $g' \in C_w \setminus G_\ell$. Since $E$ is a complete party list election, every voter ranks $C_w$'s candidates in this same canonical order, so \emph{every} voter prefers $g$ to $g'$. We thus have $\abs{\set{v\in V: g\succ_v g'}}=n>\nicefrac{n}{2}$.

For the second case, suppose $g' \notin C_w$. Let $C_i$ be the list containing $g'$, so $i \ne w$. Since $C_w$ is a weak Condorcet winner among parties, $\abs{\set{v\in V:\forall c\in C_w \ \forall d\in C_i, \ c\succ_v d}} \ge \nicefrac{n}{2}$. Taking $c = g$ and $d = g'$, this gives $\abs{\set{v\in V: g \succ_v g'}}\ge \nicefrac{n}{2}$. By assumption, the majority party preference graph $G_\mathrm{CPL}$ of $E$ is a tournament, so $\abs{\set{v\in V: g \succ_v g'}}\ne \nicefrac{n}{2}$. This implies  $\abs{\set{v\in V: g \succ_v g'}}> \nicefrac{n}{2}$.

Therefore in either case, $\abs{\set{v\in V: g\succ_v g'}}>\nicefrac{n}{2}$, so $G_\ell$ is strongly Gehrlein stable.

\end{proof}


\subsection*{Proof of Theorem \ref{thm:cwp_majjackpot_selection_function}}

The following definition will be helpful to prove Theorem \ref{thm:cwp_majjackpot_selection_function}. It is the analog of Definition \ref{defn:top_cc_preference_graph_candidates} but applied to the preference graph over parties instead of candidates.

\begin{defn}
    Consider a complete party list election whose majority party preference graph $G_\mathrm{CPL}$ is a tournament, and let $G'_\mathrm{CPL}$ be its condensation, with vertices representing the strongly connected components of $G_\mathrm{CPL}$. Since $G_\mathrm{CPL}$ is a tournament, $G'_\mathrm{CPL}$ is a transitive tournament (by a similar argument to the one used in the proof of Proposition \ref{prop:bottom_line_pnp_1}), and hence has a unique component with in-degree zero in its total ordering. We define $\mathcal{G}^*$ to be this component, and refer to it as the \emph{top strongly connected component} of $G_\mathrm{CPL}$.
\end{defn}

\begin{proof} (Of Theorem \ref{thm:cwp_majjackpot_selection_function}) We prove each part in turn.

\medskip

\noindent \textbf{(Proof of Part 1.)} We prove the claim in two steps. In the first step, we show that if $\Rule$ satisfies $(\nicefrac{1}{2},\nicefrac{1}{2})$-stability, then $\nu$ selects the weak Condorcet winning party when it exists. We proceed by the contrapositive. Suppose a weak Condorcet winning list $C_w$ exists yet $\nu(E)=i$ where $C_i\neq C_w$. Let $S$ denote the output of a majority jackpot rule $\Rule$ with winning determination function $\nu$. 
    
    Then $C_w\setminus S$ is non-empty. This follows as $\Rule$ being a majority jackpot rule gives $\abs{C_i \cap S} > \nicefrac{k}{2}$. This leaves at most $k - \abs{C_i \cap S} < \nicefrac{k}{2}$ seats for the remaining parties combined. Since $\abs{C_w} = k$, we have $\abs{C_w \setminus S} = \abs{C_w} - \abs{C_w \cap S} \ge k - \nicefrac{k}{2} = \nicefrac{k}{2} > 0$.

    Next, we let $c\in C_w\setminus S$ and $S^* = S\cap C_i$. By definition of majority jackpot, $\abs{S^*}> \nicefrac{k}{2} = (1-\nicefrac{1}{2})k$. Since $C_w$ is a weak Condorcet winning party, there is a weak majority of voters $V^*$ who prefer $c$ to every candidate in $C_i$, and in particular to every candidate in $S^*$, and thus we violate $(\nicefrac{1}{2}, \nicefrac{1}{2})$-stability.

    In the second step, we establish that $\nu(E)=\emptyset$ when no weak Condorcet winning party exists whenever $\Rule$ satisfies $(\nicefrac{1}{2},\nicefrac{1}{2})$-stability. Consider $\mathcal{G}^*$, the top strongly connected component of the majority party preference graph $G_\mathrm{CPL}$ of the election $E$. Then $\mathcal{G}^*$ must contain at least two parties. To see why, suppose to the contrary that $\mathcal{G}^* = \set{C^*}$ is a singleton for some party $C^* \in \set{C_1,\dots,C_h}$. Since $\mathcal{G}^*$ has in-degree zero, no other party beats $C^*$. Since $G_\mathrm{CPL}$ is a tournament, this means $C^*$ beats every other party directly, making $C^*$ a Condorcet winner, yielding a contradiction.

    Since the range of $\nu$ is $\set{\emptyset} \cup [h]$, there are three possible outcomes for $\nu(E)$: either $\nu(E)$ selects (i) a party outside of $\mathcal{G}^*$, (ii) a party inside of $\mathcal{G}^*$, or (iii) outputs $\emptyset$. We show the first two outcomes are impossible.

    If $\nu$ chooses a party $C_i$ outside of $\mathcal{G}^*$, then since $\mathcal{G}^*$ is the top strongly connected component, every party in $\mathcal{G}^*$ beats $C_i$ directly. Take any $C_j\in\mathcal{G}^*$ and any $c\in C_j$. Then the same argument as in the first step of the proof above shows there is a weak majority of voters who prefer $c$ to every candidate in $S\cap C_i$, violating $(\nicefrac{1}{2},\nicefrac{1}{2})$-stability. Hence, (i) is impossible.

    Further, suppose we pick $\nu(E)=i$ with $C_i\in \mathcal{G}^*$. Since $\mathcal{G}^*$ is strongly connected with at least two nodes, some node in $\mathcal{G}^*$ must have a direct edge into $C_i$ in the tournament: otherwise $C_i$ would beat every other node in $\mathcal{G}^*$ directly, and since no edge from outside $\mathcal{G}^*$ points into it, $C_i$ could never be reached by a path from another node within $\mathcal{G}^*$, contradicting strong connectedness. Let $C_j \in \mathcal{G}^*$ (with $j \ne i$) be such a node, and let $c \in C_j$. Since $G_\mathrm{CPL}$ is a tournament and $C_j$ beats $C_i$, there is a weak majority of voters who prefer all the candidates in $C_j$ to $C_i$, so picking $S^* = C_i\cap S$, we violate $(\nicefrac{1}{2},\nicefrac{1}{2})$-stability. Hence, (ii) is impossible.
    
    Thus, (iii) is the only viable option left, so $\nu(E) = \emptyset$.
\medskip
    
\noindent \textbf{(Proof of Part 2.)} Let $C_w$ be the weak Condorcet winner. Suppose $\nu(E)=w$ and $\Rule$ is candidate-wise undominated. Since every voter ranks $C_w$'s candidates in the same canonical order (as $E$ is a complete party list election), a candidate ranked higher in this order dominates one ranked lower. Candidate-wise undominatedness of $\Rule$ then forces $S \cap C_w$ to be downward-closed in this canonical order -- that is, if some $c_{w,j} \in C_w$ is also in $S$, then every $c_{w,j'}\in C_w$ with $j' < j$ (each of which dominates $c_{w,j}$) must also be in $S$. Hence $S \cap C_w$ is exactly the top $\abs{S \cap C_w}$ candidates of $C_w$ in canonical order.

Applying Lemma \ref{lem:clist_gehr} with $\ell = \abs{S \cap C_w}$, it follows that $S \cap C_w$ is a strongly Gehrlein stable set of size $\ell$. Since $\nu(E)=w$, $\Rule$ being a majority jackpot rule yields $\abs{S \cap C_w} > \nicefrac{k}{2}$. This implies $\abs{S \cap C_w} \in [\nicefrac{k}{2},k]$, so invoking Proposition \ref{prop:ginc}, we conclude that $S$ is $(\nicefrac{1}{2},\nicefrac{1}{2})$-stable.    
\end{proof}


\subsection*{Proof of Corollary \ref{cor:nfmr_mjs}}

\begin{proof}
   Suppose a weak Condorcet winning party exists in a complete party list election $E$. Since $E$ exhibits a tournament structure in $G_{\mathrm{CPL}}$, there is a unique weak Condorcet winning party. Denote this party as $C_w$. 
    
     By Theorem \ref{thm:cwp_majjackpot_selection_function}, since $\Rule$ is a majority jackpot rule satisfying $(\nicefrac{1}{2},\nicefrac{1}{2})$-stability, its winning determination function $\nu$ selects $C_w$ (i.e., $\nu(E) = w$). By definition of a majority jackpot rule, this gives $\abs{C_w \cap \Rule(E,k)} > \nicefrac{k}{2}$.

    We claim this strict seat-majority forces $C_w$ to have at least as many seats as every other party. Let $C_i \ne C_w$ be any other party list. Since all party lists are disjoint and $\abs{\Rule(E,k)} = k$,
    $$
    \abs{C_i \cap \Rule(E,k)} \le k - \abs{C_w \cap \Rule(E,k)} < k - \frac{k}{2} = \frac{k}{2} < \abs{C_w \cap \Rule(E,k)}
    $$
    Hence $\abs{C_w \cap \Rule(E,k)} > \abs{C_i \cap \Rule(E,k)}$ for every $C_i \ne C_w$, so $\Rule$ satisfies $\mathrm{CWP}_\mathrm{CPL}$. Finally, since $\mathrm{CWP}_\mathrm{CPL} \implies \mathrm{NFMR}_\mathrm{CPL}$ by Proposition \ref{prop:mpms_cwp_nfmr_chain}, $\Rule$ also satisfies $\mathrm{NFMR}_\mathrm{CPL}$.
\end{proof}

\end{appendices}


\end{document}